\documentclass[11pt]{article}
\pdfoutput=1
\usepackage[a4paper,margin=1in]{geometry}

\usepackage{microtype}
\usepackage{mathtools}
\usepackage[linesnumbered,ruled,vlined]{algorithm2e}
\SetKwFunction{Match}{Match}
\SetKwComment{Comment}{$\triangleright$\ }{}
\usepackage{standalone}
\usepackage{tikz}
\usetikzlibrary{graphs, positioning, arrows.meta, fit, backgrounds, calc}
\usepackage{xspace}
\usepackage{todonotes}
\usepackage{xcolor}
\usepackage{amsmath,amssymb,amsthm}
\usepackage{graphicx}
\usepackage{thmtools,thm-restate}
\usepackage{hyperref}
\usepackage[capitalise,noabbrev]{cleveref}
\usepackage{coqdoc}
\hypersetup{hidelinks}
\hypersetup{hypertexnames=false}
\renewcommand{\coqlibrary}[3]{} 

\newcommand{\lv}[1]{\textcolor{blue}{#1}}
\newcommand{\lc}[1]{\textcolor{teal}{#1}}
\newcommand{\mcs}[1]{\textcolor{red}{#1}}
\newcommand{\lvtodo}[1]{\todo[color=blue!50]{#1}}
\newcommand{\lctodo}[1]{\todo[color=green!50]{#1}}
\def\lv#1{}\def\lc#1{}\def\mcs#1{}\def\lvtodo#1{}\def\lctodo#1{}
\RequirePackage[normalem]{ulem} 
\RequirePackage{color}\definecolor{RED}{rgb}{1,0,0}\definecolor{BLUE}{rgb}{0,0,1} 
\definecolor{maroon}{rgb}{0.5,0,0}
\newcommand{\fpass}[1]{#1} 

\newcommand{\N}{\mathbb{N}}

\newcommand{\R}{\mathbb{R}}
\declaretheorem[name=Theorem,numberwithin=section]{theorem}
\declaretheorem[name=Lemma,sibling=theorem]{lemma}
\declaretheorem[name=Corollary,sibling=theorem]{corollary}
\declaretheorem[name=Proposition,sibling=theorem]{proposition}
\declaretheorem[name=Observation,sibling=theorem]{observation}
\declaretheorem[style=definition,name=Definition,sibling=theorem]{definition}
\declaretheorem[style=remark,name=Remark,sibling=theorem]{remark}
\declaretheorem[style=remark,name=Claim,sibling=theorem]{claim}
\DeclareMathOperator{\depth}{\mathrm{depth}}
\DeclareMathOperator{\unfold}{\mathrm{Unfold}}
\newcommand{\depthK}{\depth_{\widetilde{K}}}

\DeclareMathOperator{\Low}{\mathrm{Low}}

\DeclareMathOperator{\id}{id}
\DeclareMathOperator{\CR}{CR}
\DeclareMathOperator{\CG}{CG}
\DeclareMathOperator{\inv}{Inv}
\DeclareMathOperator{\Can}{Can}
\newcommand{\invr}{\inv^{\mathrm{rooted}}}
\newcommand{\B}{\mathcal{B}}
\newcommand{\Bup}{B_{up}}
\newcommand{\facedepth}{\mathrm{facedepth}}
\newcommand{\ro}{\mathrm{root}}
\newcommand{\iso}{\cong_{col}}
\newcommand{\Kt}{\widetilde{K}}

\definecolor{cR}{RGB}{220,50,50}
\definecolor{cG}{rgb}{0.04, 0.85, 0.32}
\definecolor{cB}{RGB}{80,130,220}
\definecolor{cY}{rgb}{1.0, 0.84, 0.0}
\definecolor{cO}{RGB}{240,150,50}
\definecolor{cP}{RGB}{240,130,170}
\definecolor{cV}{RGB}{160,100,210}

\title{Color Refinement Distinguishes Connected Rooted Outerplanar Graphs}

\author{Louann Coste\thanks{PSL University, Inria Paris, \texttt{louann.coste@inria.fr}}
    \and Mónika Csikós\thanks{IRIF, CNRS and Université Paris Cité, \texttt{csikos@irif.fr}}
    \and Laurent Viennot\thanks{Inria Paris, DI ENS, \texttt{laurent.viennot@inria.fr}}}

\begin{document}
\maketitle

\begin{abstract}
Color refinement, also known as the $1$-dimensional Weisfeiler--Leman algorithm ($1$-WL), is a classical near-linear-time heuristic for graph isomorphism, which also captures the expressive power of message-passing graph neural networks. As it is incomplete, a long line of work asks on which graph classes $1$-WL, or its higher-dimensional variant $k$-WL, identifies every graph; the least such $k$ is the WL-dimension of the class. Planar graphs have WL-dimension at most $3$ (Kiefer, Ponomarenko and Schweitzer, 2019), and whether $2$-WL already distinguishes all planar graphs is a central open problem. We bring new insight on this question for outerplanar graphs, whose WL-dimension is~$2$.

We show that color refinement, i.e. 1-WL, distinguishes connected rooted outerplanar graphs: two such graphs that are not isomorphic receive different stable colorings as soon as a single vertex, the root, is individualized. The root is necessary, as there are non-isomorphic $2$-connected outerplanar graphs that color refinement does not distinguish. Our result extends to pseudo-rooted graphs, in which the stable coloring merely contains a singleton color class, and to non-connected graphs whose components all contain such a class. As a consequence, the $(1,1)$-WL test of Rattan and Seppelt (2023) distinguishes all outerplanar graphs.

Our proof proceeds by inversion: we reconstruct a connected rooted outerplanar graph, up to isomorphism, from its color graph, the weighted quotient of the graph by its stable color classes, together with its root color. The reconstruction runs in time linear in the size of the output and is canonical, which yields a near-linear-time canonization algorithm for connected rooted outerplanar graphs. The main structural tool is a dichotomy lemma, formalized in the Rocq proof assistant: in a rooted outerplanar graph with an equitable coloring, every block either receives pairwise distinct colors or admits a color-preserving mirror symmetry fixing its vertex closest to the root.
\end{abstract}


\medskip
\noindent{\small
\textbf{AI usage} The authors used LLM-based assistants in three ways. (i) Writing: polishing the English, checking the consistency of notation, definitions and proofs across sections, and reworking the figures. (ii) Proofs: the proof of \Cref{prop:chord_above} was simplified, shortening it from 8 to 5 pages; the result was checked by the authors. (iii) Formalization: the Rocq proof of the Dichotomy Lemma was developed with such assistance. The authors take full responsibility for the content of the paper.}

\clearpage
\section{Introduction}
\label{sec:intro}

Two finite simple graphs $G$ and $H$ are \emph{isomorphic} if there is a
bijection $\varphi\colon V(G)\to V(H)$ such that
$uv\in E(G)$ if and only if $\varphi(u)\varphi(v)\in E(H)$. One of the fundamental problems of algorithmic graph theory, the \emph{graph
isomorphism problem}, asks whether such a bijection exists between two given graphs. Understanding the
complexity of this problem has been a central question for more than five
decades: graph isomorphism belongs to NP, but it is not
known to be solvable in polynomial time or to be NP-complete. The best known
worst-case algorithm for general graphs runs in quasipolynomial
time~\cite{Babai16}. For recent surveys on the graph isomorphism problem and its parameterized complexity, see~\cite{GroheS20,GroheN21,Neuen26}.

This unresolved worst-case complexity has motivated two complementary lines
of research: designing practical isomorphism testing heuristics and identifying graph classes on which the proposed efficient algorithms are provably complete.
We pursue the latter direction for one of the most widely used isomorphism testing heuristics,
 \emph{color refinement} (also known as the \emph{1-dimensional Weisfeiler--Leman algorithm} (1-WL),  or \emph{naive vertex classification}).


\begin{description}
\item[Color Refinement (CR)] is a classical heuristic for graph isomorphism
testing~\cite{CorneilG70}. Given a graph~$G$ with an initial coloring~$c_0$ (uniform unless specified otherwise), it computes a sequence of
vertex colorings by setting
\[
c_{t+1}(v)
\coloneqq
\left(
c_t(v),
\left\{\!\!\left\{c_t(u)\mid u\in N_G(v)\right\}\!\!\right\}
\right)\ \text{for every vertex }v.
\]
The process terminates once the color classes stabilize.
\end{description}

Color refinement provides an incomplete isomorphism test. When run on the disjoint union of two isomorphic input graphs $G$ and $H$ with a uniform initial coloring, CR colors them identically, i.e., $V(G)$ and $V(H)$ receive the same multiset of colors; in general, we say that $G$ and $H$ are \emph{CR-equivalent}, or that CR does not \emph{distinguish} them, when this happens. However, CR may also stabilize on identical colorings when the input graphs are non-isomorphic. For instance, for any pair $G$, $H$ of $r$-regular graphs on the same number of vertices, CR returns the initial uniform coloring, even when $G$ and $H$ are non-isomorphic.
Despite failing on very simple inputs, CR is proven to be able to distinguish almost all pairs of graphs~\cite{BabaiES80}. Moreover, as it can be implemented to run in time $O((n+m)\log n)$~\cite{CardonC82}, it is widely used in practical graph isomorphism testing methods.

Beyond graph isomorphism testing, CR connects various areas of computer science. 
In particular, for two graphs $G$, $H$, the following are equivalent:
\begin{itemize}
    \item CR does not distinguish $G$ and $H$;
    \item there exists a fractional isomorphism between $G$ and $H$~\cite{Tinhofer91,RamanaSU94};
    \item no sentence in the counting logic~$C^2$ distinguishes $G$ and $H$~\cite{ImmermanL90};
    \item $G$ and $H$ agree on their homomorphism counts for all trees~\cite{Dvorak10,DellGR18}.
\end{itemize}
 Recently, the method has received a new wave of attention due to its intricate relation to the expressivity of graph neural networks (GNN)~\cite{XuHLJ19,MorrisRFHLRG19,Zopf22a,MLMRKGFB23}. 
For more details on the applications of color refinement, see the survey~\cite{grohe2021}.

As the alternative name, the `$1$-dimensional Weisfeiler--Leman algorithm', suggests, color refinement belongs to a family of combinatorial algorithms, called WL algorithms for short, which we describe only at an intuitive level.

\begin{description}
\item[$k$-dimensional Weisfeiler--Leman algorithm ($k$-WL)] is a generalization of the CR algorithm, in which we color $k$-tuples of vertices. 
At each round, colors are refined by passing local information about the isomorphism type and the current color between tuples differing in one position.
The process terminates once the color partition stabilizes. 
\end{description}

The original algorithm
~\cite{WeisfeilerL68} is the $2$-dimensional version; soon after, the general version was developed independently by many researchers, 
see~\cite{CaiFI92}. The WL algorithms are key components of many theoretical and practical approaches to graph isomorphism~\cite{mckay1981,bollobas82,McKayP14,Babai16,Kiefer20,AndersS21}.
The algorithms form a strict hierarchy: $(k+1)$-WL is at least as powerful as $k$-WL, and for every $k$ there are pairs of graphs distinguished by $(k+1)$-WL but not by $k$-WL~\cite{CaiFI92}. Nevertheless, every fixed dimension of the WL algorithm remains incomplete as an isomorphism test: for each $k$, there exist non-isomorphic graphs that $k$-WL cannot distinguish~\cite{CaiFI92}.

On the other hand, the increase in distinguishing power comes at a computational cost. On graphs with $n$ vertices, $k$-WL can be implemented in time $O(n^{k+1}\log n)$, which becomes prohibitive on large instances even for small values of $k$. It is therefore of both theoretical and practical interest to determine the graph classes on which CR, or another low-dimensional WL algorithm, provides a complete isomorphism test.
Beyond graph isomorphism, the WL hierarchy also provides a framework for understanding the expressive power of graph neural networks. Message-passing graph neural networks are bounded in expressive power by $1$-WL, while higher-dimensional architectures have been proposed whose expressive power is characterized by higher-dimensional WL algorithms~\cite{MaronBSL19,MorrisRFHLRG19,XuHLJ19}; see~\cite{MLMRKGFB23} for a survey. These gains in expressive power likewise come at the cost of increased computation.

\subsection{Identifying and distinguishing}

We say that a graph~$G$ is \emph{identified} by $k$-WL if $k$-WL distinguishes $G$ from every non-isomorphic graph~$H$.
More generally, $k$-WL \emph{identifies} a class~$\mathcal{G}$ if it identifies every graph in~$\mathcal{G}$; the least such~$k$ is the \emph{WL-dimension} of~$\mathcal{G}$.  A landmark result on the power of the WL algorithm establishes that every proper minor-closed class has finite WL-dimension~\cite{Grohe12, Grohe17}.
It is also known that the class of graphs of treewidth at most~$k$ has WL-dimension at most~$k$~\cite{KieferN19}, and more specific results show WL-dimension~2 for cographs, interval graphs, and chordal claw-free graphs~\cite{EvdokimovPT00,BergeremGKO23}, as well as for outerplanar graphs~\cite{Kiefer20}. The class of planar graphs has WL-dimension at most~3~\cite{KieferPS19}; since it contains all outerplanar graphs, its WL-dimension is at least~2. Determining which of these bounds is tight for planar graphs remains a central open problem~\cite{Kiefer20siglog}.
Graphs identified by $1$-WL are also called \emph{amenable}, and the class of all amenable graphs was characterized independently in~\cite{KieferSS15,ArvindKRV17}. In particular, a graph $G$ is amenable if and only if its flip is a bouquet forest\footnote{Roughly, the flip of $G$ replaces every dense part of $G$ by its sparse complement: whenever two stable color classes (or one class with itself) are joined by more than half of all possible edges, these edges are complemented. A \emph{bouquet} is obtained from five vertex-disjoint isomorphic copies $(T_i,v_i)$ of a rooted tree $(T,v)$ by adding a $5$-cycle through the distinguished vertices $v_1,\ldots,v_5$. A bouquet forest is a disjoint union of vertex-colored trees and pairwise non-isomorphic vertex-colored bouquets.}~\cite{KieferSS15}.

Although graphs of WL-dimension~$1$ are well understood, they form a fairly limited family, and many natural graph classes contain graphs outside it. Indeed, identification is a demanding requirement: a graph must be distinguished from \emph{every} non-isomorphic graph, including those outside the class under consideration. In order to obtain broader graph classes in which CR, or low-dimensional WL, can serve as a provably correct isomorphism test, we relax the requirement and only ask the algorithm to distinguish every pair of non-isomorphic graphs of that class.


We say that CR \emph{distinguishes} a class~$\cal G$ of graphs if it
distinguishes every pair of non-isomorphic graphs in~$\cal G$. In other words,
CR is a complete isomorphism test within~$\cal G$, without any requirement
on pairs $G\in\mathcal G$ and $H\notin\mathcal G$. 
For example, CR distinguishes
the class of cycles $\mathcal{C} = \{C_\ell : \ell \geq 3\}$: in a cycle, all 
vertices receive the same color, so non-isomorphic cycles are distinguished
by the number of vertices of that color. Yet
CR does not identify~$\mathcal{C}$, since it fails to distinguish~$C_6$ from
the disjoint union of two copies of~$C_3$, which lies outside~$\mathcal{C}$.

Pairwise distinguishability by CR has been studied extensively through its
equivalent characterizations in counting logic~\cite{ImmermanL90}, fractional
isomorphism~\cite{Tinhofer91,RamanaSU94}, and tree homomorphism
counts~\cite{Dvorak10,DellGR18}; however, the question of finding graph classes on which CR-equivalence coincides with
isomorphism has received less systematic attention.

\subsection{Our result}

\begin{figure}[ht]
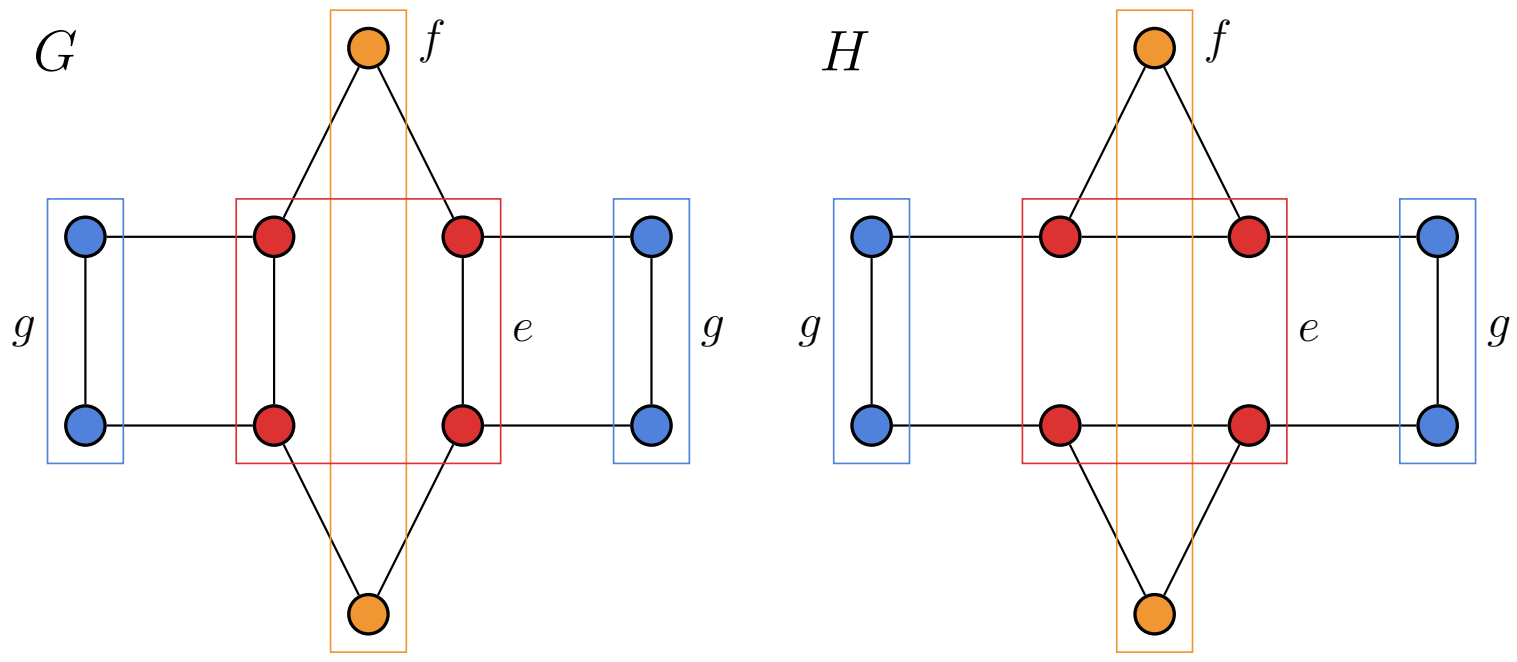

\centering
\includestandalone[scale=0.52]{figs/square_kites}
\caption{Two non-isomorphic $2$-connected outerplanar graphs $G$ and~$H$ not distinguished by CR: $H$ contains two triangles while $G$ contains none. In both graphs, the coloring with colors $e,f,g$ is equitable: each vertex of color~$f$ has two neighbors of color~$e$, each vertex of color~$g$ has one neighbor of color~$e$ and one of color~$g$, and each vertex of color~$e$ has one neighbor of each color $e$, $f$, $g$.}
\label{fig:2conn ctr}
\end{figure}

We study the distinguishing power of CR on outerplanar graphs. This class is a natural stepping stone from forests toward the planar case: outerplanar graphs remain sparse, while allowing cycles and nontrivial $2$-connected components. 
Moreover, their $2$-connected components have a unique Hamiltonian cycle~\cite{Syslo79}, paralleling the unique combinatorial embedding of $3$-connected planar graphs~\cite{Whitney32}, which plays a central role in~\cite{KieferPS19}.
Although CR does not distinguish all outerplanar graphs (see \Cref{fig:2conn ctr}), our main result shows that marking a single vertex as root (by assigning it a unique initial color) is sufficient.

\begin{restatable}{theorem}{mainthm}\label{thm:main}
 Color refinement distinguishes connected rooted outerplanar graphs.
\end{restatable}

We note two extensions of \Cref{thm:main}. Our proof uses the root only to ensure the existence of a singleton class in the stable CR partition; it does not rely on that vertex having received a distinguished initial color. Therefore, it is sufficient to assume that the stable CR partition contains a singleton color class, which can serve as a \emph{pseudo-root}. Similarly, the connectivity assumption can be dropped if every connected component contains a pseudo-root. 

Nevertheless, some form of root information is essential: there exist pairs of non-isomorphic connected outerplanar graphs that CR fails to distinguish without a root; see \Cref{fig:2conn ctr}.

The closest existing notion to rooted color refinement is the $(1,1)$-WL test introduced in~\cite{RattanS23}. Two graphs $G$ and $H$ are \emph{$(1,1)$-WL equivalent} if there exists a bijection $\pi : V(G) \to V(H)$ such that for all $v \in V(G)$ the rooted graphs $(G,v)$ and $(H,\pi(v))$ are CR-equivalent. We show that \Cref{thm:main} implies that $(1,1)$-WL distinguishes outerplanar graphs.

\begin{restatable}{theorem}{wlthm}\label{thm:1-1-WL}
$(1,1)$-WL distinguishes outerplanar graphs.
\end{restatable}

We observe that at a structural level, comparing all one-vertex individualizations followed by CR leads to the $(1,1)$-WL test, which is known to be no more powerful than $2$-WL~\cite{RattanS23}. Consequently, if CR distinguishes the rooted graphs obtained from a class~$\mathcal G$, then $2$-WL distinguishes every pair of non-isomorphic graphs in~$\mathcal G$.




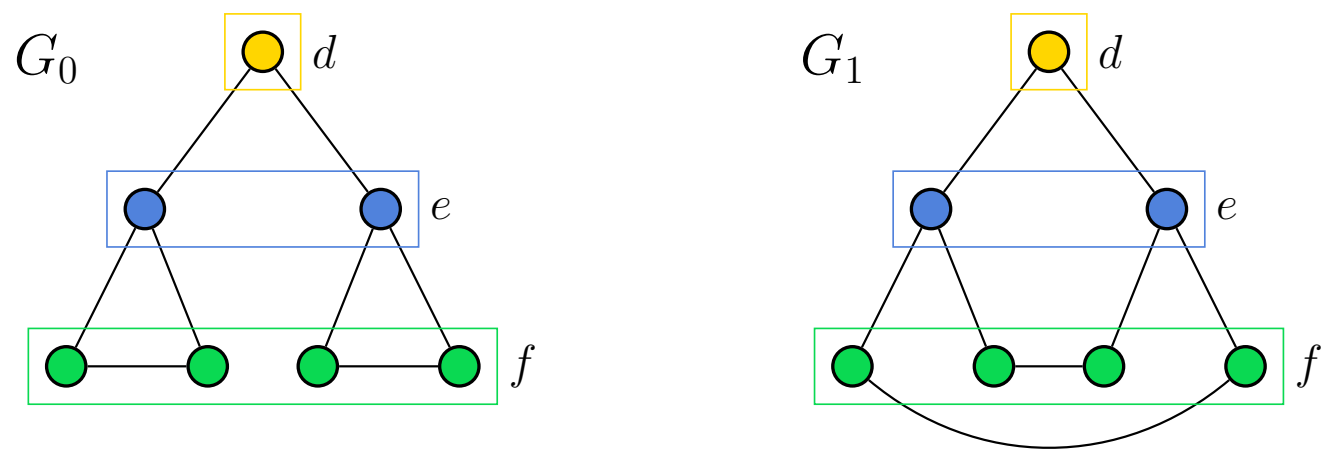
\begin{figure}[ht]
\centering
\begin{tikzpicture}[
    vertex/.style={circle, minimum size=5mm, inner sep=0pt, draw=black, very thick},
    red vertex/.style={vertex, fill=cR},
    yellow vertex/.style={vertex, fill=cY},
    blue vertex/.style={vertex, fill=cB},
    green vertex/.style={vertex, fill=cG},
    edge/.style={thick},
    every label/.style={font=\LARGE},
    class/.style={draw=#1, semithick, rectangle, inner sep=6pt},
    arr/.style={-{Stealth[length=3mm]}, thick, bend left=20},
]

\def\rx{10}

\node[yellow vertex] (r1) at (\rx-5,    0) {};
\node[blue vertex]   (b1) at (\rx-6.5, -2) {};
\node[blue vertex]   (b2) at (\rx-3.5, -2) {};
\node[green vertex]  (g1) at (\rx-7.5, -4) {};
\node[green vertex]  (g2) at (\rx-5.7, -4) {};
\node[green vertex]  (g3) at (\rx-4.3, -4) {};
\node[green vertex]  (g4) at (\rx-2.5, -4) {};

\draw[edge] (r1) -- (b1);
\draw[edge] (r1) -- (b2);
\draw[edge] (b1) -- (g1);
\draw[edge] (b1) -- (g2);
\draw[edge] (b2) -- (g3);
\draw[edge] (b2) -- (g4);
\draw[edge] (g1) -- (g2);
\draw[edge] (g3) -- (g4);

\node[yellow vertex] (r2) at (\rx+5,    0) {};
\node[blue vertex]   (b3) at (\rx+3.5, -2) {};
\node[blue vertex]   (b4) at (\rx+6.5, -2) {};
\node[green vertex]  (g5) at (\rx+2.5, -4) {};
\node[green vertex]  (g6) at (\rx+4.3, -4) {};
\node[green vertex]  (g7) at (\rx+5.7, -4) {};
\node[green vertex]  (g8) at (\rx+7.5, -4) {};

\draw[edge] (r2) -- (b3);
\draw[edge] (r2) -- (b4);
\draw[edge] (b3) -- (g5);
\draw[edge] (b3) -- (g6);
\draw[edge] (b4) -- (g7);
\draw[edge] (b4) -- (g8);
\draw[edge] (g6) -- (g7);
\draw[edge] (g5) to[bend right=40] (g8);

\node[class=cY, fit=(r1), label=right:$d$] {};
\node[class=cB, fit=(b1)(b2), label=right:$e$] {};
\node[class=cG, fit=(g1)(g4), label=right:$f$] {};
\node[class=cY, fit=(r2), label=right:$d$] {};
\node[class=cB, fit=(b3)(b4), label=right:$e$] {};
\node[class=cG, fit=(g5)(g8), label=right:$f$] {};

\node[anchor=north west] at (\rx-8.3, 0.35) {\huge $G_0$};
\node[anchor=north west] at (\rx+1.7, 0.35) {\huge $G_1$};

\end{tikzpicture}
\caption{A connected rooted outerplanar graph $G_0$ and a connected rooted
  series-parallel graph $G_1$ that is not outerplanar, not distinguished by CR
  (roots have color~$d$). Note that $G_1$ has treewidth~2.}
\label{fig:planar ctr}
\end{figure}

\smallskip
\noindent\textbf{Our approach.}
Running a canonical version of CR (colors are numbered identically for isomorphic graphs) on a graph~$G$ until stabilization yields a partition
$
V(G)=C_1\mathbin{\dot\cup}\cdots\mathbin{\dot\cup}C_k
$
into stable color classes. This partition is \emph{equitable}: for every pair of colors $i,j$, all vertices in~$C_i$ have the same number~$K_{ij}$ of neighbors in~$C_j$. We encode this information by a weighted directed graph~$K$, called \emph{color graph}, whose vertices are the stable colors, where vertex~$i$ has weight $w_i=|C_i|$ and the arc from~$i$ to~$j$ has weight~$K_{ij}$, see \Cref{fig:ex color graph} for an example; it is the quotient of~$G$ by its stable partition, with class sizes and multiplicities recorded as vertex and arc weights.  


We prove Theorem~\ref{thm:main} by solving an \emph{inversion problem}. Given the color graph~$K$ and the root color produced by CR on a connected rooted outerplanar graph~$G$, we reconstruct a graph isomorphic to~$G$, in time linear in the size of the output. 
Beyond proving distinguishability, this reconstruction yields a \emph{canonization algorithm}. The goal of canonization is to find a canonical form $\gamma$ of a class $\mathcal G$, which assigns to every graph~$G$ a graph $\gamma(G)$ isomorphic to $G$ such that for $G,H \in \mathcal G$, $\gamma(G)$ and $\gamma(H)$ are identical if and only if $G$ is isomorphic to $H$.
Our reconstruction is defined canonically from the color graph and the root color, both of which are isomorphism invariants. Consequently, isomorphic inputs produce identical labeled outputs, and so the reconstruction defines a canonical form for connected rooted outerplanar graphs.

Our canonization differs fundamentally from the canonization procedures of~\cite{ImmermanL90,KieferSS15}, which apply to graphs identified by CR. If a graph~$G$ is identified by CR, then every graph that is CR-equivalent to~$G$ is isomorphic to~$G$; consequently, reconstructing any representative of its CR-equivalence class suffices. Theorem~\ref{thm:main}, by contrast, gives only a classwise guarantee: a connected rooted outerplanar graph may be CR-equivalent to a non-isomorphic graph outside the class, as illustrated in Figure~\ref{fig:planar ctr}. Our inversion procedure therefore cannot return an arbitrary representative of the CR-equivalence class. It must exploit the outerplanar structure to recover its unique connected rooted outerplanar representative.

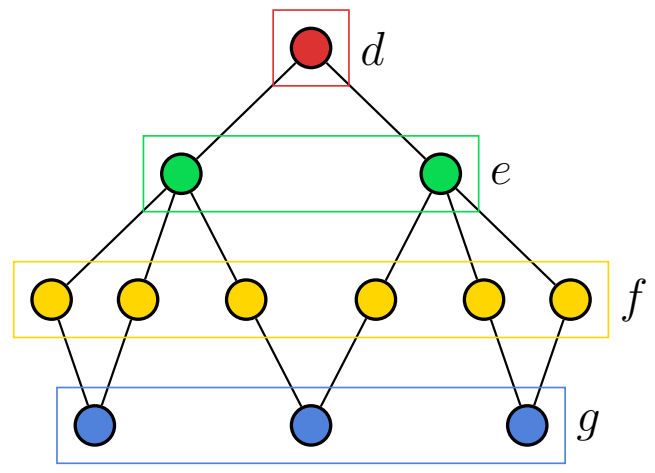
\begin{figure}
    \centering
    \begin{tikzpicture}[
    vertex/.style={circle, minimum size=5mm, inner sep=0pt,
                   draw=black, very thick},
    rv/.style={vertex, fill=cR},
    bv/.style={vertex, fill=cB},
    gv/.style={vertex, fill=cG},
    yv/.style={vertex, fill=cY},
    edge/.style={thick},
    every label/.style={font=\LARGE},
    class/.style={draw=#1, semithick, rectangle, inner sep=6pt},
  ]

  \node[rv] (a)  at (0,4.8)      {};

  \node[gv] (bl) at (-1.65,3.2)  {};
  \node[gv] (br) at ( 1.65,3.2)  {};

  \node[yv] (c1) at (-3.3,1.6)   {};
  \node[yv] (c2) at (-2.2,1.6)   {};
  \node[yv] (c3) at (-0.825,1.6) {};
  \node[yv] (c4) at ( 0.825,1.6) {};
  \node[yv] (c5) at ( 2.2,1.6)   {};
  \node[yv] (c6) at ( 3.3,1.6)   {};

  \node[bv] (d1) at (-2.75,0)    {};
  \node[bv] (d2) at (0,0)        {};
  \node[bv] (d3) at (2.75,0)     {};

  \draw[edge] (a)--(bl);
  \draw[edge] (a)--(br);

  \draw[edge] (bl)--(c1);
  \draw[edge] (bl)--(c2);
  \draw[edge] (bl)--(c3);
  \draw[edge] (br)--(c4);
  \draw[edge] (br)--(c5);
  \draw[edge] (br)--(c6);

  \draw[edge] (c1)--(d1);
  \draw[edge] (c2)--(d1);
  \draw[edge] (c3)--(d2);
  \draw[edge] (c4)--(d2);
  \draw[edge] (c5)--(d3);
  \draw[edge] (c6)--(d3);

  \node[class=cR, fit=(a), label=right:$d$] {};
  \node[class=cG, fit=(bl)(br), label=right:$e$] {};
  \node[class=cY, fit=(c1)(c6), label=right:$f$] {};
  \node[class=cB, fit=(d1)(d3), label=right:$g$] {};

\end{tikzpicture}
    \caption{A connected rooted outerplanar graph whose stable CR partition is strictly coarser than its orbit partition: the color classes $f$ and $g$ each contain two distinct orbits.}
    \label{fig:diamond}
\end{figure}

Even beyond amenable graphs, simple examples of classwise distinguishability (e.g. class of cycles or cluster graphs) are in fact implied by the stronger property of being a class of Tinhofer graphs~\cite{Tinhofer91,ArvindKRV17}, for which an isomorphism between two isomorphic graphs of the class can be constructed greedily. The situation for rooted outerplanar graphs is different: CR need not determine the vertex orbits (see \Cref{fig:diamond}), so a greedy construction may fail at the very first step.

The main structural insight guiding our inversion algorithm is the Dichotomy Lemma. It shows that outerplanarity and equitability impose a rigid structure on the CR coloring of each maximal $2$-connected subgraph of $G$ (also called a block of $G$): either all its vertices receive distinct colors, in which case we call it \emph{discrete}, or it admits a color-preserving mirror symmetry fixing the vertex closest to the root, in which case we call it \emph{symmetric}. For more details, see \Cref{lem:dichotomy}. The Dichotomy Lemma has also been formalized and proved in the Rocq proof assistant, see \Cref{app:rocq-dicho-lem}.



\subsection{Perspectives}

We note that color refinement fails to distinguish graphs in several commonly studied superclasses of the outerplanar graphs, even after individualizing a root. \Cref{fig:planar ctr} gives a connected rooted outerplanar graph that CR does not distinguish from a non-isomorphic connected rooted series-parallel graph that is not outerplanar. Since both graphs have treewidth at most~$2$, we can also conclude that \Cref{thm:main} does not extend to rooted graphs of treewidth at most~$2$, equivalently, to the class of rooted $K_4$-minor-free graphs. Two closely related extensions remain open. As outerplanar graphs are characterized by being $\{K_4,K_{2,3}\}$-minor-free, the first is whether \Cref{thm:main} extends to connected rooted $K_{2,3}$-minor-free graphs. This is a natural next class to consider, since every block of such a graph is either outerplanar or isomorphic to~$K_4$. The second is whether an analogous result holds for rooted outerplanar multigraphs when CR is modified to take edge multiplicities into account.

A natural continuation of our work is to investigate the distinguishing power of $2$-WL on planar graphs. In particular, does $2$-WL distinguish planar graphs equipped with two individualized vertices, each assigned its own initial color? More broadly, a key step toward determining the precise WL-dimension of planar graphs is to characterize the $3$-connected planar graphs for which individualizing any single vertex and then applying CR fails to produce a discrete coloring~\cite{Kiefer20siglog}. We obtain an analogous characterization for $2$-connected outerplanar graphs: for any choice of root, our dichotomy shows that the rooted CR coloring is non-discrete precisely when the graph is left--right symmetric with respect to the root. We also observe that, even in that case, its color graph may have several blocks. This
suggests studying the triconnected decompositions of the nontrivial blocks
of the underlying color graph associated with a rooted $3$-connected
planar graph.

\subsection{Structure of the paper}

\Cref{sec:prelim} introduces color refinement, color graphs and block-cut trees. \Cref{sec:structure} states the structural lemmas behind our approach, including the Dichotomy Lemma. \Cref{sec:canon} proves our main theorem by inverting color refinement: \Cref{sec:intuition} gives a high-level description of the inversion algorithm, and \Cref{sec:details} discusses its correctness. \Cref{sec:non-connected} extends our results to non-connected outerplanar graphs. Proofs are deferred to the appendices. 
\section{Preliminaries}\label{sec:prelim}
\noindent\textbf{Graphs.}
For a graph $G$, we write $V(G)$ and $E(G)$ for its
vertex and edge sets, and $N_G(v)$ for the open neighborhood of a vertex~$v$.
The running time of an algorithm taking a graph $G$ as input is expressed in terms of $n=|V(G)|$ and $m=|E(G)|$. Unless stated otherwise, all graphs are assumed to be finite, simple, and undirected. For a positive integer~$k$, we
write $[k]=\{1,\dots,k\}$, and for integers $a\le b$ we write
$[a,b]=\{a,a+1,\dots,b\}$. 
A \emph{path} of length~$\ell$ from $u_0$ to $u_\ell$, or a \emph{$u_0u_\ell$-path},
is a sequence $(u_0,\dots,u_\ell)$ of pairwise distinct vertices such that
$u_{p-1}u_p\in E(G)$ for each $p\in[\ell]$. The distance $d_G(u,v)$ is the
length of a shortest $uv$-path.

\smallskip
\noindent\textbf{Colored graphs and color refinement.}
A \emph{vertex-colored graph} $(G,c)$ is a graph with a \emph{coloring}
$c\colon V(G)\rightarrow\mathbb{N}$; we write $|c|$ for the number of colors
used and assume that these colors are $1,\dots,|c|$. The nonempty sets
$c^{-1}(1),\dots,c^{-1}(|c|)$ are the \emph{color classes}, and $|c^{-1}(i)|$
is the \emph{size} of the color~$i$. 
A \emph{color-preserving isomorphism}
from $(G_0,c_0)$ to $(G_1,c_1)$ is a bijection $\pi\colon V(G_0)\to V(G_1)$
such that $uv\in E(G_0)$ if and only if $\pi(u)\pi(v)\in E(G_1)$, and
$c_1(\pi(v))=c_0(v)$ for every $v\in V(G_0)$; if such a bijection exists, we write $(G_0,c_0)\iso(G_1,c_1)$. A \emph{rooted
graph} $(G,r)$ is a graph with a distinguished vertex~$r$, its \emph{root}; we
regard it as the vertex-colored graph in which $r$ has color~$1$ and all other
vertices have color~$2$, so that isomorphisms of rooted graphs are
color-preserving isomorphisms. 

A coloring~$c'$ \emph{refines} a coloring~$c$
if every color class of~$c'$ is contained in a color class of~$c$, and we say that $c$ is
\emph{equitable} if for every $i,j\in[|c|]$, all vertices of color~$i$ have the
same number of neighbors of color~$j$.
The \emph{color refinement} (CR) algorithm iteratively refines a vertex
coloring: at each round, two vertices keep the same color if and only if they
previously had the same color and the same multiset of neighbor colors, until
the coloring stabilizes. On input $(G,c_0)$, its output $\CR(G,c_0)=(G,c)$ is the graph~$G$ equipped with the coarsest equitable coloring~$c$ that refines
$c_0$~\cite{CardonC82}. We additionally assume that CR names the colors canonically; for example, at each round, the new colors
are numbered by the lexicographic rank of the pair consisting of the old color
and the sorted multiset of the neighbors' old colors. Consequently, if
$(G_0,c_0)\iso(G_1,c_1)$ via~$\pi$, then $\pi$ is also a color-preserving
isomorphism from $\CR(G_0,c_0)$ to $\CR(G_1,c_1)$. Note that there is an $O((n+m)\log n)$-time implementation of CR that produces a canonical coloring~\cite{BerkholzBG13}.

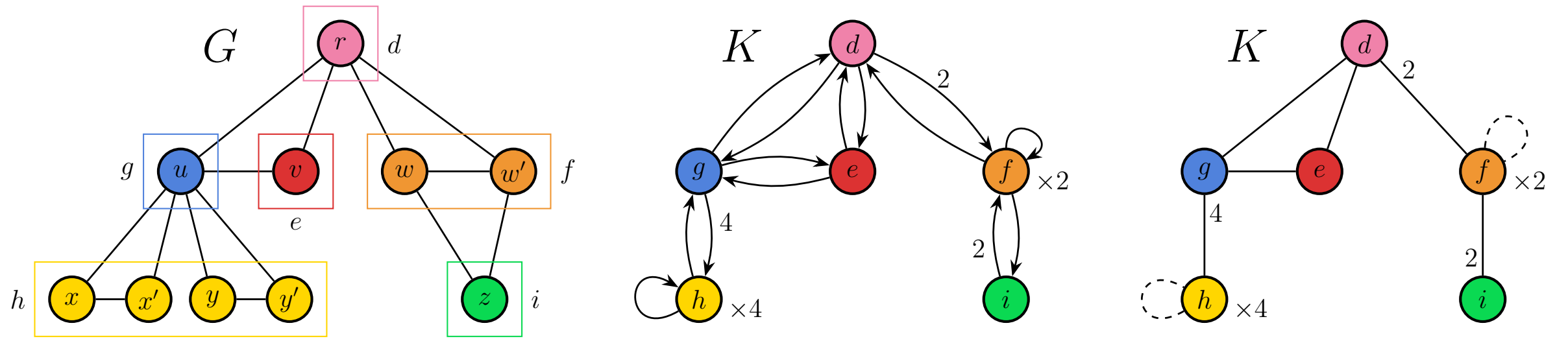
\begin{figure}[h!]
\centering
\begin{tikzpicture}[
  vertex/.style={circle, minimum size=7mm, inner sep=0pt,
                 draw=black, very thick, font=\large},
  pink vertex/.style={vertex, fill=cP},
  blue vertex/.style={vertex, fill=cB},
  red vertex/.style={vertex, fill=cR},
  orange vertex/.style={vertex, fill=cO},
  yellow vertex/.style={vertex, fill=cY},
  green vertex/.style={vertex, fill=cG},
  edge/.style={thick},
  arr/.style={-{Stealth[length=2.5mm]}, thick, bend left=15},
  biglabel/.style={font=\huge, anchor=north west},
  every label/.style={font=\large},
]

  \def\graphsep{8}

  \begin{scope}
    \node[biglabel] at (-2.3,0.35) {$G$};
    \node[pink vertex]   (r)  at (0,0) {$r$};
    \node[blue vertex]   (bl) at (-2.5,-2) {$u$};
    \node[red vertex]    (re) at (-0.7,-2) {$v$};
    \node[orange vertex] (o1) at (1.0,-2) {$w$};
    \node[orange vertex] (o2) at (2.7,-2) {$w'$};
    \node[yellow vertex] (ya) at (-4.2,-4) {$x$};
    \node[yellow vertex] (yb) at (-3.0,-4) {$x'$};
    \node[yellow vertex] (yc) at (-2.0,-4) {$y$};
    \node[yellow vertex] (yd) at (-0.8,-4) {$y'$};
    \node[green vertex]  (gr) at (2.25,-4) {$z$};

    \draw[edge] (r) -- (bl);
    \draw[edge] (r) -- (re);
    \draw[edge] (r) -- (o1);
    \draw[edge] (r) -- (o2);
    \draw[edge] (bl) -- (ya) -- (yb) -- (bl);
    \draw[edge] (bl) -- (yc) -- (yd) -- (bl);
    \draw[edge] (bl) -- (re);
    \draw[edge] (o1) -- (gr);
    \draw[edge] (o2) -- (gr);
    \draw[edge] (o1) -- (o2);

    \node[draw=cP, semithick, rectangle, fit=(r), inner sep=6pt, label=right:$d$] {};
    \node[draw=cB, semithick, rectangle, fit=(bl), inner sep=6pt, label=left:$g$] {};
    \node[draw=cR, semithick, rectangle, fit=(re), inner sep=6pt, label=below:$e$] {};
    \node[draw=cG, semithick, rectangle, fit=(gr), inner sep=6pt, label=right:$i$] {};

    \node[draw=cO, semithick, rectangle, fit=(o1)(o2),
          inner sep=6pt, label=right:$f$] {};
    \node[draw=cY, semithick, rectangle, fit=(ya)(yd),
          inner sep=6pt, label=left:$h$] {};
  \end{scope}

  \begin{scope}[shift={(\graphsep,0)}]
    \node[biglabel] at (-2.2,0.35) {$K$};
    \node[pink vertex]   (P) at (0,0) {$d$};
    \node[blue vertex]   (B) at (-2.4,-2) {$g$};
    \node[red vertex]    (R) at (0,-2) {$e$};
    \node[orange vertex] (O) at (2.4,-2) {$f$};
    \node[yellow vertex] (Y) at (-2.4,-4) {$h$};
    \node[green vertex]  (G) at (2.4,-4) {$i$};

    \draw[arr] (P) to (B);
    \draw[arr] (B) to (P);

    \draw[arr] (P) to (R);
    \draw[arr] (R) to (P);

    \draw[arr] (P) to node[pos=0.33, right=3pt] {\large$2$} (O);
    \draw[arr] (O) to (P);

    \draw[arr] (B) to (R);
    \draw[arr] (R) to (B);

    \draw[arr] (B) to node[pos=0.33, right] {\large$4$} (Y);
    \draw[arr] (Y) to (B);

    \draw[arr] (O) to (G);
    \draw[arr] (G) to node[pos=0.33, left] {\large$2$} (O);
    \draw[arr, shorten <=10pt, shorten >=10pt]
      (O) edge[loop right, min distance=10mm, out=0, in=-60] (O);
    \draw[-{Stealth[length=2.5mm]}, thick]
      (Y) to[out=210, in=150, looseness=7] (Y);

    \node[font=\large, right=-1.5pt of O, yshift=-4.5pt] {$\times 2$};
    \node[font=\large, right=-1.5pt of Y, yshift=-4.5pt] {$\times 4$};
  \end{scope}

  \begin{scope}[shift={({2*\graphsep},0)}]
    \node[biglabel] at (-2.3,0.35) {$K$};
    \node[pink vertex]   (kr)  at (0,0) {$d$};
    \node[blue vertex]   (kbl) at (-2.5,-2) {$g$};
    \node[red vertex]    (kre) at (-0.7,-2) {$e$};
    \node[orange vertex] (ko)  at (1.85,-2) {$f$};
    \node[yellow vertex] (kye) at (-2.5,-4) {$h$};
    \node[green vertex]  (kgr) at (1.85,-4) {$i$};

    \draw[edge] (kr) -- (kbl);
    \draw[edge] (kr) -- (kre);
    \draw[edge] (kbl) -- (kre);
    \draw[edge] (kbl) -- (kye);
    \draw[edge] (kr) -- (ko);
    \draw[edge] (ko) -- (kgr);

    \path (kr)--(ko) node[font=\large, inner sep=1pt, pos=0.22, above right] {$2$};
    \path (kgr)--(ko) node[font=\large, inner sep=2pt, pos=0.22, left] {$2$};
    \draw[edge, dashed] (ko) to[out=85, in=25, looseness=7] (ko);
    \node[font=\large, right=-1.5pt of ko, yshift=-4.5pt] {$\times 2$};
    \path (kbl)--(kye) node[font=\large, inner sep=2pt, pos=0.22, right] {$4$};
    \draw[edge, dashed] (kye) to[out=210, in=150, looseness=7] (kye);
    \node[font=\large, right=-1.5pt of kye, yshift=-4.5pt] {$\times 4$};
  \end{scope}

\end{tikzpicture}
\caption{A rooted outerplanar graph $G$ colored with CR coloring (left) and its color graph $K$: as a digraph (middle) and as an undirected graph (right) where weights are moved closer to the tail of the corresponding arc, e.g., the~$4$ close to $g$ indicates $K_{gh}=4$. Weights equal to~$1$ are omitted. The underlying graph $\widetilde{K}$ consists of the plain edges (right). Every color has size~$1$, except colors~$f$ and~$h$, whose sizes are $w_f=2$ and $w_h=4$ (indicated by $\times 2$ and $\times 4$, respectively).}
\label{fig:ex color graph}
\end{figure}

\smallskip
\noindent\textbf{Color graph.}
Let $c$ be an equitable coloring of~$G$. The \emph{color graph}
$K=\CG(G,c)$ of $(G,c)$ is the vertex- and arc-weighted directed graph on $[|c|]$ in which the color~$i$ carries the
vertex weight $w_i=|c^{-1}(i)|$ and there is an arc $(i,j)$ of weight
\[
  K_{ij}
  \;=\;
  \bigl|\{u\in N_G(v):c(u)=j\}\bigr|
  \qquad\text{for any vertex $v$ with $c(v)=i$}
\]
whenever this number is positive. Equitability of~$c$ ensures that $K_{ij}$ does not depend on the choice of~$v$. The color graph may have loops, and the value~$K_{ii}$ will play an important role. Since
$G$ is undirected, $K_{ij}>0$ implies $K_{ji}>0$, but the two weights may differ (see Figure~\ref{fig:ex color graph} for an example).
We write
$\widetilde{K}$ for the \emph{underlying simple graph} of~$K$, obtained by
forgetting weights, directions and loops, and $N_{\Kt}(i)$ for the neighborhood of~$i$ in~$\widetilde{K}$.
We regard $\widetilde{K}$ as a vertex-colored graph colored by the identity~$\id$, that is, each color $i\in V(\widetilde{K})$ has color $\id(i)=i$.
Given $G$ and~$c$, the color graph~$K$ is easily computed in linear time.
We say that CR \emph{distinguishes} two vertex-colored graphs $(G_0,c_0)$ and
$(G_1,c_1)$ if $\CG(\CR(G_0,c_0))\neq
\CG(\CR(G_1,c_1))$, and that it distinguishes a class of graphs
if it distinguishes every pair of non-isomorphic members of the class.
Finally, for a subgraph $H$ of $G$, we write $c(H)$ for the set $c(V(H))$ of colors occurring in $H$.
For a connected rooted graph, the color graph and the root color are sufficient to determine every color-class size and thus vertex weights. Indeed, if $C_i$ and $C_j$ are adjacent color classes, then $|c^{-1}(i)|K_{ij}=|c^{-1}(j)|K_{ji}$. Although this then allows to infer vertex weights from the root color class and arc weights, we still need them for the distinguishing test in general.

\smallskip
\noindent\textbf{Outerplanar graphs and blocks.}
A graph is \emph{outerplanar} if it admits a planar embedding in which every
vertex lies on the boundary of the outer face; equivalently, if it has
neither $K_4$ nor $K_{2,3}$ as a minor. Every 2-connected outerplanar
graph has a unique planar embedding~\cite{Syslo79}, in which all
vertices lie on a unique Hamiltonian cycle bounding the outer face; any
other edge is called a \emph{chord} of this cycle.
A \emph{cut vertex} is a vertex whose removal disconnects the graph, and a
\emph{block} of~$G$ is a maximal connected subgraph without a cut vertex of its
own. In an outerplanar graph, every block is either a $K_2$ (we call such blocks \emph{bridges}), or a
2-connected outerplanar graph. Every edge of~$G$ lies in exactly one block,
two distinct blocks share at most one vertex, which is then a cut vertex, and
consequently every path between two vertices of a block stays inside that
block. The \emph{block-cut tree} $T_G$ of $G$ is the tree whose vertex set consists of one \emph{block node} per
block and one \emph{cut-vertex node} per cut vertex, where a block node~$B$ and a cut
vertex node~$v$ are adjacent if and only if $v\in V(B)$. The block-cut tree of a graph can be computed
in linear time~\cite{HopcroftTarjan73}. For a rooted graph $(G,r)$, we add a cut-vertex node for~$r$ if $r$ is not a cut vertex, and we root $T_G$ at the node of~$r$, see top right
of Figure~\ref{fig:4types}. Then each block node~$B$ is the child of a cut
vertex node~$v$; we call $v=\ro(B)$ the \emph{root} of the block~$B$ and say that $B$ \emph{hangs} below $v$.
Equivalently, $\ro(B)$ is the unique vertex of~$B$ with minimum depth, where the \emph{depth} of a vertex~$v$ is its distance from the root: $\depth_G(v)=d_G(r,v)$.
As mentioned before, the outerplanarity of $G$ implies that each non-bridge block $B$ has a unique Hamiltonian cycle. An enumeration $v_0,\ldots,v_{k-1}$ along this cycle (either clockwise or counter-clockwise) starting from the root $v_0=\ro(B)$  is called an \emph{outer face ordering} of~$B$.

\smallskip
\noindent\textbf{Standing assumptions.}
Throughout this paper, $(G,c)$ denotes a connected outerplanar graph equipped with an
equitable coloring~$c$ that has a \emph{singleton color class}: there is a
color~$d$ with $c^{-1}(d)=\{r\}$, and we regard~$G$ as rooted at~$r$ (a pseudo-root, in the terminology of \Cref{sec:non-connected}). Such a
coloring arises when CR is applied to a rooted graph, but our results hold for
any equitable coloring with a singleton class. In the appendices, we set $d=1$ to lighten notation. As $G$ is connected, so are~$K=\CG(G,c)$ and its simple undirected underlying graph $\widetilde{K}$, and we
regard them as rooted at the color~$d$. We let $T_G$ and $T_{\Kt}$ denote the block-cut trees of $G$ and $\widetilde{K}$, respectively.
They are rooted as well (at the nodes associated with $r$ and~$d$, respectively), and each of their blocks is considered to be rooted at the cut-vertex node it is a child of. 
To avoid confusion, the blocks of $\widetilde{K}$ are called \emph{c-blocks}.

\section{Main structural properties}\label{sec:structure}

We rely on several key properties imposed by the root, equitability and outerplanarity. The detailed proofs are in Appendices~\ref{app:setting} for the first two lemmas and~\ref{app:dichotomy} for the last one.

\smallskip
\noindent\textbf{Equitability and distance from the root.}
Recall that we define the depth of a vertex $v$ as $\depth_G(v)=d_G(r,v)$. 
We similarly define the depth of a color $i$ as $\depthK(i)=d_{\Kt}(d,i)$.
Note that, by definition of $K$, an edge $ij$ of $\Kt$ means $K_{ij}>0$: every vertex of color $i$ has a neighbor of color $j$, and vice versa.
The fact that $\{r\}$ is a singleton color class then ensures that colors refine
distances to the root $r$, implying that depth in $G$ and $\widetilde{K}$ coincide:

\begin{restatable}[Depth]{lemma}{depthlemma}\label{lem:depth}
For every vertex $v\in V(G)$,\; $\depth_G(v)=\depthK(c(v))$.
\end{restatable}

As a consequence, the root $\ro(A)$ of each block $A$ of 
$G$ is the only vertex of $A$ with its color, since all other vertices of $A$ have larger depth and hence different colors.
Together with block-cut tree properties, this depth correspondence allows us to show the following lemma, which implies that all colors of any c-block appear in a common block of~$G$.

\begin{restatable}[Block maximality]{lemma}{blockmaximality}\label{lem:block_maximality}
Let $A$ be a block of~$G$, and let $S=c(A)$ be the set of
colors appearing in~$A$.  If a c-block~$B$ of~$\widetilde{K}$
contains at least two colors of~$S$, then $V(B)\subseteq S$.
\end{restatable}

\smallskip
\noindent\textbf{Equitability and outerplanarity.}
The main structural insight enabling our result is that outerplanarity and equitability leave a block very
little freedom: either all its vertices receive distinct colors, and we call the block \emph{discrete}, or it has a color-preserving
mirror symmetry fixing its block root, and we call it \emph{symmetric}. This is almost a dichotomy, stated as follows.

\begin{restatable}[Dichotomy Lemma]{lemma}{dicholemma}\label{lem:dichotomy}
Let $G$ be a connected outerplanar graph and $c : V(G) \to \N$ be an equitable coloring with a singleton color class $d$. Then for any block $B$ of $G$, the coloring $c$ restricted to $B$ is either
\begin{description}
    \item[Discrete:] the vertices in $B$ have  pairwise distinct colors; or
    \item[Symmetric:] if $v_0, v_1, \dots, v_{k-1}$ is an outer face ordering of $B$ (with $v_0=\ro(B)$), then $c(v_i) = c(v_{k-i})$ and every color appears at most twice in $B$. Moreover, the mapping $\pi(v_i)= v_{k-i}$, extended by $\pi(v_0) = v_0$, is a color-preserving automorphism of~$B$. 
\end{description}
For blocks of size $2$ (bridges), any coloring using $2$ distinct colors is both discrete and symmetric; for simplicity, we treat them as discrete.
\end{restatable}

\noindent
In \Cref{fig:ex color graph}, ${r,u,v}$ induces a discrete block, with colors $d,g,e$, and ${r,w,w',z}$ induces a symmetric block, with colors $d,f,i$, where $w$ and $w'$ are mirror images of each other (see also \Cref{fig:4types} in \Cref{sec:canon}).

The proof of this seemingly simple lemma turns out to be surprisingly difficult (more than thirteen pages); it is given in Appendix~\ref{app:dichotomy}.
On a high level, the proof is by induction on the size of the graph, using outerplanarity to select a vertex~$v$ of degree at most~2 and of maximum depth. By equitability, all vertices of color~$i=c(v)$ have \fpass{the same degree as~$v$, which we may assume to be} two. We remove all vertices of color $i$ and add an edge between any pair of non-adjacent vertices which were endpoints of an $i$-path in $G$, that is, a path whose inner vertices have color~$i$; this preserves outerplanarity. Proving that this operation also preserves equitability is the key step. A particularly delicate point is to show that if $x,y$ are connected by an $i$-path and $K_{c(x)c(y)}>0$, then $xy \in E(G)$ and so no new edges between $x$ and $y$ were added. 
The proof proceeds by contradiction, using equitability to force a pair $xy \notin E(G)$ into an alternating cycle inside a block. By outerplanarity, this cycle follows the outer-face order, yielding an edge that separates the block’s root from the rest of the cycle. This contradicts the structural properties implied by the singleton color class $d$ and the fact that $c$ refines distances to  $r$.

\smallskip
 
The proof of the Dichotomy Lemma was also formalized with the Rocq~9.1.1 proof assistant~\cite{rocq}, based on the MathComp~\cite{Mahboubi22} and GraphTheory~\cite{DoczkalPous2020} libraries, with the assistance of Claude, an LLM-based coding assistant. The formal statement for blocks of size at least~3 is given in \Cref{app:rocq-dicho-lem} and the formal proof is available at~\url{https://osf.io/vyqgm/overview?view_only=3cabd0da23e34d42a5f5078017183932}.

\section{Distinguishing through Inversion and Canonization}\label{sec:canon}
We prove our main theorem by showing that color refinement can be inverted, that is, we prove that a connected rooted
outerplanar graph can be reconstructed, up to isomorphism, from its color
graph alone. 

\begin{restatable}[Inversion]{theorem}{inversiontheorem}\label{thm:inversion}
Let $G$ be a connected outerplanar graph and let $c$ be an equitable coloring of~$G$ with color graph~$K$, having a singleton color class $c^{-1}(d)=\{r\}$.  Then a vertex-colored graph $(G^*,c^*)=\inv(K,d)$ can be deterministically computed from $(K,d)$ in time linear in the size of the output, so that $(G^*,c^*)$ is color-preserving isomorphic to $(G,c)$, i.e. $(G^*,c^*)\iso(G,c)$.
\end{restatable}

\begin{remark}
The second argument $d$ could be selected among the colors with vertex weight one in $K$. For clarity, we prefer to make this choice explicit. 
\end{remark}

Our main theorem follows by applying \Cref{thm:inversion} to the canonical CR coloring of a rooted graph.

\mainthm*


\begin{proof}
Let $(G,r)$ and $(G',r')$ be two connected rooted outerplanar graphs that color refinement does not
distinguish. By definition, $(G,c)=\CR(G,r)$ and $(G',c')=\CR(G',r')$ have the same color graph~$K$, and by canonicity of CR, their roots have the same color $d=c(r)=c'(r')$.
By Theorem~\ref{thm:inversion}, both
are color-preserving isomorphic to $\inv(K,d)$, and both isomorphisms
send the root to the unique vertex of color~$d$; composing them gives an
isomorphism between $(G,r)$ and $(G',r')$.
\end{proof}

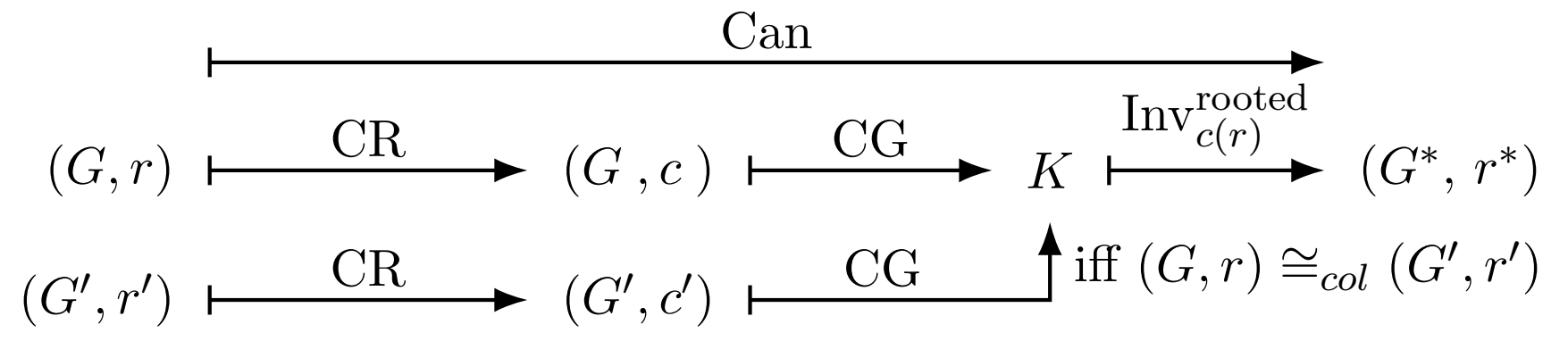
\begin{figure}[h!]
\centering
\begin{tikzpicture}[
    scale=0.8,
    font=\normalsize,
    >={Latex[length=2.5mm,width=1.8mm]},
    flow/.style={->, line width=0.8pt,
                 shorten <=7pt, shorten >=7pt},
    map/.style={|->, line width=0.8pt,
                shorten <=7pt, shorten >=7pt},
    pair/.style={anchor=east},
    every node/.style={inner sep=0pt, outer sep=0pt}
]
    \node[anchor=east] (G) at (0,0) {$(G,r)$};

    \node[pair] (pairG) at (5,0)
        {$(G^{\phantom{\prime}},c^{\phantom{\prime}})$};

    \node (Q)   at (8.1,0)  {$K$};
    \node (out) at (11.8,0) {$(G^{*},\,r^{*})$};

    \draw[map] (G.east) --
        node[above=3pt] {$\CR$} (pairG.west);
    \draw[map] (pairG.east) --
        node[above=3pt] {$\CG$} (Q.west);
    \draw[map] (Q.east) --
        node[above=3pt] {$\invr_{c(r)}$}
        (out.west);

    \coordinate (top) at (0,1);
    \draw[map] (G.east |- top) --
        node[above=3pt] {$\Can$} (out.west |- top);

    \node[anchor=east] (H) at (0,-1.2) {$(G',r')$};
    \node[pair] (pairH) at (5,-1.2) {$(G',c')$};

    \draw[map] (H.east) --
        node[above=3pt] {$\CR$} (pairH.west);
    \draw[map] (pairH.east) --
        node[above=3pt] {$\CG$}
        (Q.south |- pairH.east) --
        node[midway, right=5pt, yshift=-5pt]
            {$\text{iff } (G,r)\iso (G',r')$}
        (Q.south);
\end{tikzpicture}
\caption{The canonization pipeline: $\Can=\invr_{d}\circ\CG\circ\CR$, where $d$ is the color of $r$ in $\CR(G,r)$.}
\label{fig:pipeline}
\end{figure}

In fact, color refinement and the inversion function together also yield a canonization algorithm. For that purpose, given a color~$d$, we define the \emph{rooted inversion} $\invr_d$ as the function mapping $K$ to the rooted graph $(G^*,r^*)$, where $(G^*,c^*)=\inv(K,d)$ and $r^*$ is the unique vertex of color~$d$ in~$G^*$.
We then compose it with $\CR$ and $\CG$, as summarized in Figure~\ref{fig:pipeline} and \Cref{cor:canon}.


\begin{corollary}[Canonization]\label{cor:canon}
Let $\Can$ be the mapping that sends a rooted graph $(G,r)$ to $(G^*,r^*)=\invr_{c(r)}(\CG(G,c))$, where $(G,c)=\CR(G,r)$. Then $\Can$ is a canonization function for connected rooted outerplanar graphs: for any connected rooted outerplanar graphs $(G,r)$ and $(G',r')$, $\Can(G,r)\iso(G,r)$, and $\Can(G,r)=\Can(G',r')$ if and only if $(G,r)\iso(G',r')$.
\end{corollary}

\begin{proof}
By Theorem~\ref{thm:inversion}, $\Can(G,r)\iso(G,r)$. Isomorphic rooted graphs have the same canonical CR coloring, hence the same color graph and the same root color, and $\inv$ is
deterministic, so $\Can$ produces the same output for both. Conversely, if $\Can(G,r)=\Can(G',r')$, then $(G,r)\iso\Can(G,r)=\Can(G',r')\iso(G',r')$.
\end{proof}

\subsection{High-level description of the inversion}\label{sec:intuition}

\begin{figure}[ht]
\centering
\begin{tikzpicture}[
    vertex/.style={circle, minimum size=7mm, inner sep=0pt, draw=black, very thick, font=\Large},
    mini/.style={circle, minimum size=5mm, inner sep=0pt, draw=black, thick, font=\large},
    pink vertex/.style={vertex, fill=cP},
    blue vertex/.style={vertex, fill=cB},
    red vertex/.style={vertex, fill=cR},
    orange vertex/.style={vertex, fill=cO},
    yellow vertex/.style={vertex, fill=cY},
    green vertex/.style={vertex, fill=cG},
    mp/.style={mini, fill=cP},
    mb/.style={mini, fill=cB},
    mr/.style={mini, fill=cR},
    mo/.style={mini, fill=cO},
    my/.style={mini, fill=cY},
    mg/.style={mini, fill=cG},
    block/.style={draw, rounded corners=3mm, thick, inner sep=3mm},
    edge/.style={thick},
    tree edge/.style={thick},
    arr/.style={-{Stealth[length=2.5mm]}, thick, bend left=15},
    cdot/.style={circle, fill=#1, minimum size=4mm, inner sep=0pt, draw=black, thin, font=\tiny},
    biglabel/.style={font=\huge, anchor=north west},
    every label/.style={font=\Large},
  ]

  \node[biglabel] at (-1.3, -0.5) {$G$};
  \node[biglabel] at (-1.3, -7.85) {$K$};
  \node[biglabel, anchor=north east] at (14.5, -0.05) {$T_G$};
  \node[biglabel, anchor=north east] at (14.5, -6.95) {$T_{\Kt}$};

  \begin{scope}[xshift=2.6cm, yshift=-1.6cm]
    \node[pink vertex]   (r)  at (0, 0) {$r$};
    \node[blue vertex]   (bl) at (-2.5, -2) {$u$};
    \node[red vertex]    (re) at (-0.7, -2) {$v$};
    \node[orange vertex] (o1) at (1.0, -2) {$w$};
    \node[orange vertex] (o2) at (2.7, -2) {$w'$};
    \node[yellow vertex] (ya) at (-4.2, -4) {$x$};
    \node[yellow vertex] (yb) at (-3.0, -4) {$x'$};
    \node[yellow vertex] (yc) at (-2.0, -4) {$y$};
    \node[yellow vertex] (yd) at (-0.8, -4) {$y'$};
    \node[green vertex]  (gr) at (2.25, -4) {$z$};

    \draw[edge] (r) -- (bl);
    \draw[edge] (r) -- (re);
    \draw[edge] (r) -- (o1);
    \draw[edge] (r) -- (o2);
    \draw[edge] (bl) -- (ya) -- (yb) -- (bl);
    \draw[edge] (bl) -- (yc) -- (yd) -- (bl);
    \draw[edge] (bl) -- (re);
    \draw[edge] (o1) -- (gr);
    \draw[edge] (o2) -- (gr);
    \draw[edge] (o1) -- (o2);

    \node[draw=cP, semithick, rectangle, fit=(r), inner sep=6pt, label=above:$d$] {};
    \node[draw=cB, semithick, rectangle, fit=(bl), inner sep=6pt, label=left:$g$] {};
    \node[draw=cR, semithick, rectangle, fit=(re), inner sep=6pt, label={[label distance=2mm]below:$e$}] {};
    \node[draw=cG, semithick, rectangle, fit=(gr), inner sep=6pt, label=below:$i$] {};
    \node[draw=cO, semithick, shape=rectangle, fit=(o1)(o2), inner sep=6pt, label=right:$f$] {};
    \node[draw=cY, semithick, shape=rectangle, fit=(ya)(yd), inner sep=6pt, label=left:$h$] {};

    \draw[purple, dashed, thick, rotate around={-45:(-1.15,-1.4)}]
      (-1.3,-1.2) ellipse [x radius=1.4cm, y radius=2.2cm];
    \node[purple, font=\LARGE] at (-2.8, -0.21) {$A_1$};

    \draw[gray, dashed, thick, rotate around={33:(1.7,-2.3)}]
      (1.7,-1.85) ellipse [x radius=1.5cm, y radius=2.95cm];
    \node[gray, font=\LARGE] at (2.54, -0.21) {$A_2$};
  \end{scope}

  \begin{scope}[xshift=12.2cm, yshift=-0.4cm]
    \node[pink vertex] (Rcut) at (0, 0) {$r$};

    \node[mp] (tp) at (-3.4, -1.5) {$r$};
    \node[mb] (tb) at (-3.9, -2.2) {$u$};
    \node[mr] (tr) at (-2.9, -2.2) {$v$};
    \draw[thick] (tp) -- (tb) -- (tr) -- (tp);
    \begin{scope}[on background layer]
      \node[block, fit=(tp)(tb)(tr)] (tri) {};
    \end{scope}
    \draw[tree edge] (Rcut) -- (tri);

    \node[mp] (dp) at (3.4, -1.5) {$r$};
    \node[mo] (do1) at (2.8, -2.2) {$w$};
    \node[mo] (do2) at (4, -2.2) {$w'$};
    \node[mg] (dg) at (3.4, -2.9) {$z$};
    \draw[thick] (dp) -- (do1) -- (dg) -- (do2) -- (dp);
    \draw[thick, dashed] (do1) -- (do2);
    \begin{scope}[on background layer]
      \node[block, fit=(dp)(do1)(do2)(dg)] (dia) {};
    \end{scope}
    \draw[tree edge] (Rcut) -- (dia);

    \node[blue vertex] (Bcut) at (-3.4, -3.8) {$u$};
    \draw[tree edge] (tri) -- (Bcut);

    \node[mb] (xb) at (-4.6, -5.2) {$u$};
    \node[my] (xa) at (-5.05, -5.9) {$x$};
    \node[my] (xc) at (-4.15, -5.9) {$x'$};
    \draw[thick] (xa) -- (xb) -- (xc);
    \draw[thick, dashed] (xa) -- (xc);
    \begin{scope}[on background layer]
      \node[block, fit=(xb)(xa)(xc)] (xblock) {};
    \end{scope}
    \draw[tree edge] (Bcut) -- (xblock);

    \node[mb] (yb2) at (-2.2, -5.2) {$u$};
    \node[my] (ya2) at (-2.65, -5.9) {$y$};
    \node[my] (yc2) at (-1.75, -5.9) {$y'$};
    \draw[thick] (ya2) -- (yb2) -- (yc2);
    \draw[thick, dashed] (ya2) -- (yc2);
    \begin{scope}[on background layer]
      \node[block, fit=(yb2)(ya2)(yc2)] (yblock) {};
    \end{scope}
    \draw[tree edge] (Bcut) -- (yblock);


    \begin{scope}[on background layer]
      \node[draw=purple, dashed, thick, rounded corners=5mm, fit=(tri), inner sep=4pt] (trienc) {};
      \node[draw=gray, dashed, thick, rounded corners=5mm, fit=(dia), inner sep=4pt] (diaenc) {};
    \end{scope}
    \node[purple, font=\LARGE] at (trienc.north west) [xshift=-3mm, yshift=3mm] {$A_1$};
    \node[gray, font=\LARGE] at (diaenc.north east) [xshift=3mm, yshift=3mm] {$A_2$};
  \end{scope}

  \begin{scope}[xshift=2.6cm, yshift=-8.2cm]
    \node[pink vertex]   (Kd) at (0,0) {$d$};
    \node[blue vertex]   (Kg) at (-2.5,-2) {$g$};
    \node[red vertex]    (Ke) at (-0.7,-2) {$e$};
    \node[orange vertex] (Kf) at (1.85,-2) {$f$};
    \node[yellow vertex] (Kh) at (-2.5,-4) {$h$};
    \node[green vertex]  (Ki) at (1.85,-4) {$i$};

    \draw[edge] (Kd) -- (Kg);
    \draw[edge] (Kd) -- (Ke);
    \draw[edge] (Kg) -- (Ke);
    \draw[edge] (Kg) -- (Kh);
    \draw[edge] (Kd) -- (Kf);
    \draw[edge] (Kf) -- (Ki);

    \path (Kd)--(Kf) node[font=\Large, inner sep=1pt, pos=0.3, above right] {$2$};
    \path (Ki)--(Kf) node[font=\Large, inner sep=2pt, pos=0.3, left] {$2$};
    \draw[edge, dashed] (Kf) to[out=85, in=25, looseness=7] (Kf);
    \node[font=\Large, right=-1.5pt of Kf, yshift=-4.5pt] {$\times 2$};
    \path (Kg)--(Kh) node[font=\Large, inner sep=2pt, pos=0.3, right] {$4$};
    \draw[edge, dashed] (Kh) to[out=210, in=150, looseness=7] (Kh);
    \node[font=\Large, right=-1.5pt of Kh, yshift=-4.5pt] {$\times 4$};
  \end{scope}

  \begin{scope}[xshift=12.2cm, yshift=-7.3cm]
    \node[pink vertex] (KRcut) at (0, 0) {$d$};

    \node[mp] (ktp) at (-3.1, -1.5) {$d$};
    \node[mb] (ktb) at (-3.6, -2.2) {$g$};
    \node[mr] (ktr) at (-2.6, -2.2) {$e$};
    \draw[thick] (ktp) -- (ktb) -- (ktr) -- (ktp);
    \begin{scope}[on background layer]
      \node[block, fit=(ktp)(ktb)(ktr)] (ktri) {};
    \end{scope}
    \draw[tree edge] (KRcut) -- (ktri);

    \node[blue vertex] (KBcut) at (-3.1, -3.5) {$g$};
    \draw[tree edge] (ktri) -- (KBcut);

    \node[mb] (keb) at (-3.1, -4.8) {$g$};
    \node[my] (key) at (-3.1, -5.5) {$h$};
    \draw[thick] (keb) -- (key);
    \begin{scope}[on background layer]
      \node[block, fit=(keb)(key)] (kedgeblock) {};
    \end{scope}
    \draw[tree edge] (KBcut) -- (kedgeblock);

    \node[mp] (b1p) at (3.1, -1.5) {$d$};
    \node[mo] (b1o) at (3.1, -2.2) {$f$};
    \draw[thick] (b1p) -- (b1o);
    \begin{scope}[on background layer]
      \node[block, fit=(b1p)(b1o)] (block1) {};
    \end{scope}
    \draw[tree edge] (KRcut) -- (block1);

    \node[orange vertex] (KOcut) at (3.1, -3.5) {$f$};
    \draw[tree edge] (block1) -- (KOcut);

    \node[mo] (b3o) at (3.1, -4.8) {$f$};
    \node[mg] (b3g) at (3.1, -5.5) {$i$};
    \draw[thick] (b3o) -- (b3g);
    \begin{scope}[on background layer]
      \node[block, fit=(b3o)(b3g)] (block3) {};
    \end{scope}
    \draw[tree edge] (KOcut) -- (block3);

    \begin{scope}[on background layer]
      \node[draw=purple, dashed, thick, rounded corners=8mm, fit=(ktri), inner sep=6pt] (kenc1) {};
      \node[draw=gray, dashed, thick, rounded corners=8mm, fit=(block1)(KOcut)(block3), inner sep=6pt] (kenc2) {};
    \end{scope}
    \node[purple, font=\LARGE] at (kenc1.north west) [xshift=-3mm, yshift=3mm] {$de$-chain};
    \node[gray, font=\LARGE] at (kenc2.north east) [xshift=3mm, yshift=3mm] {$di$-chain};
  \end{scope}

\end{tikzpicture}
\caption{The colored graph $G$ of Figure~\ref{fig:ex color graph} (top left), its block-cut tree $T_G$ (top right), its color graph $K$ (bottom left), and the block-cut tree $T_{\Kt}$ of its underlying graph (bottom right). The block $A_1$ of $G$ is discrete and is isomorphic to the unique c-block of the $de$-chain.  The block $A_2$ is symmetric and is the unfolding of the $di$-chain, which consists of two c-blocks. The dashed edges of $T_G$ are those produced by the self-loops of $K$ when unfolding the corresponding chain: $ww'$ from the loop at $f$, $xx'$ and $yy'$ from the loop at $h$. The other edges correspond to edges of the c-blocks of the corresponding chains.}
\label{fig:4types}
\end{figure}
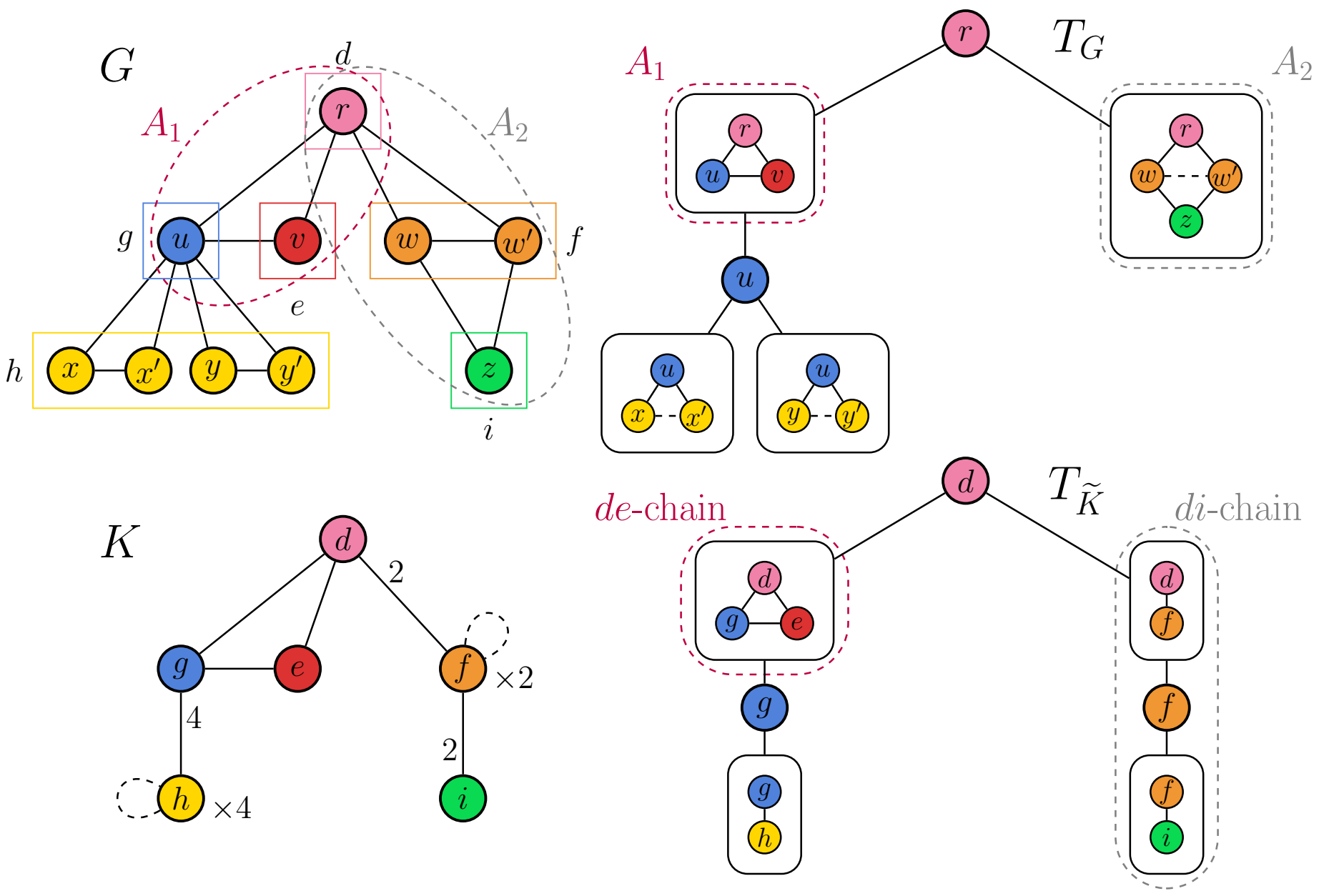

Our inversion algorithm reconstructs an isomorphic copy of $G$ from $(K,d)$ by
reconstructing $T_G$ from $T_{\Kt}$. Its correctness mainly comes from two refinements of the Dichotomy Lemma~\ref{lem:dichotomy} which are given in the next subsection (see \Cref{lem:sym_structure} and \Cref{lem:c-block-dichotomy}).

\smallskip
\noindent\textbf{Symmetric blocks and chains.}
The Dichotomy Lemma~\ref{lem:dichotomy} implies that each block of $G$ can be reconstructed from a
certain sequence of c-blocks obtained by descending in $T_{\Kt}$ (see \Cref{lem:sym_structure}). 
For that purpose, let $(i=c_0,c_1,\dots,c_\ell=j)$ be a descending path in~$\Kt$, i.e. $\depthK(c_0)<\cdots<\depthK(c_\ell)$, such that each of its edges $c_{a-1}c_a$ lies in a c-block~$B_a$ rooted at its upper end~$c_{a-1}$.  The union $C=B_1\cup\dots\cup B_\ell$ is called an \emph{$ij$-chain}, and the path its \emph{underlying path}.  
See for example the $di$-chain with underlying path $(d,f,i)$ in \Cref{fig:4types} or \Cref{fig:diamond} for a more complex chain.   
As $T_{\Kt}$ is a tree, two colors are joined by at most one chain, so a chain
can be identified with the ordered pair~$(i,j)$ of the endpoints of its
underlying path.

\smallskip
\noindent\textbf{Reconstructing a block from $K$: unfolding.}
By definition of $K$, a discrete block~$A$ of $G$ is
clearly isomorphic to the subgraph of~$\Kt$ induced by the colors of~$A$.
By the Block maximality Lemma~\ref{lem:block_maximality}, this subgraph is in fact a c-block $B$, which we view as an $ij$-chain consisting of the single c-block $B$ with root $i$, where $j$ is \fpass{the smallest} neighbor of $i$ in $B$.
A symmetric block is isomorphic to the \emph{unfolding}  of an
$ij$-chain, obtained as follows: take two copies of the chain $B_1\cup\dots\cup B_\ell$, identify the
two copies of~$i$, and add an edge between the two copies of every
color~$s$ with $K_{ss}>0$. In addition, the two copies of $j$ are identified when $K_{jj}=0$.
The resulting graph $\unfold(i,j,K)$ is colored by giving each copy $v$ of a color $s$ the color $c(v)=s$. 
Note that when $K_{jj}>0$, an edge is added between the two copies of $j$, closing the outer face of the resulting graph,
as the edges $xx'$ and $yy'$ of the two blocks hanging below $u$ in \Cref{fig:4types}.

In both cases, $i$ is the root of the first c-block of the chain and the color of the root of the block.
%
Both types are illustrated in
Figure~\ref{fig:4types}:  the discrete block $A_1$ of~$G$ is isomorphic to the
c-block induced by the colors $d,e,g$, and corresponds to the pair $(d,e)$.  The
symmetric block~$A_2$ uses the colors $d,f,i$, which induce the $di$-chain
made of the two c-blocks $\{d,f\}$ and $\{f,i\}$; taking two copies of this
path, identifying their endpoints (as $K_{ii}=0$), and adding an edge between
the two copies of~$f$ (as $K_{ff}=1$) yields a copy of~$A_2$.


\smallskip
\noindent\textbf{Block type and closing colors.}
A block of~$G$ is thus described by a pair of colors $(i,j)$, and, together with~$K$, this pair determines the block up to
color-preserving isomorphism.  Its type is determined too: it
is symmetric if and only if $j$ is what we call a \emph{closing} color, a
property that can be read off~$K$. More precisely, for any color $j\neq d$, there is a unique c-block $B$ containing $j$ whose root differs from $j$: if $j$ belongs to several c-blocks, it is a cut vertex of $\Kt$ and $B$ is the parent of its node in $T_{\Kt}$. 
If $j$ has a neighboring color $k\in B\setminus\{j\}$ with $K_{jk}>1$, $j$ is said to be \emph{solo closing} (see, e.g., color $i$ in \Cref{fig:4types}). 
If $j$ has a self-loop, that is $K_{jj}>0$, $j$ is said to be \emph{paired closing} (see, e.g., color $h$ in \Cref{fig:4types}). In both cases, $j$ is said to be \emph{closing}.

\smallskip
\noindent\textbf{Multiplicity of a c-block.}
Consider a block $A$ of~$G$ with root $v$. If $A$ is discrete, it must be isomorphic to a c-block $B$ of $\widetilde{K}$ containing $i=c(v)$ as root
by the Dichotomy Lemma~\ref{lem:dichotomy} and depth correspondence between $G$ and $\widetilde{K}$ (see Lemma~\ref{lem:depth}). Now, for any color $s$ adjacent to $i$ in $B$, $v$ has exactly one neighbor of color $s$ in $A$, and there must thus be $K_{is}$ isomorphic copies of $A$ hanging below $v$ in $T_G$. 
If $A$ is symmetric, the colors of the neighbors of $v$ in $A$ are neighbors of $i$ in $\widetilde{K}$ and all belong to a c-block $B$ of $\widetilde{K}$ containing $i$ as root (see \Cref{rem:i-neighbors-c-block} after \Cref{lem:sym_structure} in the next subsection).
 For each such color $s$, $v$ has either one or two neighbors of color $s$ in $A$ by the Dichotomy Lemma~\ref{lem:dichotomy}.
 In any case, we define the \emph{multiplicity} of a c-block $B$ with root $i$ as $\mathrm{mult}(B)=\max_{s\in V(B)\setminus\{i\}} K_{is}$. 
If $A$ is symmetric, we show that the number of isomorphic copies of $A$ hanging below $v$ in $T_G$ is exactly $\lfloor\mathrm{mult}(B) / 2\rfloor$ (see \Cref{lem:c-block-dichotomy}).

\smallskip
\noindent\textbf{Blocks as deepest well-closed chains.}
What is not apparent locally is whether a given $ij$-chain corresponds to blocks of~$G$ at all, and how many of them hang below each vertex of color~$i$: this depends on the weights of~$K$  and recovering it is one of the main difficulties.  
For that purpose, we say that an $ij$-chain $B_1\cup\dots\cup B_\ell$ is \emph{well-closed} if $j$ is
a closing color, none of the intermediate colors on its underlying path is
solo closing, and the multiplicities of $B_2,\ldots,B_{\ell}$ are all odd.
This parity condition ensures that, at each vertex colored by a root of $B_2,\ldots,B_\ell$, symmetric blocks consume neighbors in pairs and exactly one neighbor is left for the chain to continue (see the discussion after \Cref{lem:c-block-dichotomy}).
We additionally say that a well-closed $ij$-chain is \emph{deepest} if it is not a proper prefix of any other well-closed $ik$-chain. 
The deepest well-closed $ij$-chains whose first c-block has multiplicity at least two are exactly those that correspond to symmetric blocks of $G$,
while the c-blocks that do not belong to any deepest well-closed $ij$-chain are exactly those that correspond to discrete blocks of $G$ (see \Cref{lem:c-block-dichotomy}).

\smallskip
The inversion algorithm proceeds in two phases.

\smallskip
\noindent\textbf{Phase 1: a bottom-up scan of $T_{\Kt}$.}
A deepest well-closed chain is recognizable from its lower end, which carries a closing color,
while how far up it reaches depends on the weights of~$K$.  Scanning the
c-blocks of~$T_{\Kt}$ from the leaves up, we determine, for each c-block~$B$ with
root~$i$:
\begin{itemize}
\item the color~$j$ at which the deepest well-closed chains through~$B$ end, if any: it is either
      passed on from below, or a closing color of~$B$ starting a new chain
      here;
      there is at most one such~$j$ (see \Cref{lem:c-block-dichotomy});
\item the type $B$ corresponds to: symmetric if such a $j$ exists, discrete otherwise;
\item how many isomorphic blocks associated with the $ij$-chain 
      hang below a vertex of color~$i$:  $\lfloor\mathrm{mult}(B)/2\rfloor$ in the symmetric case, $\mathrm{mult}(B)$ in the discrete case
      (see \Cref{lem:c-block-dichotomy});
\item 
      in the symmetric case, $j$ is passed on to the parent of~$B$ if $\mathrm{mult}(B)$ is odd.
\end{itemize}

\smallskip
\noindent\textbf{Phase 2: a top-down construction of $T_G$ and $G$.}  We start with a
single vertex $r$ of color~$d$ for the root and expand both the tree and the graph.  At a vertex~$v$ of color~$i$,
Phase~1 tells us which pairs $(i,j)$ correspond to a block below~$v$ and with which
multiplicity. If there is at least one such block, then $v$ is a cut vertex and the corresponding node is inserted in the tree.  It remains to build the corresponding blocks --- a copy of the
c-block in the discrete case, the unfolding of the $ij$-chain in the symmetric one
--- to attach the prescribed number of copies of each to~$v$, and to recurse on
the vertices thus created. 

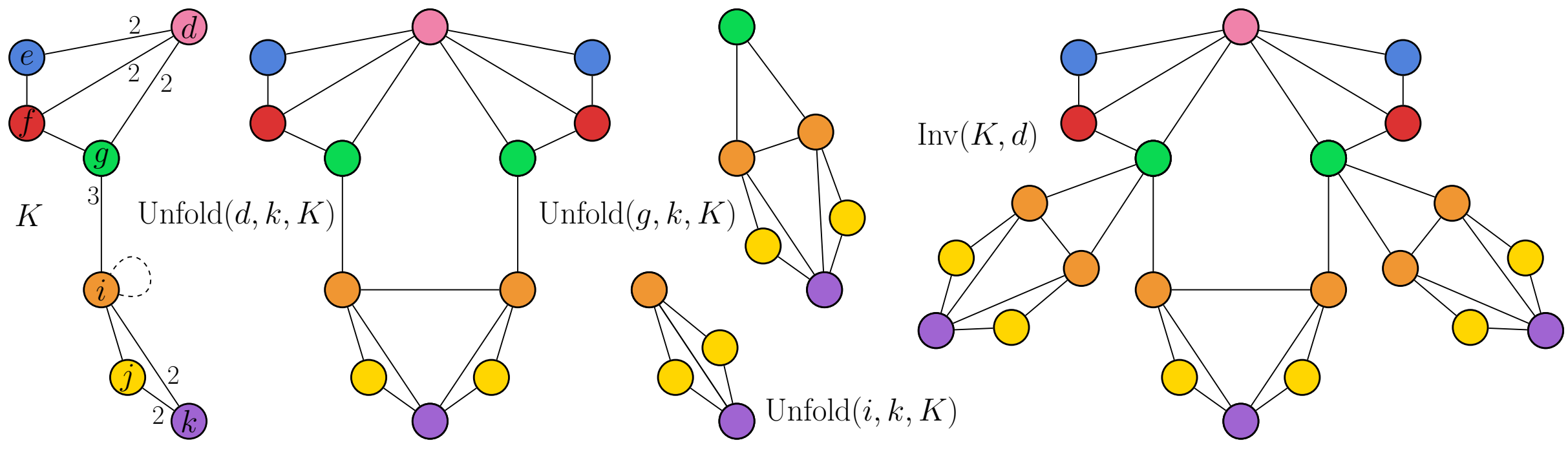
\begin{figure}[ht]
\centering
\begin{tikzpicture}[
    vertex/.style={circle, minimum size=8mm, inner sep=0pt, draw=black, very thick, font=\huge},
    edge/.style={thick},
    elabel/.style={font=\LARGE, inner sep=1pt},
  ]

  \newif\iflab
  \def\checklab#1{\labfalse\checkaux#1\relax\endcheck}
  \def\checkaux#1#2\endcheck{\if L#1\labtrue\fi}

  \def\trig#1{
    \checklab{#1}
    \node[vertex, fill=cO] (#1i) at (0,-3) {\iflab$i$\fi}; 
    \node[vertex, fill=cY] (#1j) at (0.6,-5) {\iflab$j$\fi}; 
    \node[vertex, fill=cV] (#1k) at (2,-6) {\iflab$k$\fi}; 
    \draw[edge] (#1i)--(#1j)--(#1k)--(#1i);
  }

  \def\edgetrig#1{
    \checklab{#1}
    \node[vertex, fill=cG] (#1g) at (0,0)     {\iflab$g$\fi}; 

    \trig{#1}

    \draw[edge] (#1g)--(#1i);
  }

  \def\wzUnfold#1{
    \begin{scope}[rotate=#1]
      \edgetrig{#1l}
      \begin{scope}[rotate={atan2(-6,2)}, yscale=-1, rotate={-atan2(-6,2)}]
        \edgetrig{#1r}
      \end{scope}
    \end{scope}
  }

  \def\gkUnfold#1{
    \begin{scope}
      \edgetrig{#1l}
      \begin{scope}[rotate={atan2(-6,2)}, yscale=-1, rotate={-atan2(-6,2)}]
        \edgetrig{#1r}
      \end{scope}
    \end{scope}
    \draw[edge] (#1li)--(#1ri);
  }

  \def\ikUnfold#1{
    \begin{scope}[shift={(0,3)}]
      \trig{#1l}
      \begin{scope}[shift={(0,-3)}, rotate={atan2(-3,2)}, yscale=-1, rotate={-atan2(-3,2)}, shift={(0,3)}]
        \trig{#1r}
      \end{scope}
    \end{scope}
  }

  \def\dkchain#1{
    \checklab{#1}
    \node[vertex, fill=cP] (#1d) at (2,3)    {\iflab$d$\fi}; 
    \node[vertex, fill=cB] (#1e) at (-1.7,2.3)     {\iflab$e$\fi}; 
    \node[vertex, fill=cR] (#1f) at (-1.7,0.8) {\iflab$f$\fi}; 
    \node[vertex, fill=cG] (#1g) at (0,0)     {\iflab$g$\fi}; 
    \draw[edge] (#1d)--(#1e)--(#1f)--(#1g)--(#1d);
    \draw[edge] (#1d)--(#1f);

    \edgetrig{#1}
  }

  \def\dkUnfold#1{
    \dkchain{#1l}
    \begin{scope}[xshift=4cm, xscale=-1]
      \dkchain{#1r}
    \end{scope}
    \draw[edge] (#1li)--(#1ri);
  }

  \def\K#1{
    \checklab{#1}
    \dkchain{#1}

     \draw[edge, dashed] (#1i) to[out=-15, in=60, looseness=7] (#1i);

    \iflab
      \path (#1d)--(#1e) node[elabel, pos=0.35, above right, yshift=2] {2};
      \path (#1d)--(#1f) node[elabel, pos=0.3, below, yshift=-2] {2};
      \path (#1d)--(#1g) node[elabel, pos=0.3, below right] {2};
      \path (#1g)--(#1i) node[elabel, pos=0.2, left] {3};
      \path (#1i)--(#1k) node[elabel, pos=0.8, above right] {2};
      \path (#1k)--(#1j) node[elabel, pos=0.25, below left] {2};
    \fi


    \iflab\node[font=\huge] at (-1.65,-1.3) {$K$};\fi

  }

  \K{L} 

  \begin{scope}[xshift=5.5cm]
    \dkUnfold{U}
    \node[font=\huge, anchor=east] at (-0,-1.3) {$\mathrm{Unfold}(d,k,K)$};
  \end{scope}

  \begin{scope}[shift={(14.5,3)}]
    \gkUnfold{G}
    \node[font=\huge, anchor=east] at (0.15,-4.3) {$\mathrm{Unfold}(g,k,K)$};
  \end{scope}
  \begin{scope}[shift={(12.5,-3)}]
    \ikUnfold{I}
    \node[font=\huge, anchor=east] at (7.2,-2.8) {$\mathrm{Unfold}(i,k,K)$};
  \end{scope}

  \def\Inv#1#2{
    \dkUnfold{#1}
    \begin{scope}[rotate=#2]
      \gkUnfold{#1p}
    \end{scope}
    \begin{scope}[xshift=4cm, xscale=-1]
      \begin{scope}[rotate=#2]
        \gkUnfold{#1q}
      \end{scope}
    \end{scope}
  }

  \begin{scope}[xshift=24cm]
    \Inv{N}{-70}
    \node[font=\huge] at (-4,0.5) {$\mathrm{Inv}(K,d)$};
  \end{scope}

\end{tikzpicture}
\end{document}
\caption{A color graph $K$ (left), the unfolding of the $dk$-chain, the unfoldings of the $gk$-chain and of the $ik$-chain, and the graph $G=\inv(K,d)$. Note that the weight $K_{ij}=1<2$ implies that $G$ has no block isomorphic to $\unfold(i,k,K)$.}\label{fig:unfold}    
\end{figure}

See \Cref{fig:unfold} for more complex unfoldings and an inversion example.
Phase~1 and Phase~2 take time linear in the size of $K$ and of $G$, respectively.
Phase~2 is straightforward once Phase~1 is available. 
Processing the pairs $(i,j)$ in lexicographic order and numbering lexicographically the vertices according to their color and copy number yields a canonical labeling of the reconstructed graph $\inv(K,d)$. A canonical labeling of $G$ can be obtained from the color-preserving isomorphism from $G$ to $\inv(K,d)$ which is constructed in the proof of Theorem~\ref{thm:inversion} in \Cref{app:phase2}.


\subsection{Correctness and main technical highlights}\label{sec:details}

We highlight the main lemmas behind the correctness of Phase~1.

\smallskip
\noindent\textbf{Structure of blocks.}
The main difficulty arises from symmetric blocks.
We first use the following refinement of the symmetric part of the Dichotomy Lemma~\ref{lem:dichotomy}. 

\begin{restatable}[Symmetric Block Structure]{lemma}{sym_structure}\label{lem:sym_structure}
Let $A$ be a symmetric block with outer face ordering $v_0, v_1, \dots, v_{k-1}$ (with $v_0=\ro(A)$). Define its branches as $L = \{v_0, v_1, \dots, v_{\lfloor k/2 \rfloor}\}$ and $R = \{v_0, v_{k-1}, \dots, v_{\lceil k/2 \rceil}\}$, and let $i=c(v_0)$ and $j=c(v_{\lfloor k/2 \rfloor})=c(v_{\lceil k/2 \rceil})$. If $S=c(A)$ is the set of colors appearing in~$A$, we then have:
\begin{enumerate}
\item $\widetilde{K}[S]$ is an $ij$-chain.
\item $A[L]$ and $A[R]$ are color-preserving isomorphic to~$\widetilde{K}[S]$.
\item $L\cap R=\{\ro(A)\}$ if $j$ is paired closing, 
      and $L\cap R=\{\ro(A),w\}$, where $w=v_{\lfloor k/2 \rfloor}=v_{\lceil k/2 \rceil}$ with $k$ even, if $j$ is solo closing.
\item Two vertices of~$A$ lying on opposite branches, neither of them in $L\cap R$, are adjacent if and only if they have the same color~$s$ and $K_{ss}>0$.
\end{enumerate}
\end{restatable}

The proof mainly relies on properties of block-cut trees and on the depth correspondence (\Cref{lem:depth}).
Outerplanarity is \fpass{mainly} used to show that no edge can link two vertices lying on opposite branches unless they are mirror images of each other.

The above lemma implies that $\unfold(i,j,K)$ is a color-preserving isomorphic copy of $A$.
Note that it also implies that $\widetilde{K}[S]$ is outerplanar, as $A$ is.
In particular, when an $ij$-chain corresponds to a symmetric block, all its c-blocks are outerplanar and its underlying path uses only outer face edges. Conversely, these two properties easily imply that the unfolding $\unfold(i,j,K)$ of an $ij$-chain satisfying them is outerplanar.

\begin{remark}\label{rem:i-neighbors-c-block}
Note that for any neighbor $w$ of $v_0$ in $A$, the edge $v_0w$ and the path from $w$ to $v_0$ along the outer face induce a cycle with pairwise distinct colors by the Dichotomy Lemma~\ref{lem:dichotomy}. By choosing $w=v_p$ with maximum $p\le k/2$, this implies that the colors of neighbors of $v_0$ in $A$ are all in a c-block $B$ containing $i$. Moreover, $i$ must be the root of $B$ by the depth correspondence and \Cref{lem:block_maximality}. 
\end{remark}

The fact that a discrete block is always isomorphic to a c-block is immediate from the definition of $\widetilde{K}$ and the Block maximality Lemma~\ref{lem:block_maximality}. For both types, the block can thus easily be reconstructed in linear time (in the size of the block) from an appropriate pair $(i,j)$ and the color graph~$K$ (see Appendix~\ref{app:block_structure} for the full proof).

As all colors of any c-block of $\Kt$ appear in a common block of $G$ (Block maximality Lemma~\ref{lem:block_maximality}), each c-block is color-preserving isomorphic either to a discrete block of $G$ or to a subgraph of a symmetric block (\Cref{lem:sym_structure}). In particular, we obtain the following.

\begin{restatable}{corollary}{tildeKouterplanar}\label{cor:K-outerplanar}
    Given a connected rooted outerplanar graph $(G,r)$, the underlying graph $\Kt$ of its color graph is outerplanar.
\end{restatable}

We conjecture that this still holds for any equitable coloring of a connected outerplanar graph (not necessarily with a singleton color class).

\smallskip
\noindent\textbf{Identifying and counting blocks.}
The following dichotomy for c-blocks is at the heart of Phase~1 of the inversion algorithm. It is proved in Appendix~\ref{app:block_count}.

\begin{restatable}[C-Block dichotomy and counting]{lemma}{cblockdichotomy}\label{lem:c-block-dichotomy}
    Given a c-block $B$ of $\widetilde{K}$ with root $i$, 
    \begin{itemize}
        \item either $B$ is the first c-block of a deepest well-closed $ij$-chain for a unique color $j$, and any vertex $v$ of color $i$ in $G$ has exactly $\lfloor\mathrm{mult}(B)/2\rfloor$ children in $T_G$ that are color-preserving isomorphic to $\unfold(i,j,K)$ (and thus symmetric),
        \item or $B$ does not belong to any well-closed chain, and any vertex $v$ of color $i$ in $G$ has exactly $\mathrm{mult}(B)$ children in $T_G$ that are color-preserving isomorphic to $B$ (and thus discrete).
    \end{itemize}
\end{restatable}

Note that the uniqueness of $j$ is key in the bottom-up approach of Phase~1, as at most one such $j$ is passed on to the parent.
The proof goes up the chain from its lower end~$j$, showing that every vertex~$v$ whose color $s\neq i$ lies in the chain has a symmetric up-block. The hard part is when $s$ lies on the underlying path.
We then rely on a parity argument: letting $t$ be the next color on this path, the neighbors of $v$ of color $t$ that lie in blocks hanging below $v$ come in pairs, two per symmetric block, by induction. Since the chain is well-closed, $v$ has an odd number of neighbors of color $t$, so one of them lies in the parent block of $v$, which is thus symmetric. Finally, the lowest vertices of this parent block have color $j$, as otherwise the chain could be extended beyond~$j$, contradicting that it is deepest.

\section{Non-connected outerplanar graphs}\label{sec:non-connected}

We now provide two extensions of our results to non-connected outerplanar graphs. 

\subsection{$(1,1)$-WL distinguishes outerplanar graphs}

\textbf{$(1,1)$-WL equivalence \cite{RattanS23}.}
Two graphs $G$ and $H$ are \emph{$(1,1)$-WL equivalent} if there exists a bijection $\phi:V(G)\rightarrow V(H)$ such that CR does not distinguish the rooted graphs $(G,r)$ and $(H,\phi(r))$, i.e. $\mathrm{CG}(\mathrm{CR}(G,r))=\mathrm{CG}(\mathrm{CR}(H,\phi(r)))$, for every vertex $r\in V(G)$. Otherwise, we say that $(1,1)$-WL distinguishes $G$ and $H$.

Since CR distinguishes connected rooted outerplanar graphs according to Theorem~\ref{thm:main}, it then distinguishes $(G,r)$ and $(H,r')$ for any choice of roots $r\in V(G)$ and $r'\in V(H)$ if $G$ and $H$ are non-isomorphic connected outerplanar graphs. This implies that a bijection witnessing $(1,1)$-WL equivalence cannot exist at all: not even a single vertex $r$ can be mapped to an appropriate $r'$. We thus obtain that $(1,1)$-WL distinguishes connected outerplanar graphs. We can in fact generalize this to all outerplanar graphs.

\wlthm*

\begin{proof}
    First, given two isomorphic outerplanar graphs $G$ and $H$, any isomorphism $\phi$ is a witness of their $(1,1)$-WL equivalence.

    Conversely, let $G$ and $H$ be outerplanar graphs, and suppose that a bijection $\phi:V(G)\rightarrow V(H)$ witnesses their $(1,1)$-WL equivalence. 
    For each connected component $C$ of $G$, consider a vertex $r\in V(C)$ and let $C'$ be the connected component of $H$ containing $\phi(r)$.
    The \fpass{color graphs of $(G,c)=\CR(G,r)$ and $(H,c')=\CR(H,\phi(r))$ are equal; as no edge leaves a component, the restrictions $c|_C$ and $c'|_{C'}$ are equitable and have the same color graph, namely the connected component of $\CG(G,c)$ containing the root color, in which this color is a singleton class. By Theorem~\ref{thm:inversion}, $(C,c|_C)\iso(C',c'|_{C'})$, hence $(C,r)\iso(C',\phi(r))$.}
    Thus $\phi$ maps every vertex of a component of $G$ to a vertex of an isomorphic component of $H$. Symmetrically, $\phi^{-1}$ maps every vertex of a component of $H$ to a vertex of an isomorphic component of $G$. Hence $G$ and $H$ have the same isomorphism classes of connected components.

    It remains to show that each isomorphism class has the same multiplicity  in both graphs. 
    Fix a connected component $C$ of $G$, and let  $\mathcal{C}$ and $\mathcal{C}'$ be the sets of components of $G$ and  $H$, respectively, that are isomorphic to $C$. 
    Since $\phi$ maps injectively any vertex $r$ in a component of $\cal C$ to a vertex $\phi(r)$ in $\cal C'$, we conclude that $|V(\mathcal{C})|\le |V(\mathcal{C}')|$ and thus $|\cal C|\le |\cal C'|$ since all these isomorphic components have the same number of vertices. By symmetry, we also have $|\cal C'|\le |\cal C|$, hence $|\cal C|=|\cal C'|$. We conclude that $G$ and $H$ have the same number of components from each isomorphism class, and so they are isomorphic.
\end{proof}

\subsection{CR distinguishes pseudo-rooted outerplanar graphs}

\textbf{Pseudo-rooted graphs.}
For a coloring $c$ of a graph $G$, we say that $c$ \emph{pseudo-roots} $G$ if for any connected component $C$ of $G$ there exists a color $d_C$ such that $c^{-1}(d_C)\cap V(C)$ is a singleton.
We say that a vertex-colored graph $(G,c_0)$ is \emph{CR-pseudo-rooted} if $\mathrm{CR}(G,c_0)$ pseudo-roots $G$. 
For instance, if the initial coloring $c_0$ assigns color $d$ to a single vertex in each connected component of $G$ and color $e$ to all other vertices, we get that $(G,c_0)$ is CR-pseudo-rooted. This is simply because CR can only refine the initial partition of color classes, and thus each connected component still contains exactly one element in the color class $c^{-1}(d)$. Pre-selecting a root in each connected component is only one way to obtain CR-pseudo-rootedness; even starting from a uniform coloring $c_0$, CR may isolate a vertex in each component solely from the structure of that component.
We now extend our main theorem to CR-pseudo-rooted graphs. 

\begin{theorem}\label{thm:CR-pseudo-rooted}
    CR distinguishes CR-pseudo-rooted outerplanar graphs.
\end{theorem}

We prove \Cref{thm:CR-pseudo-rooted} by extending our inversion algorithm to CR-pseudo-rooted graphs. For that purpose, we rely on the knowledge of color classes cardinalities. Recall that given a graph $G$ and an equitable coloring $c$ of $G$, we assume that its color graph $K=\mathrm{CG}(G,c)$ is equipped with vertex weights $w_i=|c^{-1}(i)|$ for each color $i\in[|c|]$.

\begin{theorem}\label{thm:inversion:CR-pseudo-rooted}
    Let $(G,c)$ be a pseudo-rooted outerplanar graph with color graph $K$ such that $c$ is equitable. Then a vertex-colored graph $(G^*,c^*)=\mathrm{Inv}_{pr}(K)$ can be deterministically computed from $K$ in linear time in the size of the output so that $(G^*,c^*)$ is color-preserving isomorphic to $(G,c)$, i.e. $(G^*,c^*)\iso(G,c)$.
\end{theorem}

The idea of the algorithm is to select a color $d$ of minimum
weight among all colors in $K$. It then considers the
connected component $K'$ of $K$ containing $d$ and applies the
inversion algorithm $\mathrm{Inv}(K',d)$ of
Section~\ref{sec:intuition}. It adds $w_d$ disjoint copies of the
resulting vertex-colored graph to the output, removes $K'$ from
$K$, and repeats until no colors remain. The following lemma
justifies both the choice of $d$ and the number of copies.
\begin{lemma}\label{lem:inversion:CR-pseudo-rooted-2}
    Let $c$ be an equitable coloring that pseudo-roots a graph $G$,
    with color graph $K$ and weights $w_i=|c^{-1}(i)|$.
    Let \fpass{$K'$ be a connected component of $K$ and let $d$ be a color of
    minimum weight among the colors of $K'$.}
    Then every connected component of $G$ containing a color
    of $K'$ contains exactly one vertex of color $d$.
    Consequently, there are precisely $w_d$ such connected
    components.
\end{lemma}

\begin{proof}
    Let $C$ be a connected component of $G$ containing a color
    of $K'$. By equitability, if a color $i$ occurs in $C$ and
    $K_{ij}>0$, then color $j$ also occurs in $C$. Together with
    the connectedness of $C$, this implies that the colors
    occurring in $C$ are exactly the vertices of $K'$.
    Since $c$ pseudo-roots $G$, there exists a color $d'$ such that
    $|c^{-1}(d')\cap V(C)|=1$. Choose a path
    $P=d_0,\ldots,d_k$ in $K'$ with $d_0=d$ and $d_k=d'$, and
    write $n_i=|c^{-1}(d_i)\cap V(C)|$.
    For each $i\in[k]$, equitability implies that every vertex
    of color $d_{i-1}$ has exactly $K_{d_{i-1}d_i}$ neighbors
    of color $d_i$, and every vertex of color $d_i$ has exactly
    $K_{d_id_{i-1}}$ neighbors of color $d_{i-1}$.
    Since no edge leaves $C$, counting these edges from each
    endpoint gives $n_{i-1}K_{d_{i-1}d_i} = n_iK_{d_id_{i-1}}$.
    Applying the same argument to the entire graph $G$ gives $w_{d_{i-1}}K_{d_{i-1}d_i} = w_{d_i}K_{d_id_{i-1}}$.
    All $n_i$ are positive, and both matrix entries are positive
    because consecutive colors are adjacent in $K'$. Hence
    \[
        \frac{n_{i-1}}{n_i}
        = \frac{K_{d_id_{i-1}}}{K_{d_{i-1}d_i}}
        = \frac{w_{d_{i-1}}}{w_{d_i}}.
    \]
    Multiplying these identities along $P$ yields $\frac{n_0}{n_k}=\frac{w_d}{w_{d'}}$. 
    Since $n_k=1$ and $w_d$ is minimum among the weights \fpass{of the colors of $K'$,}
    we obtain
    \[
        0<n_0=\frac{w_d}{w_{d'}}\le 1.
    \]
    Thus $n_0=1$, so $C$ contains exactly one vertex of color $d$.
    As this holds for every connected component containing a
    color of $K'$, their number is precisely $w_d$.
\end{proof}
We can now establish correctness. For each connected component
$K'$ of $K$, \Cref{lem:inversion:CR-pseudo-rooted-2} shows that exactly $w_d$ connected components
of $G$ have color graph $K'$, and that each contains a unique vertex
of color $d$. Theorem~\ref{thm:inversion} therefore implies that all
these components are color-preserving isomorphic to the output of
$\mathrm{Inv}(K',d)$. Taking $w_d$ copies reconstructs their disjoint
union. Processing every connected component of $K$ consequently
produces a vertex-colored graph color-preserving isomorphic to
$(G,c)$.

%


\bibliography{refs}

\clearpage
\appendix

\centerline{\Huge\bf Appendix}
\bigskip
\subsection*{Outline}\label{app:outline}

In the Appendix, we  develop the inversion algorithm and prove its correctness, together
with the tools it requires.

Recall that the algorithm takes as input a color graph~$K$ together with a distinguished root color, which we take to be the color~$1$ throughout
this appendix. The inversion proceeds in two phases.  Phase~1 scans the block-cut
tree~$T_{\widetilde{K}}$ of~$\widetilde{K}$ from the bottom up and
determines, using only~$K$, which chains correspond to blocks of~$G$ and how
many copies of each hang below a vertex of a given color.  Phase~2 starts from
a single vertex of color~$1$ and expands the graph, attaching to each vertex
the blocks prescribed by Phase~1 and recursing on the vertices created.

To make this precise, we assign to every block of~$G$ a \emph{block color}: a
pair consisting of the color of its block root and of a second
color, carefully chosen among those appearing in the block and defined
in Appendix~\ref{app:block_structure}.  We then show that this pair is
exactly the pair of endpoints of the corresponding chain in~$\widetilde{K}$,
and we establish the two lemmas on which both phases rest.
\begin{itemize}
    \item The Block Structure Lemma
(Lemma~\ref{lem:block-structure}), which states that the block color of a block,
together with~$K$, determines the block completely: the block is either
a copy of the c-block $\Bup^K(j)$ (the c-block where $j$ is not root as defined in Appendix~\ref{app:setting}), or the unfolding $\mathrm{Unfold}(i,j,K)$
of the $ij$-chain described in Section~\ref{sec:intuition}.
\item The Block Count Lemma
(Lemma~\ref{lem:block-count}) which states that, for every vertex of color~$i$ and
every color~$j$, the number of blocks of block color~$(i,j)$ hanging below that
vertex depends only on~$i$, $j$ and~$K$.
\end{itemize}

The Appendices are organized as follows.  Appendix~\ref{app:setting} gives
the definitions and structural properties used throughout, followed by a
toolkit of auxiliary lemmas.  Appendix~\ref{app:dichotomy} proves the Dichotomy
Lemma.  Appendix~\ref{app:block_structure} defines block colors,
proves the Block Structure Lemma and constructs the function $\B_K$ that
produces the block from its block color.  Appendix~\ref{app:block_count} proves
the Block Count Lemma and constructs the function $F_K$ that gives the
counts; it also proves the dichotomy for c-blocks of
Lemma~\ref{lem:c-block-dichotomy}. Appendix~\ref{app:inversion} puts everything together into the inversion
algorithm: Section~\ref{app:phase1} shows how the bottom-up scan of
Section~\ref{sec:intuition} computes~$F_K$, and Section~\ref{app:phase2} gives
the top-down construction of~$G^*$ and proves Theorem~\ref{thm:inversion}.
Finally, Appendix~\ref{app:rocq-dicho-lem} gives the statement of the Dichotomy Lemma
proven in the Rocq proof assistant.

\section{Setting and Toolkit}\label{app:setting}
\subsection{Definitions and structural properties}\label{app:definitions}

\subsubsection*{Depth}

\begin{definition}\label{def:depth}
The \emph{depth} of a vertex $v\in V(G)$ is its distance to the root,
$\depth_G(v)=d_G(r,v)$.  The \emph{depth} of a color $i\in[|c|]$ is its
distance to the color~$1$, $\depthK(i)=d_{\widetilde{K}}(1,i)$.
\end{definition}

\depthlemma*

\begin{proof}
We prove the stronger statement: for every $v\in V(G)$ and every color
$i\in[|c|]$, $\min\{d_G(v,u):c(u)=i\}=d_{\widetilde{K}}(c(v),i)$.

\emph{Upper bound on~$d_G$.}  Let $(s_0=c(v),s_1,\dots,s_m=i)$ be a shortest
path from $c(v)$ to~$i$ in~$\widetilde{K}$.  Starting at~$v$, greedily build a
walk in~$G$: from a vertex of color~$s_p$, equitability guarantees a neighbor
of color~$s_{p+1}$.  This yields a walk of length~$m$ from~$v$ to a vertex of
color~$i$, so $\min\{d_G(v,u):c(u)=i\}\le m=d_{\widetilde{K}}(c(v),i)$.

\emph{Upper bound on~$d_{\widetilde{K}}$.}  Let $u$ be a vertex of color~$i$
closest to~$v$ and let $P=(v=u_0,\dots,u_\ell=u)$ be a shortest $vu$-path.  By
equitability, two consecutive vertices of~$P$ have colors that are either equal
or adjacent in~$\widetilde{K}$.  Deleting from $(c(u_0),\dots,c(u_\ell))$ every
entry equal to its predecessor therefore yields a walk from $c(v)$ to~$i$
in~$\widetilde{K}$ of length at most~$\ell$, whence
$d_{\widetilde{K}}(c(v),i)\le\ell=d_G(v,u)$.

Setting $i=1$: the unique vertex of color~$1$ is~$r$, giving
$\depth_G(v)=\depthK(c(v))$.
\end{proof}

\begin{definition}\label{def:descending}
A path $(s_0,\dots,s_\ell)$ in~$\widetilde{K}$ is \emph{descending} if
$\depthK(s_{p+1})=\depthK(s_p)+1$ for every $p\in[0,\ell-1]$.
\end{definition}

\subsubsection*{Block Structure}

\begin{lemma}[Block root]\label{lem:highest}
Every block~$A$ of~$G$ has a unique vertex of minimum depth, its \emph{block
root} $\ro(A)$, namely the vertex corresponding to the parent of~$A$ in~$T_G$:
for every other $w\in V(A)$, $\depth_G(w)>\depth_G(\ro(A))$.
\end{lemma}

\begin{proof}
The root of~$T_G$ is a cut vertex node and block nodes and cut vertex nodes
alternate along every path of~$T_G$, so $A$ has a parent, which is a cut vertex
node; let $v$ be the corresponding vertex.  Then $v\in V(A)$, and $v$ is either
a cut vertex or~$r$.  If $v=r$, then $\depth_G(v)=0$ and, $r$ being the
only vertex of depth~$0$, the conclusion is immediate.  Otherwise $v$ is a cut
vertex and, since it corresponds to the parent of~$A$, removing~$v$ places all
vertices of $V(A)\setminus\{v\}$ in connected components not containing~$r$.
Hence $v$ lies on every $rw$-path in~$G$ for $w\in V(A)\setminus\{v\}$, giving
$\depth_G(w)=\depth_G(v)+d_G(v,w)\ge\depth_G(v)+1$.
\end{proof}

\begin{remark}\label{rk:children}
For every $v\in V(G)$, the blocks~$A$ with $\ro(A)=v$ are exactly the children
of~$v$ in~$T_G$; we call them the \emph{child blocks} of~$v$.
\end{remark}

\begin{lemma}[Unique up-block]\label{lem:up_block}
For every $v\in V(G)\setminus\{r\}$ there is a unique block~$A$ containing~$v$
with $\ro(A)\neq v$.  We call it the \emph{up-block} of~$v$ and write
$\Bup^G(v)=A$.
\end{lemma}

\begin{proof}
If $v$ is not a cut vertex, it belongs to exactly one block~$A$.  Since
$v\neq r$ and $v$ is not a cut vertex, Lemma~\ref{lem:highest}~(Block
root) gives $\ro(A)\neq v$, and uniqueness is immediate.

If $v$ is a cut vertex, it belongs to several blocks, namely those whose nodes
are adjacent to the node of~$v$ in~$T_G$.  By Remark~\ref{rk:children}, the
children of that node are exactly the blocks with block root~$v$.  Since
$v\neq r$, the node of~$v$ also has a parent~$B_p$, and $\ro(B_p)$ is the
vertex corresponding to the parent of~$B_p$ in~$T_G$, which is different
from~$v$.  Hence $B_p$ is the unique block containing~$v$ with $\ro(B_p)\neq v$.
\end{proof}

\begin{remark}\label{rem:K_rooted}
Neither Lemma~\ref{lem:highest}~(Block root) nor
Lemma~\ref{lem:up_block}~(Unique up-block) uses outerplanarity or the
coloring: both hold for any connected rooted graph.  We apply them to
$\widetilde{K}$ rooted at the color~$1$, which gives, for every c-block~$B$, a
unique color $\ro(B)\in V(B)$ of minimum depth, and, for every color $i\neq1$,
a unique c-block $\Bup^K(i)$ (its up-c-block), containing~$i$ whose block root is not~$i$.  In
particular, $\depthK(\ro(B))<\depthK(s)$ for every
$s\in V(B)\setminus\{\ro(B)\}$.
\end{remark}

\subsubsection*{Inside a block}

\begin{definition}[Face-depth, low vertices]\label{def:high_low}
Let $A$ be a block of~$G$.  The \emph{face-depth} $\facedepth_A(v)$ of a vertex
$v\in V(A)$ is the length of a shortest path from $\ro(A)$ to~$v$ along the
outer face of~$A$, that is, along its Hamiltonian cycle, ignoring chords; for a
bridge, the face-depth of the non-root vertex is~$1$.  We set
$$\Low(A)=\{v\in V(A):\facedepth_A(v)\text{ is maximum in }A\}.$$
\end{definition}

Note that face-depth and depth are not monotonically related: a vertex of
larger face-depth may have smaller depth, because of the shortcuts created by
chords.

\begin{definition}[Branches]\label{def:branch}
Let $A$ be a block of~$G$ with $k=|V(A)|\ge3$, and let
$\ro(A)=v_0,v_1,\dots,v_{k-1}$ be the vertices of the outer face of~$A$ in one
of the two traversal directions.  The \emph{left branch} of~$A$ is the path
$L=(v_0,v_1,\dots,v_{\lfloor k/2\rfloor})$ and its \emph{right branch} is the
path $R=(v_0,v_{k-1},v_{k-2},\dots,v_{\lceil k/2\rceil})$.  The choice of
the traversal direction is inconsequential, and all the statements in the sequel are
invariant under exchanging $L$ and~$R$.
\end{definition}

Note that at most two vertices share any given face-depth, one on each branch,
and the vertices in $\Low(A)$ are $v_{\lfloor k/2\rfloor},v_{\lceil k/2\rceil}$.
In other words, the \emph{branches} of~$A$ are the two paths obtained by
traversing the outer face of~$A$ from $\ro(A)$ in each of the two directions
and stopping at~$\Low(A)$.  Furthermore, if $|V(A)|=k$ is even, then
$v_{\lfloor k/2\rfloor}=v_{\lceil k/2\rceil}$ and this vertex is contained in
both branches, so that $|\Low(A)|=1$ and $L\cap R=\{v_0,v_{k/2}\}$;  if $k$ is
odd, then $|\Low(A)|=2$ and $L\cap R=\{v_0\}$.

\subsection{Auxiliary lemmas}\label{app:toolkit}

\begin{lemma}[Cut vertex separation]\label{lem:cut_vertex_sep}
Let $v\neq r$ be a cut vertex of~$G$, and let $u$ be a vertex
in a connected component of~$G\setminus\{v\}$ not
containing~$r$.  Then
\begin{enumerate}
\item $\depth_G(u)>\depth_G(v)$;
\item $c(v)$ separates~$c(u)$ from~$1$ in~$\widetilde{K}$,
      in particular, $c(v)$ is a cut vertex of~$\widetilde{K}$.
\end{enumerate}
\end{lemma}

\begin{proof}
The key observation is that every path from~$r$ to~$u$ passes through~$v$, which gives
$$
\fpass{\depth_G(u)=d_G(r,u)=d_G(r,v)+d_G(v,u)\geq\depth_G(v)+1.}
$$
To show the second property, suppose for contradiction that there exists a path from~$c(u)$
to~$1$ in~$\widetilde{K}$ avoiding~$c(v)$.  By equitability,
from~$u$ we can follow neighbors along this color sequence,
obtaining a walk from~$u$ to a vertex of color~$1$ (hence to~$r$)
in~$G$ that avoids all vertices of color~$c(v)$.  In particular,
this walk does not pass through~$v$, contradicting the fact that
every path from~$u$ to~$r$ passes through~$v$.  Hence $c(v)$ separates~$c(u)$
from~$1$, so $c(v)$ is a cut vertex of~$\widetilde{K}$.


\end{proof}

Note that the converse of second property does not hold: a color may be a cut
vertex of~$\widetilde{K}$ while some (or all) of the vertices
carrying it are not cut vertices of~$G$. For instance, the color~$f$ in
Figure~\ref{fig:4types} is a cut vertex, while the vertices~$w,w'$ are
not. 

\begin{observation}[Root color accessibility]
\label{obs:i-accessible}
Let $A$ be a block of~$G$ with $\ro(A)$ of color~$i$.  For any
vertex~$w$ in~$A$ with $w\neq\ro(A)$, every path from~$w$ to a
vertex of color~$i$ in~$G$ passes through~$\ro(A)$.
\end{observation}

\begin{proof}
\fpass{If $\ro(A)=r$, then $i=1$ and $r$ is the only vertex of color~$i$, so the
claim is immediate.  Otherwise, the vertex~$w$ lies in a connected component
of~$G\setminus\{\ro(A)\}$ not containing~$r$ (proof of
Lemma~\ref{lem:highest}).}  By
Lemma~\ref{lem:cut_vertex_sep}~(Cut vertex separation),
Part~(1), every vertex~$u$ in this component satisfies
$\depth_G(u)>\depth_G(\ro(A))$.  By Lemma~\ref{lem:depth}~(Depth),
every vertex of color~$i$ has depth~$\depth_G(\ro(A))$, so no
vertex of color~$i$ lies in this component.  Hence every path
from~$w$ to a vertex of color~$i$ passes through~$\ro(A)$.
\end{proof}

\blockmaximality*

\begin{proof}
Suppose for contradiction that some color $t\in V(B)\setminus S$ exists.
\fpass{Then $|V(B)|\ge3$, so $B$ is $2$-connected.  Let $s_0=c(\ro(A))$; by
Lemmas~\ref{lem:highest}~(Block root) and~\ref{lem:depth}~(Depth), $s_0$ is
the only color of~$S$ of minimum depth.

We first show that a color $s\in S\cap V(B)$ with $s\neq s_0$ is the block
root of~$B$ as soon as some path~$Q$ of~$B$ from~$s$ to~$t$ has no other
color in~$S$.  Let $v$ be a vertex of color~$s$ in~$A$; as $s\neq s_0$, we
have $v\neq\ro(A)$, so $\Bup^G(v)=A$ (Lemma~\ref{lem:up_block}, Unique
up-block).}  By equitability, from~$v$ we follow neighbors along the color
sequence of~$Q$, yielding a path~$P$ from~$v$ to a vertex~$u$ of \fpass{color~$t$;
as the colors of~$Q$ other than~$s$ are not in~$S$, all vertices of
$P\setminus\{v\}$ lie outside~$A$.  The first edge of~$P$ lies in a block
$A'\neq A$ containing~$v$, so $\ro(A')=v$, and $A'\setminus\{v\}$ lies in a
connected component of $G\setminus\{v\}$ not containing~$r$ (proof of
Lemma~\ref{lem:highest}).  The path $P\setminus\{v\}$ avoids~$v$ and meets
$A'\setminus\{v\}$, so $u$ lies in this component.  By
Lemma~\ref{lem:cut_vertex_sep}.2~(Cut vertex separation), $s$ separates~$t$
from~$1$ in~$\widetilde{K}$.  As $B\setminus\{s\}$ is connected and
contains~$t$, $s$ in fact separates every color of $V(B)\setminus\{s\}$
from~$1$; all these colors thus have depth larger than $\depthK(s)$, so
$s=\ro(B)$ (Remark~\ref{rem:K_rooted}).

Now pick $s_1\in S\cap V(B)$ with $s_1\neq s_0$, and a path of~$B$ from~$s_1$
to~$t$ avoiding~$s_0$ (possible by $2$-connectivity if $s_0\in V(B)$).  Let
$s$ be the last color of~$S$ on this path: the sub-path from~$s$ to~$t$ has
no other color in~$S$ and $s\neq s_0$, so $s=\ro(B)$ by the above.  Then
$s_0\notin V(B)$: otherwise $s_0\neq s$ would give $\depthK(s_0)>\depthK(s)$,
whereas $\depthK(s_0)<\depthK(s)$ as $s\in S\setminus\{s_0\}$.  Since
$|S\cap V(B)|\ge2$, there is a color $s'\in S\cap V(B)$ with $s'\neq s$, and
$s'\neq s_0$.  As $B\setminus\{s\}$ is connected, it contains a path
from~$s'$ to~$t$; let $s''$ be the last color of~$S$ on it.  Then
$s''\neq s$, $s''\neq s_0$, and the sub-path from~$s''$ to~$t$ has no other
color in~$S$, so $s''=\ro(B)=s$ by the above---a contradiction.}
\end{proof}

\begin{lemma}[Edges between colors of a block]\label{lem:edge_exists}
Let $A$ be a block of~$G$, and let $s\neq t$ be two colors
appearing in~$A$ with $K_{st}>0$.  Then $A$ contains an edge
between a vertex of color~$s$ and a vertex of color~$t$.
\end{lemma}

\begin{proof}
Suppose not.  Let $x$ be a vertex of color~$s$ in~$A$; since
$K_{st}>0$, it has a neighbor of color~$t$, and by assumption
every such neighbor lies outside~$A$.

If $x=\ro(A)$, pick any vertex~$y$ of color~$t$ in~$A$; then
$y\neq x$ since $t\neq s$, so $\depth_G(y)>\depth_G(x)$ by
Lemma~\ref{lem:highest}~(Block root), hence
$\depthK(s)<\depthK(t)$ by Lemma~\ref{lem:depth}~(Depth).

If $x\neq\ro(A)$, then $\Bup^G(x)=A$.  Let $z$ be a neighbor
of~$x$ of color~$t$; as $z\notin A$, the edge $\{x,z\}$ lies in a
block $A'\neq A$ containing~$x$, so $x=\ro(A')$ and
$\depth_G(z)>\depth_G(x)$, giving again
$\depthK(s)<\depthK(t)$.

Since $K_{st}>0$ implies $K_{ts}>0$, the same argument with
the roles of $s$ and~$t$ exchanged gives $\depthK(t)<\depthK(s)$, a
contradiction.
\end{proof}

\section{Proof of the Dichotomy Lemma}\label{app:dichotomy}

First, we recall the statement of the lemma.

\dicholemma*

\noindent

\subsubsection*{Proof}


Since $G$ is rooted, there is a distinguished root vertex $r \in V(G)$ whose color is unique, that is, no vertex of $V(G) \setminus \{r\}$ has color $c(r)$.

We prove the statement by strong induction on $n = |V(G)|$. It is easy to verify that in any coloring of $K_1, K_2, K_3,$ or  $P_3$, the color pattern of every block is either discrete or  symmetric, and that in the symmetric case, the reflection $\pi$ is a color-preserving automorphism of the block, as every block of these graphs is a single vertex, a single edge, or a rooted triangle. Thus, \Cref{lem:dichotomy} holds for any $n \leq 3 $.

Assume that $n \geq 4$. The following simple lemma will be essential for our inductive step.

\begin{lemma}\label{lem:two_deg2}
Every connected outerplanar graph on $n\ge 4$ vertices has at
least two vertices of degree at most~$2$.
\end{lemma}
\begin{proof}
We triangulate~$G$ to a maximal outerplanar graph~$G^+$, making every inner
face of~$G^+$ a triangle.  Since \fpass{$G^+$ is maximal outerplanar, its} weak dual~$T$ (one node per
inner face, with two nodes adjacent when their faces share an edge) is a \fpass{tree on $n-2 \geq 2$}
nodes and hence has at least two leaves.  Each leaf corresponds to a
triangle with two edges on the outer face; the vertex opposite the
shared edge has degree~$2$ in~$G^+$ and thus has degree at most $2$ in~$G$.  Thus, each of the
two leaves must correspond to a face containing a vertex of degree at most $2$ in $G$.
\end{proof}

By Lemma~\ref{lem:two_deg2}, $G$ has at
least two vertices of degree at most $2$.  Among these vertices, let $v^*$ be the vertex with maximum depth (recall that the depth of a vertex is its distance to the root $r$). Observe that the depth-maximality of $v^*$ ensures that
$v^*\neq r$, which implies $c(v^*)\neq c(r)$ as $r$ is the unique vertex of color $c(r)$. Let $i = c(v^*)$ and recall that, by
equitability, every vertex of color~$i$ has the same degree as~$v^*$. Note that 
\begin{equation}\label{eq-not-i-neigh}
    \text{a vertex of color $i$ must always have a neighbor of color $j_0 \neq i$.}
\end{equation} 
Otherwise, no edge of $G$ would leave $c^{-1}(i)$, which, by the connectivity of $G$, implies $V(G) = c^{-1}(i)$, contradicting $c(r)  \neq i$.

If $\deg(v^*)=1$, then every vertex of color~$i$ has degree~$1$.  Consider the graph $G'$ obtained from $G$ by removing all vertices of
color~$i$; the initial coloring can naturally be restricted to $G'$, giving an equitable coloring (color~$i$ is removed
entirely, and no other adjacency count changes).  Since $i\neq c(r)$, the
root and its singleton class are preserved. By induction, every block of $G'$ is symmetric or discrete, and in the symmetric case, the corresponding reflection $\pi$ is a color-preserving automorphism. Observe that a block $B$ of $G$ is either a bridge with colors $\{i,j_0\}$ (which is a discrete coloring) or a block of $G'$, in which case it has the same edge set, the same root, and the same outer-face ordering in both graphs. 
In both cases, the coloring must be discrete or symmetric, with the reflection being a color-preserving automorphism in the symmetric case. Thus, \Cref{lem:dichotomy} holds for $G$.


Therefore, we can assume that $\deg(v^*) = 2$. Let $j_0,j_1$ denote the colors of the neighbors of any vertex of color $i$, with $j_0 \neq i$. Moreover, let $I,J_0,$ and $J_1$ denote the sets of vertices of $G$ of colors $i,j_0,$ and $j_1$, respectively. Note that we may have $j_1=i$ or $j_0 = j_1$, but $j_0=j_1=i$ cannot happen by \cref{eq-not-i-neigh}.

Recall that, by \Cref{lem:depth}, all vertices of a given color $a$ have the same depth, which we denote by $\depthK(a)$. We will also use the following immediate consequence of the preliminaries.

\begin{observation}[Uniqueness of root color]\label{obs:root-color-unique}
For any block $B$, the root $\ro(B)$ is the unique vertex of its color in $B$; equivalently, no block rooted at a vertex of color $a$ contains another vertex of color $a$.
\end{observation}
\begin{proof}
By the unique highest vertex property (\Cref{lem:highest}), every $w \in V(B) \setminus \{\ro(B)\}$ satisfies $\depth_G(w) > \depth_G(\ro(B))$; hence $c(w) \neq c(\ro(B))$ by \Cref{lem:depth}.
\end{proof}

Since $v^*$ was chosen as a deepest vertex of degree at most $2$, we can immediately deduce some simple but useful properties.

\begin{claim}[Basic Properties]\label{claim:prop-color-i}
~
   \begin{enumerate}
       \item \label{p1} No vertex of color $i$ can be the root of a block, or equivalently, $c(r) \neq i$ and no vertex of color $i$ is a cut vertex.
       \item \label{p2} For any vertex $u \in I$, its neighbors belong to the same block, in particular, the neighbors of $u$ on the outer face of the block are vertices of color $j_0$ and $j_1$.
       \item \label{p3} For any block $B$ and any vertex $v \in B \cap (J_0 \cup J_1)$, $v$ has at most two neighbors of color $i$ within $B$, that is, $|N(v)\cap B \cap I | \leq 2.$
       \item \label{p4} There is no pair of vertices  $x,y \in V(G) \setminus I$ with $xy \notin E(G)$ that are joined by two internally disjoint paths whose inner vertices are all in $I$. 
       \item \label{p5} At most one of the colors $j_0$ and $j_1$ can be the color of the root of a block containing a vertex of color $i$. Moreover, if $x \in \{0,1\}$ is such that some block rooted at a vertex of color $j_x$ contains a vertex of color $i$, then $\depthK(j_{1-x}) > \depthK(j_x)$ and $K_{j_{1-x}j_x} \leq 1$.
   \end{enumerate}
\end{claim}
\begin{proof}
 \begin{enumerate}
     \item The two formulations are equivalent since the root of a block is always a cut vertex or the root $r$, and every cut vertex $v$ is the root of at least one block in every connected component of $G \setminus \{v\}$ that does not contain $r$. We will prove the second formulation. We have already argued that the depth-maximality of $v^*$ ensures that $v^*\neq r$, and since $r$ is the unique vertex of color $c(r)$, we get that $i \neq c(r)$. Now assume for the sake of contradiction that $v\in I$ is a cut vertex and let $C$ be a connected component of $ G \setminus \{v\}$ that does not contain $r$. Then \fpass{$C$ contains a vertex $w$ with $\deg_G(w) \leq 2$: if $|V(C)| \leq 2$, every vertex of $C$ has at most two neighbors, all in $V(C) \cup \{v\}$; otherwise, $G[V(C)\cup \{v\}]$ is a connected outerplanar graph on at least $4$ vertices, so by \Cref{lem:two_deg2} it has at least two vertices of degree at most $2$, one of which is different from $v$, and a vertex of $C$ has the same degree in $G$ and in $G[V(C)\cup \{v\}]$. By} \Cref{lem:cut_vertex_sep}.1,   $\depth_G(w) > \depth_G(v) = \depthK(i) = \depth_G(v^*) $ (the equalities holding by \Cref{lem:depth}), contradicting the depth-maximality of $v^*$.
 \item The two neighbors of $u$ are in different blocks if and only if $u$ is a cut vertex, which is impossible by Basic Property~\ref{p1}. 
 The second part of the claim is immediate from the equitability of $c$ and the fact that the outer face of a block of size at least $3$ is a cycle.
 \item Since any vertex of $B$ has exactly two neighbors on the outer face of $B$, and every neighbor of $v$ in $I\cap B$ must be next to it on the outer face of $B$, we get that $v$ can have at most two neighbors in $I \cap B$.
 \item Assume, for the sake of contradiction, 
 that there are two internally disjoint $xy$-paths $P_1,P_2$ such that $V(P_1 \cup P_2) \subseteq I \cup \{x,y\}$. Then, since every vertex in $I$ has degree $2$ and $xy \notin E(G)$, $P_1 \cup P_2$ induces a cycle, which by outerplanarity is a block $B$ of $G$, with only $x$ and $y$  as possible cut vertices. This implies that $x$ or $y$ must be the root of this block, giving $c(x) \neq c(y)$ by \Cref{lem:highest}. Note that since both $x$ and $y$ have neighbors in $I$, we must have $x,y \in J_0 \cup J_1$, and thus by $c(x) \neq c(y)$, we get $j_0 \neq j_1$. Assume without loss of generality that $x$ is the root of the block and $u$ is one of its neighbors of color $i$. 
 \fpass{Then every path from $y$ to $r$ passes through $x = \ro(B)$ (see the proof of \Cref{lem:highest}), so that $\depth_G(y) = \depth_G(x) + d_G(x,y) \geq \depth_G(x) + 2 = \depth_G(u) + 1 > \depth_G(u) = \depth_G(v^*)$ as $xy \notin E(G)$, and either $\deg_G(y)= \deg_B(y)$ or $y$ is a cut vertex. In the first case, we get that $\deg_G(y)= 2$ while $\depth_G(y) > \depth_G(v^*)$, a contradiction. In the second case, since $y$ is not the root of $B$, we can apply \Cref{lem:cut_vertex_sep}.1 and \Cref{lem:two_deg2} to a block $B' \neq B$ pending at $y$ and get a contradiction in a way analogous to the proof of  Basic Property~\ref{p1}.}
 \item Let $C$ be a block rooted at a vertex $z$ of color $j_x$ and containing a vertex $u$ of color $i$. By \Cref{obs:root-color-unique} (Uniqueness of root color), $z$ is the unique vertex of color $j_x$ in $C$. The two neighbors of $u$ lie in $C$ (Basic Property~\ref{p2}), so the neighbor of $u$ of color $j_x$ is $z$ itself, while its neighbor $w$ of color $j_{1-x}$ is a non-root vertex of $C$; hence $\depthK(j_{1-x}) = \depth_G(w) > \depth_G(z) = \depthK(j_x)$ by \Cref{lem:highest,lem:depth}. Since the colors $j_0$ and $j_1$ cannot each have greater depth than the other, this proves the first part and the depth inequality. Finally, by \Cref{obs:i-accessible} (Root color accessibility) applied to the block $C$, whose root has color $j_x$, every path from $w$ to a vertex of color $j_x$ passes through $z$, in particular, $z$ is the only possible neighbor of $w$ of color $j_x$, so $K_{j_{1-x}j_x} = |N(w) \cap J_x| \leq 1$.
 
 \end{enumerate}
 
\end{proof}

\newcommand{\caprop}{Chord Above Property\xspace}

 \subsubsection*{The inductive step}

Our plan is to remove all vertices of color $i$ 
and then apply induction on the resulting graph. In particular,  we will apply induction on a graph $G'$ defined in the following way.
   
\begin{definition}[$i$-path, the graph $G'$]
    We say that an $xy$-path $P$ is an $i$-path if $|V(P)| \geq 3$, $x,y \notin I$, and $V(P) \setminus \{x,y\} \subseteq I$.
    Let $G'$ be the simple graph obtained from $G$ by deleting all vertices of color $i$ and adding edges between all pairs $x,y \in V(G')$ for which $x \neq y$, $xy \notin E(G)$, and $G$ has an $i$-path between $x$ and $y$.
\end{definition}

    Note that by the equitability of the coloring $c$, for all edges $xy \in E(G') \setminus E(G)$, we must have  $x,y \in J_0 \cup J_1$. 
In order to apply induction on $G'$,  we will need to show that it is outerplanar and that the original coloring can naturally be extended to an equitable coloring of the new graph. For the second part, the following statement will play a crucial role.

     \begin{restatable}[\caprop]{proposition}{chordabove}\label{prop:chord_above}
     Suppose that $G$ has an $i$-path with distinct endpoints $x$ and $y$.  If $K_{c(x)c(y)}\ge 1$, then $xy\in E(G)$ must hold.
     \end{restatable}
     
     The proof of \Cref{prop:chord_above} is significantly more involved than the proofs of the Basic Properties; therefore, it is presented separately and postponed to \Cref{subsec:chord-above-proof}.
\subsection{Proof of the Dichotomy Lemma assuming the \caprop}
We start by proving the following claim, which will allow us to apply induction to $G'$.
\begin{claim}
$G'$ is a rooted, connected outerplanar graph and the coloring $c|_{V(G')}$ is equitable on $G'$.
\end{claim}
\begin{proof}
Since we only deleted vertices of $G$ and potentially added edges between non-adjacent vertices that were connected on the outer face via a path with all internal vertices of degree $2$, $G'$ is outerplanar.

 If $u,v \in V(G')$ are two distinct vertices that are connected in $G$ by a path $P$, then the path obtained from $P$ by short-cutting every contiguous segment of $P \cap I$ by a direct edge is a $uv$-path in $G'$; therefore, $u$ and $v$ are connected in $G'$, which proves the connectivity of $G'$.
 
To show equitability, first observe that if no new edges were added in $G'$, that is, $G'=G[V\setminus I]$, then since all vertices of color $i$  were deleted and no adjacency counts changed between pairs of other colors, equitability is straightforward. Now assume that some new edges have been added during the construction of $G'$. Since new edges were only added between vertices of $J_0 \cup J_1$, for any pair of colors $a \in [n] \setminus \{i,j_0,j_1\}$ and $b \in [n]$, we still have the property that for any vertex of color $a$ (respectively, color $b$), the number of its neighbors of color $b$ (respectively, color $a$) is $K_{ab}$ (respectively, $K_{ba}$). Therefore, we only need to focus on the adjacency counts between vertices with colors $j_0$ and $j_1$.
\begin{itemize}
    \item If $j_1 \neq i$, then we have two scenarios. If $K_{j_0j_1} \geq 1$, then the \caprop (\Cref{prop:chord_above}) implies that no new edge was added during the construction of $G'$ and so the adjacency counts are preserved. If $K_{j_0j_1} =0$, then every vertex of color $j_0$ gets $K_{j_0i}$ new neighbors of color $j_1$ (note that Basic Property~\ref{p4} ensures that two new neighbors cannot collide). Similarly, every vertex of color $j_1$ gets $K_{j_1i}$ new neighbors of color $j_0$, and so the coloring is equitable.
    \item If $j_1 = i$ and  $K_{j_0j_0} \geq 1$, then similarly to the previous case, the \caprop (\Cref{prop:chord_above}) implies that no new edge was added during the construction of $G'$ and so the adjacency counts are preserved. Thus, we can assume that  $K_{j_0j_0} =0$. We need to show that there is a number $K'_{j_0j_0}$ such that for any vertex $u \in J_0$, the number of its neighbors of color $j_0$ in $G'$ is $K'_{j_0j_0}$. We will show that 
    \begin{equation*}
        K'_{j_0j_0} = \begin{cases} 1 \quad &\text { if  $K_{j_0i}$ is odd } \\ 0 \quad &\text { if  $K_{j_0i}$ is even } \end{cases}\, 
    \end{equation*}
    is the right value.
    By \Cref{obs:root-color-unique} (Uniqueness of root color), no block rooted at a vertex $u \in J_0$ can contain another vertex from $J_0$. Therefore, any block that is rooted at a vertex of color $j_0$ and contains a vertex from $I$ must consist of $3$ vertices only, of colors $j_0,i,i$. Thus, we get that 
    \begin{equation}\label{eq:block-parity-j0}
        \text{for any block $B'$ rooted at a vertex $u\in J_0$, we have $|N(u) \cap B' \cap I| \in \{0,2\}$}.
    \end{equation}
    Recall that, by the construction rules of $G'$, all these pending $j_0,i,i$ triangles disappear in $G'$. In particular, we get that if $u=r$, then $u$ is the only vertex of color $j_0$ and $K'_{j_0j_0} = 0$ is well defined. Thus, we can assume that $u \neq r$. The only new edges introduced were between vertices $u,v \in J_0$, $u \neq v$, for which there is a path $uxyv$ in $G$ with $x,y \in I$. From our previous observation on the structure of blocks rooted at $u$, we can conclude that all these paths must be contained in the unique block $\Bup(u)$ that contains $u$ and is not rooted at $u$, that is, in the up-block of $u$ (\Cref{lem:up_block}). 
    By the construction rules of $G'$, the number of neighbors of $u$ of color $j_0$ in $G'$ is exactly the number of neighbors of $u$ of color $i$ in $\Bup(u)$. By Basic Property~\ref{p3}, we have $|N(u) \cap \Bup(u) \cap I| \in \{0,1,2\}$. First, we show that $|N(u) \cap \Bup(u) \cap I|  = 2$ cannot happen. Assume that this is the case, and consider the ordering $w_0,w_1, \dots, w_k$ of the vertices of $\Bup(u)$ along its outer face, starting with its root $w_0$. Note that $c(w_0) \notin \{i,j_0\}$ by Basic Property~\ref{p1} and \Cref{obs:root-color-unique} (Uniqueness of root color).  Let $w_f$ be the first vertex in the ordering with $N(w_f) \cap \Bup(u) \cap I \neq \emptyset$. By equitability, $w_f \in J_0 \cup I$. Furthermore, observe that $w_f \in J_0$ must hold: if $w_f$ had color $i$, then its predecessor $w_{f-1}$, being one of its two neighbors, would already have a neighbor in $\Bup(u) \cap I$ (the predecessor is well-defined since the root of $\Bup(u)$ is not of color $i$). Observe that, since every vertex in $I \cap \Bup(u)$ is adjacent to its neighbors of color $i$ and $j_0$ along the outer face (Basic Property \Cref{p2}), we get that if $|N(w_f) \cap \Bup(u) \cap I| =2$, then $c(w_{f-1}) = i$ must hold, which implies $c(w_{f-2}) = i$ and $c(w_{f-3}) = j_0$, contradicting the minimality of $w_f$. Here we again used that the predecessors are well-defined since the root of $\Bup(u)$ is not of color $i$ or $j_0$. Therefore, $|N(w_f) \cap \Bup(u) \cap I| =1$, which implies that $K_{j_0i} = |N(w_f) \cap I|$ is odd by \Cref{eq:block-parity-j0}. This contradicts the assumption $|N(u) \cap \Bup(u) \cap I| = 2$, which by the same counting would force $K_{j_0i} = |N(u) \cap I|$ to be even.
    In summary, we have shown that for any vertex $u \in J_0$, $|N(u) \cap \Bup(u) \cap I| \in \{0,1\}$ and for any block $B'$ rooted at $u$, $|N(u) \cap B' \cap I| \in \{0,2\}$. Therefore, if $K_{j_0i}$ is odd, then $|N(u) \cap \Bup(u) \cap I| = 1$ must hold for every $u \in J_0$, implying that every vertex of color $j_0$ has exactly one neighbor of color $j_0$ in $G'$; thus, $c$ is equitable in $G'$. Similarly, if $K_{j_0i}$ is even, then $|N(u) \cap \Bup(u) \cap I| = 0$ must hold for every $u \in J_0$, implying that $G' = G[V \setminus I]$, where the coloring $c$ is clearly equitable.
    
 \end{itemize}
\end{proof}

Therefore, we can apply induction on $G'$ and get that the coloring of any block of $G' $ is either discrete or symmetric, and in the symmetric case, we also get that the reflection $\pi'$ of a symmetric block is a color-preserving automorphism.
Based on this, we will show that the blocks of $G$ are also discrete or symmetric, with the reflection $\pi$ being a color-preserving automorphism in the symmetric case.  \\

If a block $B$ of $G$ does not contain any vertex of color $i$, then  $B$ is also a block in $G'$, with the same edge set, the same root, and the same outer-face ordering (no added edge has both endpoints in $B$: the endpoints of an added edge lie in a common block of $G$ containing vertices of $I$, and two distinct blocks share at most one vertex). Therefore, $c$ is symmetric or discrete on $B$, and in the symmetric case, $\pi = \pi'$ is a color-preserving automorphism of $B$ directly by induction. 

Let $B$ be a block of $G$ with $B \cap I \neq \emptyset$ and consider  $B' = G'[V(B) \setminus I]$. First observe the following property of the construction of $G'$, implied by the fact that the vertices of color $i$ are contained in the same block as their neighbors and they lie between their neighbors on the outer face of their block (Basic Property~\ref{p2}).

\begin{observation}\label{observation:new-edges-are-outer}
For any pair of vertices $p, q \in V(B) \setminus I = V(B')$ connected in $G$ by an $i$-path, the two vertices are adjacent on the outer face of $B'$.
\end{observation}

If $B'$ is a singleton, then $B$ is a triangle rooted at a vertex of $J_0$ and containing two vertices from $I$, which implies that the coloring of $B$ is symmetric and $\pi$, which fixes the root and exchanges the two vertices of color $i$, is a color-preserving automorphism of $B$. 

Therefore, we can assume that $|V(B')|\geq 2$. We continue by arguing that in this case, $B'$ is a block of $G'$.

\begin{claim}
If $|V(B')| \geq 2$, then $B'$ is a block of $G'$ with root $\ro(B)$. 
\end{claim}
\begin{proof}
We will prove the following statement, which essentially implies our claim: two vertices $x,y\in V(G')$ are connected by two internally disjoint paths $P',Q'$ in $G'$ if and only if they are  connected by two internally disjoint paths $P,Q$ in $G$.
Indeed, we can obtain $P',Q'$ from $P,Q$ by short-cutting every maximal $i$-path they contain.  Similarly, we can obtain $P,Q$ by following the paths $P'$ and $Q'$ and inserting an $i$-path in place of every edge $xy \in E(P')\setminus E(G)$ or $xy \in E(Q')\setminus E(G)$; since every internal vertex of an $i$-path has degree $2$, the resulting paths are internally disjoint. 

The preceding observation and the maximality of the block $B$ imply that $B'$ must be a block of $G'$.
\fpass{Finally, $\ro(B) \in V(B) \setminus I$ by Basic Property~\ref{p1}, and it is the root of $B'$: if $\ro(B) \neq r$, it separates $V(B) \setminus \{\ro(B)\}$ from $r$ in $G$, hence $V(B') \setminus \{\ro(B)\}$ from $r$ in $G'$, since a path of $G$ avoiding a vertex of $V(G')$ short-cuts to a path of $G'$ avoiding it; so $\ro(B)$ is the vertex of $B'$ closest to $r$ in $G'$.}
\end{proof}

Thus, the induction hypothesis gives us that for any block $B$ of $G$, the coloring $c'$ of  $B' = G'[V(B) \setminus I]$ is either discrete or symmetric. 
 We will conclude the proof by different arguments based on the relationships between the colors $i,j_0,j_1$. In several cases of this analysis, we will apply the following claim, which reduces the verification of the automorphism property of $\pi$ to that of $\pi'$. 

\begin{claim}\label{claim:autom-lift}
Assume that $|V(B')| \geq 2$ and that we have already proved that the coloring of $B$ is symmetric. Then the restriction of $\pi$ to $V(B')$ coincides with the reflection $\pi'$ of $B'$; moreover, if $\pi'$ is a color-preserving automorphism of $B'$, then $\pi$ is a color-preserving automorphism of $B$.
\end{claim}
\begin{proof}
Since the coloring of $B$ is symmetric, $\pi$ is color-preserving, in particular, $\pi$ maps $I \cap B$ to itself. The outer face of $B'$ is obtained from the outer face of $B$ by deleting the vertices of $I$ (recall that the edges added during the construction of $G'$ short-cut paths along the outer face), and the root of $B'$ is $\ro(B)$. Since we assumed that the coloring of $B$ is symmetric, the restriction of $\pi$ to $V(B')$ fixes the root of $B'$ and reverses its outer-face ordering, that is, it coincides with the reflection $\pi'$ of $B'$.

Assume now that $\pi'$ is a color-preserving automorphism of $B'$. Since $\pi$ is a color-preserving bijection of $V(B)$, it suffices to show that $\pi$ maps every edge of $B$ to an edge of $B$. As $\pi$ reverses the outer-face ordering of $B$, it maps outer-face edges of $B$ to outer-face edges of $B$. Furthermore, every vertex of color $i$ has degree $2$ and both of its incident edges lie on the outer face of $B$. Hence, every remaining edge of $B$, that is, every chord of $B$, has both endpoints in $V(B')$. Let $pq$ be a chord of $B$. Since $p,q \notin I$ and $pq \in E(G)$, the edge $pq$ is also present in $G'$, and thus $pq \in E(B')$, which yields $\pi(p)\pi(q) = \pi'(p)\pi'(q) \in E(B')$. Moreover, $\pi'(p)\pi'(q)$ cannot be one of the edges added during the construction of $G'$: the endpoints $x,y$ of any added edge satisfy $xy \notin E(G)$ and are joined by a path with all internal vertices in $I$, so $K_{c(x)c(y)} = 0$ by the \caprop, whereas $pq \in E(G)$ gives $K_{c(p)c(q)} \geq 1$, which implies $K_{c(\pi'(p))c(\pi'(q))} \geq 1$ as $\pi'$ is color-preserving. Therefore, $\pi'(p)\pi'(q) \in E(G)$, and since a block contains every edge of $G$ joining two of its vertices, we conclude that $\pi(p)\pi(q) \in E(B)$.
\end{proof}

Now we conclude the proof of the Dichotomy Lemma (\Cref{lem:dichotomy}) assuming the \caprop, using different arguments depending on the relationships between the colors $i,j_0,j_1$.
Note that if $|V(B')| = 2$, then $B'$ is a bridge of colors $j_0,j_1$ with $j_1 \neq i $ and $j_1\neq j_0$ as the root of a block is the only vertex of its color
  in the block (\Cref{obs:root-color-unique}), so a bridge with two endpoints of the same color is impossible. Thus, in the first two cases, we can assume $|V(B')| \geq 3$.

\begin{itemize}
\item Case 1: $j_0 = j_1$ (note that this forces $j_1 \neq i$, as $j_0 \neq i$). In this case, every vertex of $I \cap B$ has two distinct neighbors of color $j_0$, both contained in $B$ (Basic Property~\ref{p2}). Thus, $B'$ contains at least two vertices of color $j_0$, in particular, the coloring of $B'$ cannot be discrete; hence it is symmetric by the inductive hypothesis, and so $B'$ contains exactly two vertices $p$ and $q$ of color $j_0$. 
By \Cref{obs:root-color-unique} (Uniqueness of root color), neither of them is $\ro(B)$, so they occupy mirrored positions on the outer face of $B'$.
Moreover, $|I \cap B| = 1$: two vertices of $I \cap B$ would both lie between $p$ and $q$ on the outer face of $B$, forcing the outer face to be a $4$-cycle on these four vertices. However, we have already argued that $\ro(B) \notin \{p,q\}$ and by Basic Property~\ref{p1}, no vertex of $I \cap B$ can be $\ro(B)$, a contradiction.
Let $v$ be the vertex in $I\cap B$. Since $v$ is adjacent to both $p$ and $q$, they are joined by an $i$-path in $G$ and so by \Cref{observation:new-edges-are-outer}, $p$ and $q$ are adjacent on the outer face of $B'$. 
Since $B'$ is symmetric, two vertices of the same color can only be adjacent on the outer face of $B'$ if they are its two low vertices; thus, we get that $\mathrm{Low}(B') = \{p,q\}$ (see \cref{def:high_low}) and that $\mathrm{Low}(B) = \{v\}$.  
Hence, $B'$ was obtained from $B$ by deleting its low vertex $v$ of color $i$ and adding the edge $pq$ if it was not yet present in $G$. 
Thus, since the coloring $c'$ of $B'$ is symmetric, we obtain that the coloring $c$ of $B$ is symmetric as well. Moreover, the induction hypothesis on $B'$ also implies that its reflection $\pi'$ is a color-preserving automorphism of $B'$, and hence by \Cref{claim:autom-lift} we get that $\pi$ is a color-preserving automorphism of $B$. In the rest of the case analysis, we can therefore assume that $j_0 \neq j_1$.

    \item Case 2: $j_1 = i$. 
    Since we have already excluded the case when $B$ is a triangle of vertices of colors $j_0,i,i$ (corresponding to $|V(B')| =1$), Basic Property~\ref{p2} implies that   $B'$ must contain two vertices $p,q$ of color $j_0$. 
    Thus, as in the previous case, the coloring of $B'$ must be symmetric and \Cref{obs:root-color-unique} (Uniqueness of root color) implies that $\ro(B) \notin \{p,q\}$.
    Moreover, $p$ and $q$ are joined by an $i$-path in $G$ and so they are adjacent on the outer face (\Cref{observation:new-edges-are-outer}). This implies that the two low vertices of the block $B'$ are $p$ and $q$. \fpass{Moreover, $|I \cap B| = 2$: the vertices of $I \cap B$ come in pairs $x,y$ of adjacent vertices whose other neighbors are two distinct vertices of color $j_0$ of $B'$, that is, $p$ and $q$; two such pairs would give two internally disjoint $i$-paths from $p$ to $q$, made of outer-face edges of $B$, so the outer face of $B$ would be the $6$-cycle they form, whose root would be $p$ or $q$ by Basic Property~\ref{p1}, a contradiction. Hence, $B$ is obtained from $B'$ by reinserting two vertices of color $i$ between its two low vertices, which} does not break the symmetry of the coloring, and so the coloring of $B$ \fpass{is} symmetric as well. This and the induction hypothesis allow us to apply \Cref{claim:autom-lift} and conclude that $\pi$ is a color-preserving automorphism of $B$.
    In the rest of the case analysis, we can therefore assume that $j_1 \neq i$.
    
     \item Case 3: $i,j_0,j_1$ are $3$ distinct colors and $B'$ is a bridge.
     Let us denote the two vertices of $B'$ by $a$ and $b$. Since $j_1 \neq i$, the two neighbors of any vertex $u \in I \cap B$ have colors $j_0$ and $j_1$ and lie in $V(B) \setminus I = \{a,b\}$ (Basic Property~\ref{p2}); consequently, every vertex of $I \cap B$ is adjacent to both $a$ and $b$. This implies that $|I \cap B| \leq 2$, as three common neighbors of $a$ and $b$ would form a $K_{2,3}$, contradicting the outerplanarity of $G$.
     Without loss of generality, we may assume that $c(a) = j_0$ and $c(b) = j_1$. Since $\ro(B) \notin I$ by Basic Property~\ref{p1}, the root of $B$ is either $a$ or $b$. 
     If $|I \cap B| = 1$, then $B$ is a triangle whose vertices have the pairwise distinct colors $j_0, i, j_1$, so the coloring of $B$ is discrete. If $|I \cap B| = 2$, then applying Basic Property~\ref{p4} to the two disjoint paths joining $a$ and $b$ through the two vertices of $I \cap B$ yields $ab \in E(G)$, so $B$ is a $4$-cycle $a,x,b,y$ plus the chord $ab$ with $c(x) = c(y) =i$. As $\ro(B) \in \{a,b\}$, the outer face of $B$ in counterclockwise order is $\ro(B)= v_0, v_1, v_2,v_3$ with $\{v_1,v_3\} = \{x,y\}$ and $v_2 \in \{a,b\}\setminus\{\ro(B)\}$; hence  $c(v_1) = c(v_3) = i$ and the coloring of $B$ is symmetric. Finally, since $B'$ is a bridge, its reflection $\pi'$ is the identity map on $\{a,b\}$, which is trivially a color-preserving automorphism of $B'$ and so $\pi$ is a color-preserving automorphism of $B$ by \Cref{claim:autom-lift}.

\item Case 4: $i,j_0,j_1$ are $3$ distinct colors and $B'$ is not a bridge; in particular, $|V(B')| \geq 3$.    
 Suppose first that the coloring of $B'$ is discrete. The two neighbors of any vertex of $I \cap B$ have colors $j_0$ and $j_1$ and lie in $B$ (Basic Property~\ref{p2}), so $B'$ contains a vertex of color $j_0$ and a vertex of color $j_1$; by discreteness, it contains exactly one of each, say $p$ and $q$, and every vertex of $I \cap B$ is adjacent to both $p$ and $q$. Then $|I \cap B| = 1$: two such vertices, each consecutive to both $p$ and $q$ on the outer face of $B$, would force this outer face to be a $4$-cycle on these four vertices, making $B'$ a bridge. Hence, $B$ is obtained from $B'$ by inserting a single vertex of color $i$, and since the color $i$ does not appear in $B'$, the coloring of $B$ is discrete as well.
 
 Suppose now that the coloring of $B'$ is symmetric, and let $\ro(B)= v_0, \dots, v_{k-1}$ and $\ro(B)= v'_0, \dots, v'_{k'-1}$ denote the outer faces of $B$ and of $B'$, respectively, in counterclockwise order (so that, as in the statement of \Cref{lem:dichotomy}, $\pi(v_s) = v_{k-s}$, and similarly $\pi'(v'_s) = v'_{k'-s}$ for the reflection $\pi'$ of $B'$).

\begin{claim}\label{claim:j_a-has-i-neigh}
There is a value $a \in \{0,1\}$ such that no block rooted at a vertex of color $j_a$ contains a vertex of color $i$. Moreover, for any such color $j_a$,
\begin{equation}\label{eq:j_a-has-i-neigh}
     \text{every vertex $w \in B \cap J_a$ has at least one neighbor in $I \cap B$.}
 \end{equation}
\end{claim}
\begin{proof}
The first part of the statement is a direct consequence of Basic Property~\ref{p5}. For such a value $a$, we have $c(\ro(B)) \neq j_a$, since $B$ itself contains a vertex of color $i$; hence every vertex $w \in B \cap J_a$ is a non-root vertex of $B$, so the blocks containing $w$ other than $B$ are rooted at $w$ (\Cref{lem:up_block}) and thus contain no vertex of color $i$, that is, $N(w) \cap I \subseteq B \cap I$. Since $K_{j_ai} \geq 1$, we get that
every vertex $w \in B \cap J_a$ has at least one neighbor in $I \cap B$.
 \end{proof}
  Similarly to the previous cases, we have that $B \cap I \neq \emptyset$ and the two neighbors of a vertex of $I \cap B$ lie in $B$ (Basic Property~\ref{p2}), have colors $j_0$ and $j_1$, and become adjacent on the outer face of $B'$ (\Cref{observation:new-edges-are-outer}). Therefore, the block $B'$ has an outer edge $e = y_0 y_1$ with $c(y_0) = j_0$ and $c(y_1) = j_1$, and by the symmetry of the coloring of $B'$, its mirror image $f = \bar y_0 \bar y_1$, where $\bar y_t = \pi'(y_t)$ for $t \in \{0,1\}$, is a second such edge. The edges $e$ and $f$ are distinct: otherwise $\pi'$ would either exchange $y_0$ and $y_1$, which is impossible as they have different colors, or fix both of them, which is impossible as the only fixed points of $\pi'$ are $\ro(B)$ \fpass{and, when $|V(B')|$ is even, the unique vertex of $\mathrm{Low}(B')$, which are not adjacent on the outer face of $B'$ as $|V(B')| \geq 3$.} The edges $e$ and $f$ might, however, share an endpoint; in that case the shared endpoint is a fixed point of $\pi'$ (as $\pi'$ is color-preserving, it cannot be the mirror image of the other endpoint of $e$), and the two non-shared endpoints are mirror images of each other.
  
  It remains to locate the vertices of $I \cap B$. First, observe that $y_0$ and $\bar y_0$ are the only vertices of color $j_0$ in $B$, and $y_1$ and $\bar y_1$ are the only vertices of color $j_1$ in $B$ (by the symmetry of the coloring of $B'$, every color appears at most twice in $B'$, which settles the case $y_t \neq \bar y_t$; and if $y_t = \bar y_t$, a further vertex of color $j_t$ would either come with a distinct mirror image, making $j_t$ appear three times in $B'$, or be a second fixed point of $\pi'$, in which case one of the two vertices of color $j_t$ would be $\ro(B)$, contradicting \Cref{obs:root-color-unique} (Uniqueness of root color)). \fpass{Every vertex $u'$} of $I \cap B$ is consecutive, on the outer face of $B$, to one vertex of color $j_0$ and to one vertex of color $j_1$ (Basic Property~\ref{p2}), \fpass{which are therefore adjacent on the outer face of $B'$ (\Cref{observation:new-edges-are-outer}). These two vertices form one of the pairs $\{y_0,y_1\}$, $\{\bar y_0,\bar y_1\}$, $\{y_0,\bar y_1\}$, $\{\bar y_0,y_1\}$, and the last two are excluded unless they coincide with one of the first two: if $y_0 \neq \bar y_0$, $y_1 \neq \bar y_1$ and $y_0 \bar y_1$ were an outer-face edge of $B'$, so would be its mirror image $\bar y_0 y_1$, and $y_0, y_1, \bar y_0, \bar y_1$ would form the whole outer face of $B'$, so that $\ro(B)$ would share its color with another vertex of $B'$, contradicting \Cref{obs:root-color-unique} (Uniqueness of root color). Hence, $u'$} lies between the two endpoints of $e$ or between the two endpoints of $f$; moreover, each of these two positions hosts at most one vertex of $I \cap B$, since two consecutive vertices of color $i$ would contradict $j_1 \neq i$.
  We will show that each of these positions contains a vertex of $I \cap B$, showing the symmetry of the coloring of $B$.
  
  \begin{description}
      \item[Case A] The edges $e$ and $f$ are vertex-disjoint, or they share an endpoint and the color $j_\alpha$ of the non-shared endpoints $y_\alpha \neq \bar y_\alpha$ satisfies the condition of \Cref{claim:j_a-has-i-neigh}. \fpass{In the former case, let $\alpha = a$ be a value given by \Cref{claim:j_a-has-i-neigh}. In both cases, by} \Cref{eq:j_a-has-i-neigh}, $y_\alpha$ and $\bar y_\alpha$ each have a neighbor in $I \cap B$; the neighbor of $y_\alpha$ lies in the position delimited by $e$, and likewise the neighbor of $\bar y_\alpha$ lies in the position delimited by $f$, so both positions are occupied. Note that the situation where the shared endpoint of $e$ and $f$ is the root $\ro(B)$ falls under this case: then $B$ is a block rooted at a vertex of color $c(\ro(B)) \in \{j_0, j_1\}$ containing a vertex of color $i$, so by the first part of Basic Property~\ref{p5} no block rooted at a vertex of the other color -- the color of the non-shared endpoints -- contains a vertex of color $i$.
      \item[Case B] The edges $e$ and $f$ share an endpoint $w$,  and the color $j_\alpha$ of the non-shared endpoints $y_\alpha \neq \bar y_\alpha$ does not satisfy the condition of \Cref{claim:j_a-has-i-neigh}. 
      This means that there is some block rooted at a vertex of color $j_\alpha$ that contains a vertex of color $i$. 
      Then by \Cref{claim:j_a-has-i-neigh}, the color $j_{1-\alpha} = c(w)$  satisfies \Cref{eq:j_a-has-i-neigh}, and the second part of Basic Property~\ref{p5} implies  $K_{j_{1-\alpha}j_\alpha} \leq 1$. By \Cref{eq:j_a-has-i-neigh}, $w$ has a neighbor $u \in I \cap B$, which occupies one of the two positions, say the one between $y_\alpha$ and $w$ (the other case being symmetric). If  $w$ and $\bar y_\alpha$ did not have a common neighbor in $G$ of color $i$, then $w$ and $\bar y_\alpha$ would be consecutive on the outer face of $B$, so that $w \bar y_\alpha \in E(G)$ and thus $K_{j_{1-\alpha}j_\alpha} \geq 1$; applying the \caprop to the path $y_\alpha, u, w$ would then yield $y_\alpha w \in E(G)$ as well, giving $K_{j_{1-\alpha}j_\alpha} = |N(w) \cap J_{\alpha}| \geq 2$, a contradiction. Hence, both positions are occupied.
  \end{description}
  
  Consequently, $|I \cap B| = 2$ and the coloring of $B$ is obtained from the symmetric coloring of $B'$ by inserting one vertex of color $i$ into each of two positions that are exchanged by the mirror symmetry of the outer face, which gives that the coloring of $B$ is symmetric. Once again, since the coloring of $B'$ is symmetric, the induction hypothesis implies that its reflection $\pi'$ is a color-preserving automorphism of $B'$, and thus \Cref{claim:autom-lift} implies that $\pi$ is a color-preserving automorphism of $B$.

\end{itemize}

In all cases, the coloring of $B$ is discrete or symmetric with $\pi$ a color-preserving automorphism of $B$, which completes the proof of \Cref{lem:dichotomy} (assuming the \caprop, which we prove below).
\subsection{Proof of the \caprop ~(\Cref{prop:chord_above})}\label{subsec:chord-above-proof}

In this subsection, we prove the following statement.

\chordabove*

Throughout the proof, we keep the notation introduced in the proof of the Dichotomy Lemma: $v^*$ denotes a deepest vertex of degree at most $2$, $i = c(v^*)$, the colors $j_0$ and $j_1$ (with $j_0 \neq i$) are the colors of the two neighbors of any vertex of color $i$, and $I, J_0, J_1$ denote the color classes $c^{-1}(i),c^{-1}(j_0),c^{-1}(j_1)$, respectively. Recall that for two colors $a$ and $b$, we write $K_{ab}$ for the number of neighbors of color $b$ of a vertex of color $a$. By the equitability of the coloring $c$, the value of $K_{ab}$ does not depend on the choice of the vertex of color $a$, and $K_{ab} \geq 1$ holds if and only if $K_{ba} \geq 1$ does; both conditions are equivalent to the existence of an edge between the two color classes. Note that the proofs of \Cref{obs:root-color-unique} (Uniqueness of root color) and the Basic Properties 1-5 (\Cref{claim:prop-color-i}) do not rely on the \caprop and so they remain at our disposal.

The proof of \Cref{prop:chord_above} is organized as follows.
\begin{description}
    \item[Step 1.] We show that an $i$-path $P$ cannot intersect several blocks of $G$ and we collect structural properties of the colors $i, j_0, j_1$ within blocks of $G$ (\Cref{observations:j0}).
    \item[Step 2.] Within a block $B$, we highlight all the pairs $\{x,y\}$ that satisfy the assumption of \Cref{prop:chord_above}.
    \item[Step 3.] We show that each of the highlighted pairs must be covered by an edge of the graph.
\end{description}

\begin{proof}[Proof of \Cref{prop:chord_above}]
~

\smallskip
\noindent\textbf{Step 1: setup and structural observations.}

Let $P$ be an $i$-path with endpoints $x \neq y$. 
By the definition of an $i$-path, $P$ has at least one internal vertex and at most two, since, with three or more, some internal vertex would have both of its neighbors in $I$, contradicting $j_0 \neq i$.

In both cases, all vertices of $P$ lie in a common block $B$ of $G$: the two neighbors of a vertex of color $i$ lie in a common block by Basic Property~2 in \Cref{claim:prop-color-i}, and this is the unique block containing that vertex by Basic Property~1 in \Cref{claim:prop-color-i}. Therefore, there is a block $B$ of $G$ that contains $P$.


Let $\ro(B)=v_0, v_1, v_2, \dots, v_{k-1}$ be the sequence of vertices of $B$ counterclockwise along the outer face, starting from the root of $B$.
Let $v_f$ be the first vertex in this order that has a color in $\{i, j_0, j_1\}$ (it exists since $P$ lies in $I \cap B$).

First, observe that $c(v_f) \neq i$. Recall that $\ro(B) \notin I$ by Basic Property~1 and $c(r) \neq i$; therefore, $v_{f-1}$ is well defined. If $v_f$ had color $i$, then its predecessor $v_{f-1}$ would have a color in $\{j_0, j_1\}$, contradicting the minimality of $f$. 

\begin{claim}
If $v_f = \ro(B)$, then the \caprop holds for $P$.
\end{claim}
\begin{proof}
Since $\ro(B) \notin I$,  $v_f = \ro(B)$ implies $c(\ro(B)) \in \{j_0,j_1\}$. Let $t \in \{0,1\}$ be such that $c(\ro(B)) = j_t$. By \Cref{obs:root-color-unique} (Uniqueness of root color), $\ro(B)$ is the unique vertex of color $j_t$ in $B$. 

\fpass{If $P$ has two internal vertices, then $j_1 = i$ and both endpoints $x \neq y$ of $P$ have color $j_0$; as $\ro(B) \notin I$, we get $j_t = j_0$, so the two vertices $x \neq y$ of $B$ would both have color $j_t = c(\ro(B))$, which is impossible.} 

Thus, we can suppose that $P$ has one internal vertex. If $j_0 = j_1$, we get the same contradiction as before. 
Therefore, $j_0 \neq j_1$ and we may assume without loss of generality that $x = \ro(B)$. The other endpoint $y$ then has color $j_{1-t}$ and is a non-root vertex of $B$. By \Cref{obs:i-accessible}, every path from $y$ to a vertex of color $j_t$ passes through $\ro(B)$, in particular, the only possible neighbor of $y$ of color $j_t$ is $\ro(B)$. Since $K_{j_{1-t}j_t} \geq 1$, the vertex $y$ does have such a neighbor, which implies that $y$ must be adjacent to $\ro(B) = x$.
\end{proof}
Therefore, for the rest of the proof we can assume that
\begin{equation}\label{eq:root-color-assumption}
    v_f \neq \ro(B), \qquad \text{or equivalently,} \qquad c(\ro(B)) \notin \{i, j_0, j_1\}.
\end{equation}

Without loss of generality, we can assume that $c(v_f) = j_0$ (this is automatic when $j_1 = i$ since $c(v_f) \neq i$, and the two labels are interchangeable when $j_1 \neq i$). Recall that by Basic Property~2, every neighbor of $v_f$ of color $i$ within $B$ is consecutive to $v_f$ on the outer face of $B$. Since the predecessor of $v_f$ is not of color $i$, we get that
\begin{equation}\label{eq:v_f-one-i-neigh}
    |N(v_f) \cap I \cap B| \leq 1.
\end{equation}
The following structural claim collects some useful properties of the colors $i,j_0,j_1$ within $B$.
     \begin{claim}\label{observations:j0}
      \begin{itemize}
        \item[i)] Let $C$ be the connected component of $G - \ro(B)$ that contains $B \setminus \{\ro(B)\}$. For any vertex $u \in C \cap J_0$, there is a path $P$ from $u$ to $\ro(B)$ which does not use any of the colors in $\{i,j_0,j_1\}$ internally, that is, $P \cap (I \cup J_0 \cup J_1) = \{u\}$.
          \item[ii)] For any chord $v_pv_q$ in $B$ with $v_p,v_q \in (J_0\cup J_1)\cap B$, the vertices $v_s$ with $p < s < q$ cannot have color $j_0$.
          \item[iii)] No vertex $u\in B$ of color $c(u) \in \{i,j_0,j_1\}$ can have a neighbor $v$ of color $j_0$ that does not belong to the block $B$, and no vertex $u \in B\cap J_1$ can have a neighbor $w \in I \setminus B$.
      \end{itemize}
      \end{claim}
      \begin{proof}
      \begin{itemize}
          \item[i)] 
        Let $S = c^{-1}(c(\ro(B))) \cup I \cup J_0 \cup J_1$. We start from the path 
        $$
        Q= \ro(B)=v_0, v_1, \dots, v_{f-1}, v_f.
        $$
       Since $v_f$ is the first vertex of $I \cup J_0 \cup J_1$ in the outer-face order and $\ro(B)$ is the unique vertex of color $c(\ro(B))$ in $B$, none of $v_1, \dots, v_{f-1}$ lies in $S$.  As $c(u) = j_0 = c(v_f)$, the equitability of $c$ yields a walk $W$ with the vertex sequence  $u = u_f, u_{f-1}, \dots, u_0$ in $G$ with $c(u_s) = c(v_s)$ for every $s$ (once $u_s$ has been constructed, the edge $v_sv_{s-1}$ gives $K_{c(v_s)c(v_{s-1})} \geq 1$, so $u_s$ has a neighbor of color $c(v_{s-1})$, which we take as $u_{s-1}$). In particular, $c(u_0) = c(\ro(B))$ and the vertices $u_{f-1}, \dots, u_1$ all lie outside of $S$. Thus, $W$ is a walk from $u$ to a vertex of color $c(\ro(B))$ without any internal vertex from $S$. \fpass{As shown in the proof of \Cref{obs:i-accessible} (Root color accessibility), no vertex of color $c(\ro(B))$ lies in $C$; as $u \in C$, the walk $W$ must leave $C$, hence} contain $\ro(B)$, implying that $u_0 = \ro(B)$. After removing any cycle occurring in $W$, we obtain the required path.
          \item[ii)] The statement follows from part i) combined with outerplanarity, which implies that any path from $v_s$ to $\ro(B)$ would need to contain either $v_p$ or $v_q$ as an internal vertex. (Note that a simple path between two vertices of a block cannot leave that block and re-enter it at another vertex; otherwise, the outside subpath would belong to the same block.)
          \item[iii)] We will use the same argument for both cases. Assume to the contrary that there exists a vertex $u \in B$ violating the condition and consider a vertex $v \in J_0$ that is contained in a block rooted at $u$. In the first case, $v$ is simply the neighbor of $u$ in $J_0 \setminus B$, and in the second case, we can take the neighbor of $w \in I \setminus B$ of color $j_0$ to be $v$. Since $v \notin B$, there is no path from $v$ to $\ro(B)$ that does not use $u$. Since $c(u) \in \{i,j_0,j_1\}$, this contradicts  the first part of the claim.
      \end{itemize}
      \end{proof}
\smallskip
\noindent\textbf{Step 2: highlighted pairs and interesting edges.}
We define a graph $H$ on $J_0 \cup J_1$ with
\begin{equation*}
    xy \in E(H) \iff x\neq y, \text{ there is an } i \text{-path between } x \text{ and } y \text{ and } K_{c(x)c(y)} \geq 1.
\end{equation*}
We refer to the edges of $H$ as highlighted pairs. Note that the \caprop is equivalent to saying that every highlighted pair must be covered by an edge of $G$. In other words, it suffices to show that $E(H) \subseteq E(G)$.

Recall that the endpoints of $i$-paths always lie in $J_0 \cup J_1$. More precisely, they lie in $J_0$ and $J_1$ when
$j_1\notin\{i,j_0\}$, and both lie in $J_0$ when
$j_1\in\{i,j_0\}$.  

\newcommand{\Eint}{\tilde{E}}

We define the set of \emph{interesting edges} $\Eint \subseteq E(G)$ as those edges of $G$ that can cover highlighted pairs; formally, 
 $\Eint$ is the set of edges induced by $J_0$ when
$j_1\in\{i,j_0\}$ and the set of edges between $J_0$ and $J_1$ otherwise.


\smallskip
\noindent\textbf{Step 3: conclusion.}

Recall that, in Step 1, we picked an arbitrary $i$-path $P$ and argued that it is contained in a block $B$. In this final step, we argue that the endpoints of $P$ are adjacent in $G$.
We distinguish two cases depending on the color $j_1$.

\smallskip
\noindent\emph{\bf Case 1: $j_1\in\{i,j_0\}$.}
In this case, we will not only show that the endpoints of $P$ are adjacent but also argue that every highlighted pair in $B$ is covered by an interesting edge.
\begin{claim}\label{claim:highlight-interesting-in-path}
Every highlighted pair and interesting edge in $B$ consists of two consecutive vertices of $J_0\cap B$ in the outer-face order. In addition, if $|J_0 \cap B| \geq 3$ and $u,v$ are consecutive vertices of $J_0 \cap B$ that are connected by an outer-face path $P_{uv}$ containing $\ro(B)$ and no vertex of $J_0 \cap B$, then $\{u,v\}$ is neither a highlighted pair nor an interesting edge.
\end{claim}
\begin{proof}
Since every $i$-path occurs consecutively on the outer face and does not contain any vertex of $J_0 \cap B$ in its interior, the endpoints of highlighted pairs must be consecutive vertices of $J_0 \cap B$.  
If $|J_0 \cap B| \geq 3$ and we consider the special consecutive pair $u,v $ described in the statement of the claim, then  $\{u,v\}$ cannot be a highlighted pair, as its witnessing $i$-path would need to be $P_{uv}$, which is impossible as $c(\ro(B)) \neq i$.

The endpoints of any interesting edge in $B$ are also consecutive; otherwise, that edge is a chord with a $J_0$-vertex strictly between its endpoints, contradicting \Cref{observations:j0}.ii).
Similarly, if the special consecutive pair $\{u,v\}$ were an interesting edge, then it would form a chord separating $J_0 \cap B \setminus \{u,v\}$ from the root, contradicting \Cref{observations:j0}.ii). 
\end{proof}
 \Cref{claim:highlight-interesting-in-path} implies that both
the highlighted pairs and the interesting edges within $B$ are subgraphs of the same spanning path $\Pi$ of $J_0\cap B$, corresponding to their outer-face ordering.


\begin{claim}\label{claim:chord-highlighted-degree}
Every $v \in J_0\cap B$ is contained in exactly one highlighted pair within $B$, that is, $|N_H(v) \cap B| = 1$.
\end{claim}

\begin{proof}
Suppose first that $j_1=j_0$. In this case, \Cref{observations:j0}.iii) gives that every neighbor of $v$ of color $i$ lies in $B$.   Therefore, each $v \in J_0\cap B$ has exactly $K_{j_0i}$ neighbors in $I \cap B$. We know that $K_{j_0i}\geq 1$ and \Cref{eq:v_f-one-i-neigh} shows that its value at $v_f$ is at most one; thus, $K_{j_0i} = 1$. The edge connecting $v$ to its neighbor of color $i$ can always be completed to an $i$-path with endpoint $v$; therefore, $v$ is contained in exactly one highlighted pair.

Now suppose that $j_1=i$.  Observe that any block rooted at a vertex
$v$ that contains a vertex of color $i$ is a triangle of the form
$z-I-I-z$.  Indeed, Basic Property~2 in
\Cref{claim:prop-color-i} implies that both $I$-vertices and their $J_0$-neighbors must be contained in that block, and \Cref{obs:root-color-unique} forces both $J_0$ vertices to
be its root $v$.  Each such triangle contributes two to
$|N_G(v)\cap I|$.  Setting
$
    d_{B\cap I}(v)=|N_G(v)\cap I\cap B|,
$
we get that
\begin{equation}\label{eq:chord-parity}
    d_{B\cap I}(v)\equiv K_{j_0i}\pmod 2
    \qquad (v\in J_0\cap B),
\end{equation}
and Basic Property~3 in \Cref{claim:prop-color-i} gives $d_{B\cap I}(v)\leq 2$. Analogously to the previous case, it is sufficient to prove that $d_{B\cap I}(v) = 1$ and apply the argument that the edge $v - I$ can be completed to an $i$-path. We show that $d_{B\cap I}(z) = 1$ by proving that $K_{j_0i}$ is odd and applying \Cref{eq:chord-parity}.

Since $B$ contains the $i$-path $P$, there exists a vertex $z\in J_0\cap B$ such that $d_{B\cap I}(z)>0$.  Choose the first such $z$ in the outer-face order.  By our previous arguments, we have that $d_{B\cap I}(z)\in\{1,2\}$.   If $d_{B\cap I}(z)=2$, then the two outer-face neighbors
of $z$ lie in $I$.  Let $u$ be the predecessor of $z$.  The other neighbor
$u'$ of $u$ lies in $I$, and the other neighbor $z'$ of $u'$ lies in $J_0$.
If $z'\neq z$, then $z'$ occurs before $z$ and has an $I$-neighbor in $B$,
contradicting the choice of $z$.  If $z'=z$, then $z-u-u'-z$ is a triangle;
the degree-$2$ vertices $u,u'$ force it to be the whole block $B$, contrary
to \Cref{eq:root-color-assumption}.  Hence, $d_{B\cap I}(z)=1$.
Since \Cref{eq:chord-parity} holds for any $v \in J_0 \cap B$, applying it to $z \in J_0 \cap B$, we get that $K_{j_0i}$ is odd. 
\end{proof}

\begin{claim}\label{claim:chord-interesting-degree}
Every vertex $v \in J_0\cap B$ is incident to one interesting edge within $B$.
\end{claim}

\begin{proof}
\Cref{observations:j0}.iii) implies that every $J_0$-neighbor of $v$ must be in $B$.  Hence, every $J_0$-neighbor of $v$ gives an interesting edge and so $v$ is incident to $K_{j_0j_0}$ interesting edges. In the case $j_1 \in \{i,j_0\}$, the global assumption $K_{c(x)c(y)} \geq 1$ gives $K_{j_0j_0}\geq 1$.  As observed above, the interesting edges form a subgraph of the path $\Pi$ on the ordered set $J_0\cap B$.  An endpoint
of that path is incident to at most one interesting edge, so
$K_{j_0j_0}=1$.
\end{proof}

By \Cref{claim:chord-highlighted-degree} and
\Cref{claim:chord-interesting-degree}, the highlighted pairs and the
interesting edges spanned by $B$ are two perfect matchings of $J_0 \cap B$ that are subgraphs of the same spanning path $\Pi$ of $J_0 \cap B$.  A path has at
most one perfect matching, so the two matchings are the same, that is, $E(H) = \Eint$.
Hence, every highlighted pair in $B$ is covered by an edge of $G$.

\smallskip
\noindent\emph{\bf Case 2: $j_1\notin\{i,j_0\}$.}

In this case, we set the more modest goal of showing that the highlighted endpoints of $P$ form an interesting edge.

The three colors are now distinct.  For every $u\in I$, let
$z(u)\in J_0$ and $w(u)\in J_1$ be its two neighbors. Now every $i$-path is of the form
$z(u)-u-w(u)$ yielding the highlighted pair
$\{z(u),w(u)\}$. 
The interesting edges $\Eint$ are the edges \fpass{of $G$ between $J_0$ and $J_1$ (in the whole graph $G$, not only in $B$).}

In this case, we consider $H$ as a multigraph, that is, if there are $u,u' \in I \cap B$ such that $z(u) = z(u')$ and $w(u) = w(u')$, then the pair $\{w(u),z(u)\}$ appears twice as a highlighted pair. 
We set the constants
\[
  h_0=K_{j_0i},\quad h_1=K_{j_1i},\qquad
  e_0=K_{j_0j_1},\quad e_1=K_{j_1j_0}.
\]
Thus, $h_t$ is the number of highlighted-pair occurrences at a vertex of
$J_t$, while $e_t$ is the number of interesting edges incident to a vertex of $J_t$. Note that our assumptions give  $h_0,h_1,e_0,e_1>0$. 
Double counting first the highlighted pairs and then the interesting edges gives
\[
  |J_0|h_0=|I|=|J_1|h_1,
  \qquad
  |J_0|e_0=|J_1|e_1,
\]
and therefore
\begin{equation}\label{eq:balance-ratio}
  e_0h_1=h_0e_1.
\end{equation}

We will use these equalities to force the highlighted pair supplied by $P$ onto
an alternating cycle.  To this end, we define a directed multigraph $D$ on $J_0 \cup J_1$ as
follows:
\begin{itemize}
\item for every highlighted-pair occurrence $\{z,w\}$, put $e_0$ parallel
      \emph{highlighted arcs} $z \to w$ from $z\in J_0$ to $w\in J_1$; and
\item for every interesting edge $zw\in \Eint$, put $h_0$ parallel
      \emph{interesting arcs} $w \to z$ from $w\in J_1$ to $z\in J_0$.
\end{itemize}

\begin{claim}
Every arc of $D$ lies on a directed simple cycle.
\end{claim}
\begin{proof}
The key observation is that $D$ is \emph{balanced}, that is, for any $v \in J_0 \cup J_1$, $\deg^{\rm in}_D (v) =\deg^{\rm out}_D  (v) $.
Indeed, at a $J_0$-vertex, indegree and outdegree are both $e_0h_0$, and at a
$J_1$-vertex, indegree and outdegree are, respectively, $e_0h_1$ and
$h_0e_1$, which are equal by
\eqref{eq:balance-ratio}. 

Every arc of a finite balanced directed multigraph lies on a directed
simple cycle.  For an arc $a\to b$, let $S$ be the set of vertices
reachable from $b$.  No arc leaves $S$.  Summing
indegree \(=\) outdegree over $S$ shows that no arc enters $S$ either.
Hence, $a\in S$, and a shortest directed path from $b$ to $a$, followed by
$a\to b$, is a directed cycle.
\end{proof}

Observe that, by the definition of $D$, every directed cycle in $D$ must alternate between highlighted arcs and interesting arcs. In particular, every directed cycle in $D$ has even length.

Let $\Gamma$ be a cycle containing one of the highlighted arcs $z\to w$
associated with $P$.  If $\Gamma$ is a directed $2$-cycle, its reverse
arc $w\to z$ would be an interesting arc and would certify that $zw\in \Eint$.
Therefore, we can assume that $\Gamma$ has length at least $4$.

\begin{claim} 
$\Gamma$ is contained in the block $B$ of $P$. 
\end{claim}
\begin{proof}
An interesting arc from a vertex of $J_1\cap B$ stays in
$B$ by \Cref{observations:j0}.iii).  Suppose that a highlighted arc
$z\to w$ leaves the block $B$, giving a vertex $z\in J_0\cap B$ that is the endpoint of an $i$-path which is contained in a block $A$ pending at $z$.  Basic Property~5 applied to
$A$ gives
\[
  \depth(j_1)>\depth(j_0),
  \qquad K_{j_1j_0}\le1.
\]
The block $A$ is the up-block of $w$.  Every other block containing $w$
is rooted at $w$; if such a block contained a $J_0$-vertex, it would give
$\depth(j_0)>\depth(j_1)$, a contradiction.  Thus, every $J_0$-neighbor of
$w$ lies in $A$.  Since $e_1>0$, such a neighbor exists; since $A$ has
only one vertex of its root color $j_0$, that neighbor is $z$ and is
unique.  Therefore, the highlighted arc $z\to w$ must be followed on $\Gamma$
by an interesting arc $w\to z$.  This immediately closes a directed $2$-cycle,
impossible because $\Gamma$ is a simple cycle of length at least $4$.
\end{proof}

Now we expand $\Gamma$ back into $G$: replace every highlighted arc $z(u)\to w(u)$ by its
witnessing $i$-path $z(u)-u-w(u)$, and replace every interesting arc by its underlying edge. Since no two highlighted arcs of the cycle $\Gamma$ can have
the same endpoint pair, we add distinct vertices of
$I$ during their expansion. These new vertices are also distinct from all vertices of $\Gamma$; thus, the expansion yields a simple cycle $\widehat\Gamma$ contained in
$B$.

We now locate $\ro(B)$ relative to this cycle. First, observe that $\ro(B)\notin V(\widehat\Gamma)$ as every vertex of
$\widehat\Gamma$ has a color in $\{i,j_0,j_1\}$ and we already assumed that $\ro(B)$ has none of these colors (see \Cref{eq:root-color-assumption}). In an outerplanar
embedding, the vertices of a simple cycle occur in the same cyclic order
on the outer face as on the cycle; otherwise two edges of the cycle would
have interlacing endpoints and would cross.  Thus
$\ro(B)$ lies in one of the outer-face intervals
between two consecutive vertices $v,v'$ of $\widehat\Gamma$.

The edge $vv'$ cannot be one of the two edges of an inserted $i$-path as
 every edge incident with an $I$-vertex lies on the outer face of
$B$, so its endpoints are consecutive on the outer face and its intervening
interval cannot contain $\ro(B)$.  Thus, $e=vv'$ comes from an interesting arc of
$\Gamma$, implying that $e\in E(G)$. In particular, $e$ is a chord of $B$ that separates $\ro(B)$ from the complementary path $\widehat\Gamma-e$. 

The directed cycle $\Gamma$ has length at least $4$, so it contains at
least two $J_0$-vertices.  Consequently $\widehat\Gamma-e$ contains a $J_0$-vertex
$z\notin e$ on the opposite side of $e$ from $\ro(B)$, which contradicts  \Cref{observations:j0}.ii).

Therefore, the cycle $\Gamma$ can only have length two, which implies that the endpoints of the highlighted pair supplied by $P$ must be joined
by an actual edge of $G$.

\end{proof}

\section{Block Structure}\label{app:block_structure}

\subsection{Block colors}
\label{subsec:block_colors}

Section~\ref{sec:intuition} describes a block of~$G$ by a pair of colors~$(i,j)$, the endpoints of the chain of~$\widetilde{K}$ it corresponds to, and claims that this pair, together with~$K$, also determines the type of the block, through closing colors.  We now define this pair directly from the block, as its \emph{block color}, and prove both facts: the block color is the pair of endpoints of the chain, and its second component indeed determines the type.  We also give the precise definition of closing colors.

\begin{definition}[Block color]\label{def:block_color}
The \emph{block color} of a block~$A$ of~$G$ is the pair $\tilde{c}(A)=(i,j)$,
where $i=c(\ro(A))$ and
\[
  j=\begin{cases}
    c(w)\ \text{for any } w\in\Low(A), & \text{if $A$ is symmetric,}\\[2pt]
    \min\{c(w):w\in N_A(\ro(A))\},     & \text{if $A$ is discrete.}
  \end{cases}
\]
In the symmetric case the choice of~$w$ does not matter, since all vertices
of~$\Low(A)$ have the same color by the Dichotomy Lemma.
\end{definition}
\begin{remark}\label{rem:bridge_discrete}
The two vertices of a bridge have different depths, hence different colors by
Lemma~\ref{lem:depth}~(Depth), so a bridge is discrete.  Bridges are
also the only blocks that could be both discrete and symmetric. We treat
bridges as discrete blocks throughout.
\end{remark}

By Lemmas~\ref{lem:highest}~(Block root) and~\ref{lem:depth}
(Depth), the block root of~$A$ is the unique vertex of color~$i$
in~$A$, and it is the vertex of~$A$ of smallest depth.  The second component of the block color
stores more information than it might appear  to: we will see (Corollary~\ref{cor:block_type})
that a block of block color~$(i,j)$ is symmetric if and only if $j$ is a
\emph{closing} color, a property of~$j$ that can be read off~$K$.  The type of
a block is therefore determined by its block color, and neither phase of the
algorithm has to determine it separately.

We show in this appendix that the pair $(i,j)$ is also the pair of
endpoints of the chain the block corresponds to, in the sense of
Section~\ref{sec:intuition}: a discrete block of block color~$(i,j)$ is a copy
of the c-block $\Bup^K(j)$, whose block root is~$i$, and which is an
$ij$-chain consisting of a single c-block; and a symmetric block of block
color~$(i,j)$ is the unfolding $\unfold(i,j,K)$ of the
$ij$-chain.  More precisely, we show that the structure of a
block of~$G$ whose block color is~$(i,j)$ is completely determined by~$K$.  
The proof
proceeds in three steps:
\begin{enumerate}
\item \emph{Type recognition}
      (Section \ref{subsec:block_type}). A block of color~$(i,j)$ is
      symmetric if and only if $j$ is a so-called closing color; otherwise
      it is discrete (Corollary~\ref{cor:block_type}).
\item \emph{Discrete case}
      (Section \ref{subsec:discrete_block}). A discrete block of
      color~$(i,j)$ is color-preserving isomorphic
      to~$\Bup^K(j)$ (Lemma~\ref{lem:asym_iso}).
\item \emph{Symmetric case}
      (Section \ref{subsec:symmetric}). A symmetric block of
      color~$(i,j)$ is obtained by \emph{unfolding} a subgraph
      of~$\widetilde{K}$ into two copies, glued at the root
      (color~$i$), closed at the bottom (color~$j$), and joined
      by some chords between vertices of equal color.  We
      identify this subgraph and describe how the two copies are
      attached to each other
      (Lemmas~\ref{lem:color_set} and~\ref{lem:sym_structure}).
\end{enumerate}
These three steps are summarized in the Block Structure
Lemma~\ref{lem:block-structure}, in Section~\ref{subsec:bsl_proof}.

\subsection{Block types recognition}\label{subsec:block_type}

\begin{definition}[Closing color]\label{def:closing}
A color~$j$ is a \emph{closing color} if it satisfies one of the
following two conditions:
\begin{itemize}
\item \emph{Solo closing:} there exists a color $j'\neq j$ in the up-c-block
      of~$j$ with $K_{jj'}\ge2$.
\item \emph{Paired closing:} $K_{jj}>0$.
\end{itemize}
\end{definition}


Intuitively, the color of the vertices at face-lowest position of
symmetric blocks are closing colors.  If the block has an even number of
vertices, the face-lowest vertex is unique and its color is solo closing
(e.g., color~$i$ in Figure~\ref{fig:ex color graph}).  If the block has an odd
number of vertices, the two face-lowest vertices are adjacent,
have the same color, and that color is paired closing (e.g., color~$h$ in Figure~\ref{fig:ex color graph}).  We show that while a
solo closing color must appear at the face-lowest position of its up-block, a
paired closing color can also appear at higher face-depths
(connected by a chord to its partner on the opposite branch, see color~$f$ in
Figure~\ref{fig:ex color graph}). Also see Figure~\ref{fig:ex free chain} for more
examples of solo/paired closing colors.
These claims are made precise in the following lemma.

\begin{lemma}[Closing colors and symmetric blocks]
\label{lem:bottom_symmetric}
\begin{enumerate}
\item If $j$ is a closing color, then for every vertex $u$ of
      color~$j$, its up-block $\Bup^G(u)$ is a symmetric block.
      Furthermore, if $j$ is solo closing, then $\Low(\Bup^G(u))= \{u\}$.
\item For every symmetric block~$A$ of $G$, let $j$ be the color of the
      vertices in $\Low(A)$.  Then $j$ is a closing color: solo
      closing if $|\Low(A)|=1$, and paired closing if
      $|\Low(A)|=2$.
\end{enumerate}
\end{lemma}

\begin{proof}
\emph{Part~(1).}
Let $A=\Bup^G(u)$.
Suppose $j$ is solo closing: there exists a color
$j'\in V(\Bup^K(j))$, $j'\neq j$, such that $K_{jj'}\ge2$.
We claim that every neighbor~$w$ of~$u$ of color~$j'$ lies in~$A$.
Suppose for contradiction that
$w\notin A$.  Then the edge $\{u,w\}$ belongs to some block~$A'$
with $A'\neq A$.  Since $A=\Bup^G(u)$, we must have $\ro(A')=u$.
By \fpass{Lemma~\ref{lem:cut_vertex_sep}.2~(Cut} vertex separation),
$j=c(u)$ separates~$1$ from $j'=c(w)$ in~$\widetilde{K}$.  But
$j'\in V(\Bup^K(j))$ means that there exists a path from~$j'$
to~$\ro(\Bup^K(j))\neq j$ (either because $\Bup^K(j)$ is 2-connected or
$\ro(\Bup^K(j))=j'$) and then to~$1$ without passing by~$j$, so $j$
cannot separate~$j'$ from~$1$---contradiction. Hence $w\in A$.
As $K_{jj'}\ge2$, the block~$A$ thus contains at least two vertices of
color~$j'$, so it is not discrete, and it is symmetric by the Dichotomy Lemma.
A symmetric block contains at most two vertices of each color, so $u$ has
exactly two neighbors of color~$j'$, which we call $w$ and~$w'$, and
$K_{jj'}=2$.
We show that $u$ is the unique vertex of color~$j$ in~$A$.  Suppose
for contradiction that another vertex~$u'$ of color~$j$ exists
in~$A$.  By the same reasoning as for~$u$, both neighbors
of~$u'$ of color~$j'$ lie in~$A$, and since $w,w'$ are the only
vertices of color~$j'$ in~$A$, both $u$ and~$u'$ are adjacent to both $w$
and~$w'$.  As $j$ and~$j'$ each occur twice in~$A$, none of these four
vertices is $\ro(A)$ or a unique lowest vertex; so, by the Dichotomy Lemma,
$u,u'$ lie on opposite branches at the same face-depth, and so do $w,w'$, at a
different face-depth as $j\neq j'$.  Let $\{x,x'\}$ be the one of the pairs
$\{u,u'\}$ and $\{w,w'\}$ with smaller face-depth, and $\{y,y'\}$ the other.
The segments of the two branches from~$x$ and from~$x'$ to~$\ro(A)$ contain
only vertices of face-depth smaller than that of $y$ and~$y'$, and meet only
at~$\ro(A)$.  Together with the four edges between $\{x,x'\}$ and
$\{y,y'\}$, they form a subdivision of~$K_{2,3}$ with parts $\{x,x'\}$ and
$\{y,y',\ro(A)\}$, contradicting the outerplanarity of~$A$.
Hence $u$ is the unique vertex of color~$j$ in~$A$.  Since $A$
is symmetric and $u\neq\ro(A)$ (by definition of up-block), and
every face-depth level except the root has two vertices of the
same color (by the Dichotomy Lemma), the only way $u$ can be
unique is if $u\in\Low(A)$ and $|\Low(A)|=1$.

Now suppose $j$ is paired closing (and not solo closing).  Then
$K_{jj}>0$, so $u$ has a neighbor~$u'$ of color~$j$.  If
$u'\notin\Bup^G(u)$, then $\{u,u'\}$ belongs to a block rooted at~$u$, giving
by Lemma~\ref{lem:cut_vertex_sep} (Cut vertex separation) that
$\depthK(j)>\depthK(j)$, a contradiction.  Hence
$u'\in\Bup^G(u)$.  Since $u$ and~$u'$ have the same color, $A$
is symmetric (two vertices of the same color means it cannot be
discrete).

\medskip
\emph{Part~(2).}
Let $A$ be a symmetric block and $j$ the color of the vertices
in~$\Low(A)$.

\emph{Case $|\Low(A)|=2$:}
The two vertices $u,u'$ in $\Low(A)$ have the same color~$j$
(Dichotomy Lemma) and are adjacent (they are outer-face neighbors
at maximum face-depth).  Hence $K_{jj}>0$, so $j$ is paired
closing.

\emph{Case $|\Low(A)|=1$:}
Let $u$ be the unique vertex in $\Low(A)$, of color~$j$.  Let
$j'$ be the color of the two outer-face neighbors of~$u$ in~$A$
(they have the same color by the Dichotomy Lemma since they are
face-depth partners). Then $K_{jj'}\ge 2$, and $j'\neq j$, as
these two neighbors lie at a smaller face-depth than~$u$.

It remains to show that $j'\in V(\Bup^K(j))$.  The colors $j$
and~$j'$ are neighbors in~$\widetilde{K}$, so they share a c-block~$B$.  If
$j=\ro(B)$, then $j$ separates~$j'$ from~$1$ in~$\widetilde{K}$.
However, consider the path along one branch of the outer face of~$A$ from
$\ro(A)$ to a vertex~$w$ of color~$j'$, avoiding~$u$.  Since $A$ is
symmetric and $u$ is the unique vertex of color~$j$ in~$A$, this
path avoids all vertices of color~$j$.  Hence $c(\ro(A))$
and~$j'$ are connected in~$\widetilde{K}$ without passing
through~$j$, and since $\depthK(c(\ro(A)))<\depthK(j)$, color~$1$
can reach~$c(\ro(A))$ and then~$j'$ without passing through~$j$---a
contradiction.  Hence $B=\Bup^K(j)$ and $j'\in V(\Bup^K(j))$, so $j$ is solo
closing.
\end{proof}

\begin{corollary}[Block type determination]
\label{cor:block_type}
A block of color~$(i,j)$ is symmetric if and only if $j$ is a
closing color.  If $j$ is not a closing color, the block is
discrete.
\end{corollary}

\begin{proof}
\fpass{Let $A$ be a block of block color~$(i,j)$.  If $A$ is symmetric, $j$ is
closing by Lemma~\ref{lem:bottom_symmetric}(2).  Conversely, in both cases of
Definition~\ref{def:block_color}, $j$ is the color of a vertex~$w$ of~$A$
other than~$\ro(A)$, so that $\Bup^G(w)=A$; if $j$ is closing, $A$ is
therefore symmetric by Lemma~\ref{lem:bottom_symmetric}(1).  The Dichotomy
Lemma~\ref{lem:dichotomy} gives that blocks that are not symmetric are}
discrete.
\end{proof}

\subsection{Discrete blocks}\label{subsec:discrete_block}

We regard $K$ as a colored graph via the identity coloring, where the
vertex~$i$ of~$K$ has color~$i$.

\begin{lemma}[Discrete block isomorphism]
\label{lem:asym_iso}
Let $u$ be a vertex of color~$j$.  If $\Bup^G(u)$ is a discrete
block, then $\Bup^G(u)$ is color-preserving isomorphic to the
c-block $\Bup^K(j)$.
\end{lemma}

\begin{proof}
Since $\Bup^G(u)$ is discrete, all its vertices have pairwise
distinct colors.  Let $S$ be the set of colors appearing
in~$\Bup^G(u)$.

Every edge of $\Bup^G(u)$ connects vertices of distinct colors,
inducing an edge in~$\widetilde{K}[S]$. Conversely, if $s\neq t$
are colors of~$S$ with $K_{st}>0$, Lemma~\ref{lem:edge_exists}
(Edges between colors of a block) gives the existence of an
$st$-edge in~$\Bup^G(u)$; as the coloring is injective there, it joins the
unique vertices of colors $s$ and~$t$.  Hence $\Bup^G(u)$ is color-preserving
isomorphic to~$\widetilde{K}[S]$.

Now, we show that~$\widetilde{K}[S]=\Bup^K(j)$. $\widetilde{K}[S]$
is either a bridge or 2-connected (since $\Bup^G(u)$ is), so it is contained
in a unique c-block.  Since $j\in S$ and $j\neq c(\ro(\Bup^G(u)))$
(by definition of up-block), and the isomorphism preserves depth
(Lemma~\ref{lem:depth}~(Depth), $u$ is not the highest vertex in
$\Bup^G(u)$ implies $j$ is not the highest color in $\widetilde{K}[S]$),
this c-block is $\Bup^K(j)$.  Hence $S\subseteq V(\Bup^K(j))$.

Since $|S|\ge 2$ and $S\subseteq V(\Bup^K(j))$, the c-block
$\Bup^K(j)$ contains at least two colors of~$S$.  By
Lemma~\ref{lem:block_maximality}~(Block maximality),
$V(\Bup^K(j))\subseteq S$.  Hence $S=V(\Bup^K(j))$ and
$\Bup^G(u)$ is color-preserving isomorphic to~$\Bup^K(j)$.
\end{proof}

\subsection{Symmetric Blocks}\label{subsec:symmetric}

\subsubsection{Auxiliary lemmas}
We first establish two structural properties of symmetric blocks
that are used throughout this subsection.

\begin{observation}[No crossing between branches]
\label{obs:no-crossing}
Let $A$ be a symmetric block of~$G$.  Let $u,v$ be vertices of~$A$ with
$\facedepth_A(u)\neq\facedepth_A(v)$ that do not lie on a common
branch.  Then $u$ and~$v$ are not adjacent.
\end{observation}

\begin{proof}
Suppose for contradiction that $u$ and~$v$ are adjacent, with $u$
on the left branch at face-depth~$d$ and $v$ on the right branch
at face-depth~$d'$, $d<d'$.  Since $u$ and~$v$ lie on no common
branch, neither belongs to the intersection of the two branches,
so their face-depth partners $u'$ (right branch, depth~$d$) and
$v'$ (left branch, depth~$d'$) satisfy $u'\neq u$ and $v'\neq v$;
in particular $u,u',v,v'$ are pairwise distinct.  By the
left--right automorphism (Dichotomy Lemma~\ref{lem:dichotomy}),
$\{u',v'\}$ is also an edge of~$A$.

The four vertices $u,u',v,v'$ are then joined pairwise by six
internally disjoint paths: the edges $\{u,v\}$ and $\{u',v'\}$;
the segment of the left branch from~$u$ to~$v'$ and the segment of
the right branch from~$u'$ to~$v$; the segment of the outer face
from~$u$ to~$u'$ through $\ro(A)$; and the segment from~$v'$
to~$v$ through $\Low(A)$.  Hence $A$ contains a subdivision
of~$K_4$, contradicting outerplanarity.
\end{proof}

\begin{lemma}[Face-depth ordering in~$T_{\widetilde{K}}$]
\label{lem:facedepth_upblock}
Let $A$ be a block of~$G$.  Let $i,j$ be two colors
appearing in~$A$ with vertices $u$ of color~$i$ and $v$ of
color~$j$ that are outer-face neighbors in~$A$ and
$\facedepth_A(v)=\facedepth_A(u)+1$.  Then either \fpass{$i\neq1$ and
$\Bup^K(i)=\Bup^K(j)$, or $i$ is the block root of $\Bup^K(j)$, that is, its
parent in~$T_{\widetilde{K}}$ (the root~$1$ or a cut vertex
of~$\widetilde{K}$).}

More generally, if $A$ is symmetric and $i,j$ are two colors
in~$A$ with $i\neq1$ and 
$\facedepth_A(i)<\facedepth_A(j)$, then $\Bup^K(i)$
is an ancestor of or equal to $\Bup^K(j)$ in~$T_{\widetilde{K}}$.
\end{lemma}

\begin{proof}
\fpass{As $u$ and~$v$ have distinct face-depths, $i\neq j$: in a discrete block
all colors are distinct, and in a symmetric block two vertices of the same
color are mirror images, at the same face-depth.}  Since $u$ and~$v$ are
neighbors, $i$ and~$j$ are neighbors in~$\widetilde{K}$.  Let $B$ be the
unique c-block of~$\widetilde{K}$ containing both $i$ and~$j$.  \fpass{Either
$\ro(B)\notin\{i,j\}$, so that $B=\Bup^K(i)=\Bup^K(j)$ (first alternative),
or $\ro(B)=i$, so that} $B=\Bup^K(j)$ and $i$ is the parent of~$B$
in~$T_{\widetilde{K}}$ (second alternative), or \fpass{$\ro(B)=j$ and
$B=\Bup^K(i)$.}

We show the third case cannot happen.  Suppose $j=\ro(B)$ and
$B=\Bup^K(i)$.  Then $j$ is a cut vertex
of~$\widetilde{K}$ whose removal separates $V(B)\setminus\{j\}$,
and in particular~$i$, from every color of depth smaller
than~$\depthK(j)$.  Let $t=c(\ro(A))$.  Since
$\facedepth_A(j)\ge 1$, we have $j\neq t$, so
$\depthK(t)<\depthK(j)$ by Lemmas~\ref{lem:highest}~(Block root)
and~\ref{lem:depth}~(Depth).  Hence $j$ separates~$i$ from~$t$
in~$\widetilde{K}$.
However, consider the path along the outer face of~$A$ from
$\ro(A)$ to $u$ on the branch where $u$ and $v$
appear.  This path has strictly increasing face-depth and thus
pairwise-distinct colors; in particular, no vertex on this path
has color~$j$ (since by the Dichotomy Lemma, any vertex of
color~$j$ has the same face-depth as~$v$, i.e., strictly greater
face-depth than~$u$, and so cannot appear between $\ro(A)$
and~$u$ on a branch of increasing face-depth).  This gives a
color path from~$t$ to~$i$ in~$\widetilde{K}$ avoiding~$j$,
contradicting that $j$ separates~$i$ from~$t$.

The general statement follows by induction along the outer face:
if $\facedepth_A(j)>\facedepth_A(i)$, there is a sequence of
consecutive face-depth pairs from~$i$ down to~$j$, and at each
step $\Bup^K$ either stays the same or moves one level deeper
in~$T_{\widetilde{K}}$.
\end{proof}

\subsubsection{Color set}

To describe the colors appearing in a symmetric block, we recall the chains of Section~\ref{sec:intuition}.

\begin{definition}[$ij$-chain]\label{def:ij-chain}
Let $i,j$ be two colors, and let $(i=c_0,c_1,\dots,c_\ell=j)$, with
$\ell\ge1$, be a path in~$\widetilde{K}$ such that, for every $a\in[1,\ell]$,
its $a$-th edge $c_{a-1}c_a$ lies in a c-block~$B_a$ whose block root is its
upper end~$c_{a-1}$. We call the union $C=B_1\cup\dots\cup B_\ell$ an
\emph{$ij$-chain}, the path $(c_0,\dots,c_\ell)$ its \emph{underlying path},
and $V(C)$ its \emph{color set}.
\end{definition}

Since $c_{a-1}$ is the block root of~$B_a$, it is its only color of minimum
depth (Remark~\ref{rem:K_rooted}), so the underlying path is descending. In particular, the colors $c_a$ are pairwise distinct and so are the
c-blocks~$B_a$, which have distinct block roots.

\begin{lemma}[Chains in~$T_{\widetilde{K}}$]\label{lem:chain_unique} 
Let $C=B_1\cup\dots\cup B_\ell$ be an $ij$-chain with underlying path
$(i=c_0,c_1,\dots,c_\ell=j)$.  For every $a\in[1,\ell]$, we have
$B_a=\Bup^K(c_a)$, which implies the following properties:
\begin{enumerate}
\item the $ij$-chain, if it exists, is unique, and its underlying path is
      obtained by starting at~$j$ and repeatedly moving from the current
      color~$s$ to the block root of~$\Bup^K(s)$, until $i$ is reached.
      Conversely, if this process reaches~$i$ and each color it meets is
      adjacent to the next one, the colors met form the underlying path of
      an $ij$-chain;
\item each color~$c_a$ with $a\in[0,\ell-1]$ is either~$1$ or a cut vertex
      of~$\widetilde{K}$ separating $c_{a+1}$ from~$1$;
\item for $0\le a<a'\le\ell$, the union $B_{a+1}\cup\dots\cup B_{a'}$ is a
      $c_ac_{a'}$-chain, with underlying path $(c_a,\dots,c_{a'})$.
\end{enumerate}
\end{lemma}

\begin{proof}
Let $a\in[1,\ell]$.  The c-block~$B_a$ contains $c_a$, which is not its block
root~$c_{a-1}$, so $B_a=\Bup^K(c_a)$ (Remark~\ref{rem:K_rooted}).  Therefore
$i=c_0,B_1,c_1,\dots,c_{\ell-1},B_\ell=\Bup^K(j)$ is a path
in~$T_{\widetilde{K}}$; as $T_{\widetilde{K}}$ is a tree, it is the unique
path from~$i$ to~$\Bup^K(j)$, so the $c_a$ and the~$B_a$ are determined by~$i$
and~$j$.  More precisely, $c_{a-1}$ is the block root of $B_a=\Bup^K(c_a)$,
which is the process of~(1).  Conversely, if this process reaches~$i$ and each
color it meets is adjacent to the next one, write $(i=c_0,\dots,c_\ell=j)$ for
the colors met, in reverse order, and $B_a=\Bup^K(c_a)$: the edge $c_{a-1}c_a$
lies in~$B_a$, whose block root is~$c_{a-1}$, so $B_1\cup\dots\cup B_\ell$ is
an $ij$-chain.  This gives~(1).  Since $c_a=\ro(B_{a+1})$, the color~$c_a$ is
either~$1$ or the cut vertex that is the parent of~$B_{a+1}$
in~$T_{\widetilde{K}}$, which separates $V(B_{a+1})\setminus\{c_a\}$, and in
particular~$c_{a+1}$, from~$1$; this gives~(2).  Finally,
$B_{a+1}\cup\dots\cup B_{a'}$ satisfies the conditions of
Definition~\ref{def:ij-chain} with underlying path $(c_a,\dots,c_{a'})$; this
gives~(3).
\end{proof}

Examples of $ij$-chains and of their underlying paths are given in Figure~\ref{fig:ex free chain}; other examples of chains appear in Figure~\ref{fig:4types}.

\begin{lemma}[Color set of a symmetric block]
\label{lem:color_set}
Let $A$ be a symmetric block of~$G$ with $\tilde{c}(A)=(i,j)$, and let
$S$ be the set of colors appearing in~$A$.  Then there exists an
$ij$-chain~$C$ such that $S=V(C)$.
\end{lemma}

\begin{proof}
\emph{Construction.}
Let $w\in\Low(A)$.  Let $P$ be a shortest path in~$A$ from
$\ro(A)$ to~$w$, and let $(i=c_0,c_1,\dots,c_\ell=j)$ be its color
sequence.  Since consecutive vertices of~$P$ are adjacent,
$K_{c_ac_{a+1}}>0$ for all $a\in[0,\ell-1]$.

\emph{The path is descending.}
Since $\ro(A)$ is either~$r$ itself or a cut vertex separating
$A\setminus\{\ro(A)\}$ from~$r$ in~$G$, every shortest path
from~$r$ to a vertex~$v$ in~$A$ passes through~$\ro(A)$, so
$\depth_G(v)=\depth_G(\ro(A))+d_A(\ro(A),v)$.  Since $P$ is a
shortest path in~$A$, the distance $d_A(\ro(A),\cdot)$ strictly
increases at each step, hence
$\depthK(c_0)<\depthK(c_1)<\cdots<\depthK(c_\ell)$.

\begin{claim}\label{cl:chain_cut}
For every $a\in[0,\ell-1]$, the edge $c_ac_{a+1}$ lies in a c-block
whose block root is~$c_a$.
\end{claim}

\begin{proof}[Proof of Claim]
Let $B$ be the c-block containing the edge $c_ac_{a+1}$.

For $a=0$: $B$ contains the colors $i$ and~$c_1$ of~$A$, so all its
colors appear in~$A$ by Lemma~\ref{lem:block_maximality}~(Block maximality).
As $i$ is the color of~$\ro(A)$, it is the color of~$A$ of smallest depth
(Lemmas~\ref{lem:highest}, Block root, and~\ref{lem:depth}, Depth), so
$i=\ro(B)$.

For $a\in[1,\ell-1]$: we first show that $c_a$ is a cut vertex
of~$\widetilde{K}$ separating $j$ from~$1$.  Note that $c_a\neq1$ since
$\depthK(c_0)<\depthK(c_a)$.
Let $v_a$ and~$v'_a$ denote the two vertices
of color~$c_a$ in~$A$ (one on each branch at the same
face-depth; by the Dichotomy Lemma). We show that every simple path
from~$w$ to~$\ro(A)$ in~$G$ passes through~$v_a$ or~$v'_a$.
(Since $A$ is maximally 2-connected, any such simple path stays
within~$A$.)
Suppose for contradiction that a path~$Q$ from~$w$ to~$\ro(A)$
in~$A$ avoids both~$v_a$ and~$v'_a$.  Then
some edge $\{u,v\}$ of~$Q$ satisfies
$\facedepth_A(u)>\facedepth_A(v_a)>\facedepth_A(v)$. By
Observation~\ref{obs:no-crossing}~(No crossing between branches),
$u$ and~$v$ are on the same branch.  Without loss of generality, say $P$
passes through~$v_a$.  Let $\{u',v'\}$ be the edge $\{u,v\}$ if $u$
and~$v$ lie on the branch of~$v_a$, and its image by the left--right
automorphism otherwise.  As this automorphism exchanges the two branches and
preserves face-depths, $u'$ and~$v'$ lie on the branch of~$v_a$, with
$\facedepth_A(v')<\facedepth_A(v_a)<\facedepth_A(u')$.  In the outerplanar
embedding of~$A$, the chord $\{u',v'\}$ separates the vertices of this branch
lying strictly between $v'$ and~$u'$, among which~$v_a$, from the other
vertices of~$A$ except $v'$ and~$u'$.  As $\ro(A)$ and~$w$ do not lie strictly
between $v'$ and~$u'$, the path~$P$, which goes from~$\ro(A)$ to~$w$
through~$v_a$, passes \fpass{through both $v'$ and~$u'$, one before~$v_a$ and the
other} after it (possibly $v'=\ro(A)$ or $u'=w$).  Replacing its sub-path from~$v'$ to~$u'$,
of length at least~$2$ as it contains~$v_a$, by the edge~$\{v',u'\}$ gives a
shorter path from~$\ro(A)$ to~$w$, contradicting the minimality of~$P$.
Every path from~$w$ to $r$ must pass through~$\ro(A)$, hence
through~$v_a$ or~$v'_a$.  Therefore every color path from~$j$
to~$1$ in~$\widetilde{K}$ passes through~$c_a$, making $c_a$ a
cut vertex of~$\widetilde{K}$ separating $1$ from $j$.

Finally, suppose for contradiction that $\ro(B)\neq c_a$.  Then
$\depthK(\ro(B))<\depthK(c_a)$ (Remark~\ref{rem:K_rooted}), and $B$ is not a
bridge, whose block root would be~$c_a$, so $B\setminus\{c_a\}$ is connected.
The color~$j$ then reaches~$c_{a+1}$ along $(c_\ell,\dots,c_{a+1})$, whose
colors are deeper than~$c_a$, then $\ro(B)$ inside $B\setminus\{c_a\}$, then~$1$
along a path of decreasing depth, which also avoids~$c_a$; this contradicts that
$c_a$ separates $j$ from~$1$.  Hence $c_a=\ro(B)$.
\end{proof}

Hence, writing $B_{a+1}$ for the c-block of the claim,
$C=B_1\cup\dots\cup B_\ell$ is an $ij$-chain, with underlying path
$(i=c_0,c_1,\dots,c_\ell=j)$.

\medskip
\emph{Containment: $S\subseteq V(C)$}.
Since $i=\ro(B_1)\in V(B_1)$ and $j\in V(B_\ell)$, it suffices to show
$s\in V(C)$ for every $s\in S$ with $s\notin\{i,j\}$.  By
Lemma~\ref{lem:facedepth_upblock}~(Face-depth ordering), as $s$ appears in~$A$
at a face-depth strictly between those of $i$ and~$j$, the c-block
$\Bup^K(s)$ is an ancestor of or equal to~$B_\ell=\Bup^K(j)$
in~$T_{\widetilde{K}}$, and, if $i\neq1$, a descendant of or equal
to~$\Bup^K(i)$.  So $\Bup^K(s)$ lies on the path of~$T_{\widetilde{K}}$ from
$\Bup^K(i)$ (or from the root~$1$, if $i=1$) to~$B_\ell$.  This path is
$\Bup^K(i),i,B_1,c_1,\dots,c_{\ell-1},B_\ell$ by
Lemma~\ref{lem:chain_unique}~(Chains in~$T_{\widetilde{K}}$).  If
$\Bup^K(s)\in\{B_1,\dots,B_\ell\}$, then $s\in V(\Bup^K(s))\subseteq
V(C)$.  Otherwise, $\Bup^K(s)=\Bup^K(i)$: the colors $s$ and~$i$ belong
to the same c-block, with $i=\ro(B_1)$ a cut vertex separating
$s$ from~$j$ in~$\widetilde{K}$.  But in~$A$, both $s$ and~$j$
appear at face-depth ${}>0$, so the branch of~$A$ from the vertex
of color~$s$ to a vertex of color~$j$ avoids~$\ro(A)$, giving a
color path from~$s$ to~$j$ in~$\widetilde{K}$ avoiding~$i$---a
contradiction.

\emph{Maximality: $V(C)\subseteq S$.}
Each $B_a$ with $a\ge 1$ contains at least two colors of~$S$:
namely $c_{a-1}$ and~$c_a$.  By
Lemma~\ref{lem:block_maximality}~(Block maximality),
$V(B_a)\subseteq S$ for all $a\ge 1$.  Hence $V(C)\subseteq S$.
\end{proof}

\begin{remark}\label{rem:chain_shortest_path}
The proof of Lemma~\ref{lem:color_set} applies to any shortest path
from~$\ro(A)$ to any vertex of~$\Low(A)$: the color sequence of each such path
is the underlying path of an $ij$-chain.  As the $ij$-chain is unique
(Lemma~\ref{lem:chain_unique}, Chains in~$T_{\widetilde{K}}$), all these
color sequences are equal, to the underlying path of the $ij$-chain.
\end{remark}

\subsubsection{Structure of a symmetric block}

Now we are ready to present how to reconstruct a symmetric block from a corresponding chain in the color graph.
The picture to have in mind is that a symmetric block is obtained
by \emph{unfolding} the chain into two copies of
$H=\widetilde{K}[S]$, the subgraph of~$\widetilde{K}$ induced by $S$
the set of colors of~$A$, each of the two branches of~$A$ will have one full
copy of~$H$, with one vertex per color of~$S$.  The two copies are
attached at the root of the block (color~$i$), and the outer face
closes up at the bottom in one of two ways: if $j$ is solo closing,
the two copies are also attached at the unique face-lowest vertex
(color~$j$); if $j$ is paired closing, they remain distinct there
but are joined by an edge between the two vertices of color~$j$,
which exists precisely because $K_{jj}>0$.  No other vertices of the two copies are identified, and the only edges joining them are chords
between face-depth partners, which are present exactly for the colors~$s$
with $K_{ss}>0$. We first define this unfolding formally; the lemma that follows then
shows that a symmetric block of block color~$(i,j)$ is exactly the unfolding
of the $ij$-chain.

\begin{definition}[Unfolding]\label{def:unfold}
Let $i,j$ be two colors with $j$ closing.  If no $ij$-chain exists,
$\unfold(i,j,K)$ is the empty graph.  Otherwise, let $C$ be the
$ij$-chain.
Take two disjoint copies $C_1$ and~$C_2$ of~$C$, and give the
vertex of~$C_1$ (resp.~$C_2$) corresponding to a color
$s\in V(C)$ the color~$s$; thus each copy contains exactly one vertex
of each color of~$V(C)$.  Then $\unfold(i,j,K)$ is obtained
from $C_1\cup C_2$ by:
\begin{itemize}
\item identifying the two vertices of color~$i$;
\item if $j$ is solo closing, identifying the two vertices of
      color~$j$;
\item adding, for every color $s\in V(C)$ that still has two copies
      after these identifications and satisfies $K_{ss}>0$, an
      edge between its two copies.
\end{itemize}
The result is taken as a simple graph, rooted at the vertex of color $i$; in particular, if the identifications
create a pair of parallel edges, they are replaced by a single
edge.
\end{definition}

The unfolding identifies the two vertices of color~$j$ when $j$ is solo
closing, while Section~\ref{sec:intuition} does so when $K_{jj}=0$.  The two
conditions agree, by the following remark.

\begin{remark}\label{rem:solo_paired}
A solo closing color~$j$ satisfies $K_{jj}=0$: a neighbor of color~$j$
of a vertex~$u$ of color~$j$ has the same depth as~$u$
(Lemma~\ref{lem:depth}, Depth), so it would lie in~$\Bup^G(u)$, where $u$ is
the only vertex of color~$j$ (Lemma~\ref{lem:bottom_symmetric}(1), Closing
colors and symmetric blocks).  So for a closing color~$j$, being solo closing
amounts to $K_{jj}=0$.
\end{remark}

\begin{lemma}[Structure of a symmetric block, reformulation of Lemma~\ref{lem:sym_structure}]
Let $A$ be a symmetric block with $\tilde{c}(A)=(i,j)$ and $S$ be the set
of colors appearing in~$A$, let $H=\widetilde{K}[S]$, and
let $L,R$ be the vertex sets of the branches of~$A$. Then the following statements are true.
\begin{enumerate}
\item $H$ is an $ij$-chain.
\item $A[L]$ and $A[R]$ are color-preserving isomorphic to~$H$.
\item $L\cap R=\{\ro(A)\}$ if $j$ is paired closing, and
      $L\cap R=\{\ro(A),w\}$, where $w$ is the unique vertex
      of~$\Low(A)$, if $j$ is solo closing.
\item Two vertices of~$A$ lying on opposite branches, neither of
      them in $L\cap R$, are adjacent if and only if they have the
      same color~$s$ and $K_{ss}>0$.
\end{enumerate}
\end{lemma}

\begin{proof}
\emph{Part~(1).}  By Lemma~\ref{lem:color_set}~(Color set), there is an
$ij$-chain $C=B_1\cup\dots\cup B_\ell$ with $V(C)=S$, so $C$ is a subgraph of
$H=\widetilde{K}[S]$ with the same vertices; we show that $H$ has no other
edge.  Let $st$ be an edge of~$\widetilde{K}$ with $s,t\in S$.  As $C$ is
connected, it contains a path from~$s$ to~$t$; if this path is not the edge
$st$ itself, it forms a cycle together with~$st$.  This cycle lies in a single
c-block~$B$.  The first edge of the path is an edge of~$C$, so it lies in
some~$B_a$, and as each edge lies in a unique c-block, $B=B_a$.  So $st$ lies
in~$B_a\subseteq C$.  Hence $H=C$.

\medskip
\emph{Part~(2).}
By definition of a symmetric block, two vertices of~$A$ have the same color if
and only if they lie at the same face-depth; in particular each branch carries
exactly one vertex of each color of~$S$.  Fix a branch $X\in\{L,R\}$ and, for
$s\in S$, let $x_s$ be the vertex of color~$s$ on~$X$.  Then $x_s\mapsto s$ is
a color-preserving bijection from~$X$ onto $V(H)=S$, and a graph
morphism from~$A[X]$ to~$H$ by equitability and the definition
of~$\widetilde{K}$.  It remains to show that every edge of~$H$ comes from an
edge of~$A[X]$.
Let $s,t\in S$ be distinct with $K_{st}>0$.
Then Lemma~\ref{lem:edge_exists} (Edges between colors of a block)
gives that $A$ contains an $st$-edge $\{u,v\}$ with $c(u)=s$ and
$c(v)=t$.  As $s\neq t$, the vertices $u$ and~$v$ lie at distinct
face-depths, so by
Observation~\ref{obs:no-crossing}~(No crossing between branches)
they must lie on a common branch.  If this branch is~$X$, then
$\{u,v\}=\{x_s,x_t\}$ and we are done.  Otherwise the left--right
automorphism of~$A$, which is color-preserving and exchanges the
two branches, maps $\{u,v\}$ to an $st$-edge of~$X$, which is
again $\{x_s,x_t\}$.

\medskip
\emph{Part~(3).}
The only vertices lying on both branches of~$A$ are $\ro(A)$ and
if $A$ has even size, the unique vertex with face-depth $|A|/2$
(i.e. the unique vertex in~$\Low(A)$).
Lemma~\ref{lem:bottom_symmetric}(2)~(Closing colors and symmetric blocks)
gives that~$|\Low(A)|=1$ if and only if~$j$, the color of the elements
in $\Low(A)$, is solo closing (otherwise, $j$ is paired closing).

\medskip
\emph{Part~(4).}
Let $x,y$ lie on opposite branches with $x,y\notin L\cap R$.

If $x$ and~$y$ are adjacent, then
Observation~\ref{obs:no-crossing}~(No crossing between branches)
forces them to have the same face-depth, hence the same color~$s$,
and $K_{ss}>0$ follows from the edge $\{x,y\}$.

Conversely, suppose $c(x)=c(y)=s$ with $K_{ss}>0$.  Since
$x\notin L\cap R$, we have $x\neq\ro(A)$, so $\Bup^G(x)=A$.  As
$K_{ss}>0$, the vertex~$x$ has a neighbor~$y'$ of color~$s$, and
$\depth_G(y')=\depth_G(x)$ by Lemma~\ref{lem:depth}~(Depth).  If
$y'\notin A$, the edge $\{x,y'\}$ would lie in a block distinct from~$A$ and
thus rooted in $x$. Hence $\depth_G(y')>\depth_G(x)$, a contradiction.  So
$y'\in A$, and the only two vertices of color~$s$ in~$A$ are $x$ and~$y$; as
$y'\neq x$ (we consider simple graphs), we get $y'=y$, so $x$ and~$y$ are
adjacent.
\end{proof}

\begin{remark}\label{rem:shared_edge}
If $j$ is solo closing and satisfies $K_{ij}>0$, then $\ro(A)$ and the
unique vertex~$w$ of~$\Low(A)$ both lie in $L\cap R$, and the edge
$\{\ro(A),w\}$ lies on both branches: it is a single edge of~$A$,
even though it corresponds to an edge of~$C$ in each of the two
copies (multiple edges are not allowed). In this case, the $ij$-chain is made of a single c-block.
\end{remark}

\subsection{Summary: the Block Structure Lemma}\label{subsec:bsl_proof}

We summarize the results of Appendix~\ref{app:block_structure} in the following lemma, which describes every
block of~$G$ from its block color.

\begin{lemma}[Block Structure]\label{lem:block-structure}
For two colors $i,j$, let $\B_K(i,j)$ be the rooted colored graph defined
as follows.
\begin{itemize}
\item \textbf{If $j$ is not a closing color:} (Discrete case).\; A copy
      of the c-block $\Bup^K(j)$, colored by the identity and rooted at its
      block root.
\item \textbf{If $j$ is a closing color:} (Symmetric case).\; The rooted colored graph
      $\unfold(i,j,K)$ (Definition~\ref{def:unfold}).
\end{itemize}
Then every block of~$G$ of block color~$(i,j)$ is color-preserving
isomorphic to~$\B_K(i,j)$.
\end{lemma}

\begin{proof}
Let $A$ be a block of~$G$ with $\tilde{c}(A)=(i,j)$.  If $j$ is not
closing, $A$ is discrete by Corollary~\ref{cor:block_type}~(Block type
determination), and color-preserving isomorphic to~$\Bup^K(j)$ by
Lemma~\ref{lem:asym_iso}~(Discrete block isomorphism).  If $j$ is closing, $A$
is symmetric, an $ij$-chain exists by Lemma~\ref{lem:color_set}~(Color set),
and Lemma~\ref{lem:sym_structure}~(Structure of a symmetric
block) shows that it is color-preserving isomorphic to
$\unfold(i,j,K)$: Part~(1) identifies~$H$ with the $ij$-chain,
Part~(2) each branch with a copy of~$H$, Part~(3) the vertices shared by the
two branches with those identified in the unfolding, and Part~(4) the edges
between the branches with those added at the end of the unfolding.  In both
cases, the isomorphism sends $\ro(A)$, the only vertex of color~$i$ in~$A$, to
the root of $\B_K(i,j)$.  The computation of $\B_K(i,j)$ is described below.
\end{proof}

Note that~$\mathcal{B}_K(i,j)$ can be non-empty even
if no block of color~$(i,j)$ can exist.

\smallskip

We can now prove Corollary~\ref{cor:K-outerplanar}, stated in
Section~\ref{sec:details}.

\tildeKouterplanar*

\begin{proof}[Proof of Corollary~\ref{cor:K-outerplanar}]
A graph is outerplanar if and only if all its blocks are, so it suffices to
show that every c-block~$B$ is outerplanar.  Let $st$ be an edge of~$B$.  A
vertex of color~$s$ has a neighbor of color~$t$, and the edge between them
lies in a block~$A$ of~$G$, which contains the colors $s$ and~$t$ of~$B$.  By
Lemma~\ref{lem:block_maximality}~(Block maximality), every color of~$B$
appears in~$A$, so $B$ is a subgraph of~$\widetilde{K}[c(A)]$.  If $A$ is
discrete, $\widetilde{K}[c(A)]$ is color-preserving isomorphic to~$A$, as
shown in the proof of Lemma~\ref{lem:asym_iso}~(Discrete block isomorphism).
If $A$ is symmetric, $\widetilde{K}[c(A)]$ is color-preserving isomorphic to
the subgraph of~$A$ induced by a branch, by
Lemma~\ref{lem:sym_structure}(2)~(Structure of a symmetric block).  In both
cases, $B$ is a subgraph of an outerplanar graph, hence outerplanar. 
\end{proof}

\begin{remark}\label{rem:outer_path}
When an $ij$-chain $B_1\cup\dots\cup B_\ell$ is the chain of a symmetric
block~$A$, each edge $c_{a-1}c_a$ of its underlying path lies on the outer
face of~$B_a$.  Indeed, let $X$ be a branch of~$A$.  By
Lemma~\ref{lem:sym_structure}(1) and~(2), $A[X]$ is a copy of the chain, in
which the copy of~$B_a$ is the part of the branch between its vertices of
colors $c_{a-1}$ and~$c_a$ (as $c_1,\dots,c_{\ell-1}$ are the cut vertices of
the chain).  The edges of the branch lie on the outer face of~$A$, so the
outer cycle of this copy of~$B_a$ is this part of the branch, closed by the
edge between its two ends, of colors $c_{a-1}$ and~$c_a$.
\end{remark}

\subsection{Computation of $\B_K$}\label{subsec:compute_B}

We assume that $T_{\widetilde{K}}$ has been computed beforehand, together with
the c-block $\Bup^K(s)$ of every color~$s$.  If $j$ is not a closing color, we output a copy of~$\Bup^K(j)$, colored by the identity and rooted at its
block root.  Otherwise we first need to find the $ij$-chain,
following Lemma~\ref{lem:chain_unique}(1)~(Chains in~$T_{\widetilde{K}}$): we
start at~$j$ and repeatedly move from the current color~$s$ to the block root
of~$\Bup^K(s)$, until the color~$i$ is reached; if the root
of~$T_{\widetilde{K}}$ is reached first, or if two consecutive colors met are
not adjacent, then there is no $ij$-chain.  Otherwise, the c-blocks $\Bup^K(s)$ of
the colors~$s\neq i$ met are those of the chain, and their union is the
$ij$-chain~$C$ used in Definition~\ref{def:unfold}.  It only remains to create
two copies of~$C$, merge the vertices that have to be merged, and add an edge
between the two copies of each color~$s$ with $K_{ss}>0$.  All these
computations are linear in the size of the output.

\section{Block Count}\label{app:block_count}

In this section, we show that all the vertices of a given color~$i$ have the same
multiset of block colors of their child blocks, possibly empty, and that this
multiset, denoted~$F_K(i)$, can be read off~$K$. As it was for the Block Structure Lemma, the main result of this section, the Block Count Lemma (Lemma~\ref{lem:block-count}), will be crystallized and proved only later, in
Section~\ref{subsec:bcl_proof}. In the earlier subsections, we set the scene for the proof, which has two parts.
The first one is \emph{type determination}: given a neighbor~$w$ of a
vertex~$v$ of color~$i$, decide from~$K$ alone whether $\Bup^G(w)$ is
symmetric or discrete, and what its block color is.  The second one is
\emph{counting}: once this is known, count the neighbors of~$v$ in each
block, and deduce the number of blocks of each block color. Type determination is the difficult part, for which we need the notion of
\emph{well-closed chains}, a key structure to formalize the bottom-up scan
described in Section~\ref{sec:intuition}.

\subsection{Definitions and Preliminary Lemmas}
\label{subsec:bcl_def}

\begin{definition}[Multiplicity]\label{def:mult}
The \emph{multiplicity} of a c-block~$B$ with block root~$i$ is
$\mathrm{mult}(B)=\max_{s\in V(B)\setminus\{i\}}K_{is}$. The
\emph{multiplicity} of an $ij$-chain is the multiplicity of its first
c-block.
\end{definition}

As $i$ always has a neighbor in~$B$, $\mathrm{mult}(B)\ge1$.

\begin{definition}[Opening c-block]\label{def:opening}
A c-block~$B$ is an \emph{opening c-block} if $\mathrm{mult}(B)$ is
even.
\end{definition}

\begin{definition}[Well-closed $ij$-chain]\label{def:free-chain}
An $ij$-chain $B_1\cup\dots\cup B_\ell$ with underlying path
$(i=c_0,c_1,\dots,c_\ell=j)$ is a \emph{well-closed $ij$-chain} if $j$ is a
closing color and, for $\ell \geq 2$:
\begin{enumerate}
\item $B_a$ is not an opening c-block for any $a\in[2,\ell]$, and
\item $c_a$ is not a solo closing color for any $a\in[1,\ell-1]$.
\end{enumerate}
\end{definition}

\begin{definition}[Deepest well-closed $ij$-chain]\label{def:deepest-chain}
A well-closed $ij$-chain is \emph{deepest} if no well-closed $ij'$-chain with
$j'\neq j$ has $j$ on its underlying path. In other words, the
underlying path of the $ij$-chain is not a strict prefix of the
underlying path of another well-closed chain.
\end{definition}

The intuition behind those definitions is that a symmetric block of color
$(i,j)$ can exist only if there exists a deepest well-closed $ij$-chain.

\begin{figure}[ht]
\centering
\begin{tikzpicture}[
  lvertex/.style={circle, minimum size=4.5mm, inner sep=0pt, draw=black, thick},
  kvertex/.style={circle, minimum size=7mm, inner sep=0pt, draw=black, very thick},
  edge/.style={thick},
  biglabel/.style={font=\Large, anchor=north west},
  layerbox/.style={draw=#1, semithick, rectangle, inner sep=6pt},
  weight/.style={font=\scriptsize, inner sep=1pt},
  mult/.style={font=\scriptsize, text=black!50, anchor=west},
  ring/.style={draw=annRed, thick, circle, minimum size=9.5mm, inner sep=0pt},
]

  \providecolor{cR}{RGB}{220,50,50}
  \providecolor{cG}{rgb}{0.04,0.85,0.32}
  \providecolor{cB}{RGB}{80,130,220}
  \providecolor{cY}{rgb}{1.0,0.84,0.0}
  \providecolor{cO}{RGB}{240,150,50}
  \providecolor{cP}{RGB}{240,130,170}
  \providecolor{cV}{RGB}{160,100,210}
  \colorlet{cd}{cR}
  \colorlet{ce}{cB}
  \colorlet{cf}{cP}
  \colorlet{cg}{cG}
  \colorlet{ch}{cY}
  \colorlet{ci}{cV}
  \colorlet{cj}{cO}
  \definecolor{annGreen}{RGB}{100,180,60}
  \definecolor{annRed}{RGB}{230,50,40}

  \def\showannotations{1}

  \def\yd{0}
  \def\ye{-1.2}
  \def\yf{-2.6}
  \def\yg{-4.2}
  \def\yh{-5.6}
  \def\yi{-7.0}
  \def\yj{-8.3}
  \def\kx{7.5}
  \def\mx{8.75}

  \node[biglabel] at (-4.3,0.45) {$G$};

  \node[lvertex, fill=cd] (d)  at (0,\yd) {};
  \node[lvertex, fill=ce] (e1) at (-1.1,\ye) {};
  \node[lvertex, fill=ce] (e2) at ( 1.1,\ye) {};
  \node[lvertex, fill=cf] (f1) at (-1.95,\yf) {};
  \node[lvertex, fill=cf] (f2) at ( 1.95,\yf) {};
  \node[lvertex, fill=cg] (g1) at (-1.95,\yg) {};
  \node[lvertex, fill=cg] (g2) at ( 1.95,\yg) {};
  \foreach \n/\x in {1/-3.15, 2/-2.2, 3/-1.1, 4/1.1, 5/2.2, 6/3.15}
    \node[lvertex, fill=ch] (h\n) at (\x,\yh) {};
  \foreach \n/\x in {1/-2.7, 2/0, 3/2.7}
    \node[lvertex, fill=ci] (i\n) at (\x,\yi) {};
  \foreach \n/\x in {1/-3.3, 2/-2.1, 3/-0.6, 4/0.6, 5/2.1, 6/3.3}
    \node[lvertex, fill=cj] (j\n) at (\x,\yj) {};

  \draw[edge] (d) -- (e1) -- (f1) -- (g1) -- (g2) -- (f2) -- (e2) -- (d);
  \draw[edge] (d) to[bend left=25]  (f1);
  \draw[edge] (d) to[bend right=25] (f2);
  \foreach \n in {1,2,3} \draw[edge] (g1) -- (h\n);
  \foreach \n in {4,5,6} \draw[edge] (g2) -- (h\n);
  \foreach \u/\v in {1/1, 2/1, 3/2, 4/2, 5/3, 6/3}
    \draw[edge] (h\u) -- (i\v);
  \foreach \u/\v/\w in {1/1/2, 2/3/4, 3/5/6}
    \draw[edge] (j\v) -- (i\u) -- (j\w) -- (j\v);

  \node[layerbox=cd, fit=(d),       label=left:$d$] {};
  \node[layerbox=ce, fit=(e1)(e2), label=left:$e$] {};
  \node[layerbox=cf, fit=(f1)(f2), label=left:$f$] {};
  \node[layerbox=cg, fit=(g1)(g2), label=left:$g$] {};
  \node[layerbox=ch, fit=(h1)(h6), label=left:$h$] {};
  \node[layerbox=ci, fit=(i1)(i3), label=left:$i$] {};
  \node[layerbox=cj, fit=(j1)(j6), label=left:$j$] {};

  \node[biglabel] at (\kx-1.3,0.45) {$K$};

  \node[kvertex, fill=cd] (kd) at (\kx,\yd)     {$d$};
  \node[kvertex, fill=ce] (ke) at (\kx+1.2,\ye) {$e$};
  \node[kvertex, fill=cf] (kf) at (\kx,\yf)     {$f$};
  \node[kvertex, fill=cg] (kg) at (\kx,\yg)     {$g$};
  \node[kvertex, fill=ch] (kh) at (\kx,\yh)     {$h$};
  \node[kvertex, fill=ci] (ki) at (\kx,\yi)     {$i$};
  \node[kvertex, fill=cj] (kj) at (\kx,\yj)     {$j$};

  \draw[edge] (kd) -- (ke) -- (kf) -- (kd);
  \draw[edge] (kf) -- (kg) -- (kh) -- (ki) -- (kj);

  \draw[edge, dashed] (kg) to[out=30, in=-30, looseness=6] (kg);
  \draw[edge, dashed] (kj) to[out=30, in=-30, looseness=6] (kj);

  \node[weight] at ($(kd)+(0.55,-0.25)$)  {$2$};
  \node[weight] at ($(kd)+(-0.18,-0.7)$)  {$2$};
  \node[weight] at ($(kg)+(-0.15,-0.6)$)  {$3$};
  \node[weight] at ($(ki)+(-0.14,0.66)$)   {$2$};
  \node[weight] at ($(ki)+(-0.15,-0.5)$) {$2$};

  \node[mult] at (\kx+2.0,\ye) {$\times 2$};
  \node[mult] at (\mx,\yf) {$\times 2$};
  \node[mult] at (\mx,\yg) {$\times 2$};
  \node[mult] at (\mx,\yh) {$\times 6$};
  \node[mult] at (\mx,\yi) {$\times 3$};
  \node[mult] at (\mx,\yj) {$\times 6$};

  \ifnum\showannotations=1
    \draw[annGreen, thick, rounded corners=10pt]
      ($(kd)+(-0.5,0.5)$) -- ($(kd)+(0.4,0.5)$) -- ($(ke)+(0.65,0.1)$)
      -- ($(kf)+(0.4,-0.5)$) -- ($(kf)+(-0.5,-0.5)$) -- cycle;
    \node[draw=annGreen, thick, rounded corners=4pt,
          fit=(ki)(kj), inner sep=5pt] {};
    \foreach \n in {kg,ki,kj} \node[ring] at (\n) {};
  \fi

\end{tikzpicture}
\caption{On the left a rooted colored graph and its color graph on the right.
The colors $g$ and $j$ are paired closing colors while $i$ is solo closing. The underlying path of the $dh$-chain is $(d,f,g,h)$. Closing colors are circled in red and opening c-blocks are outlined in green.}
\label{fig:ex free chain}
\end{figure}
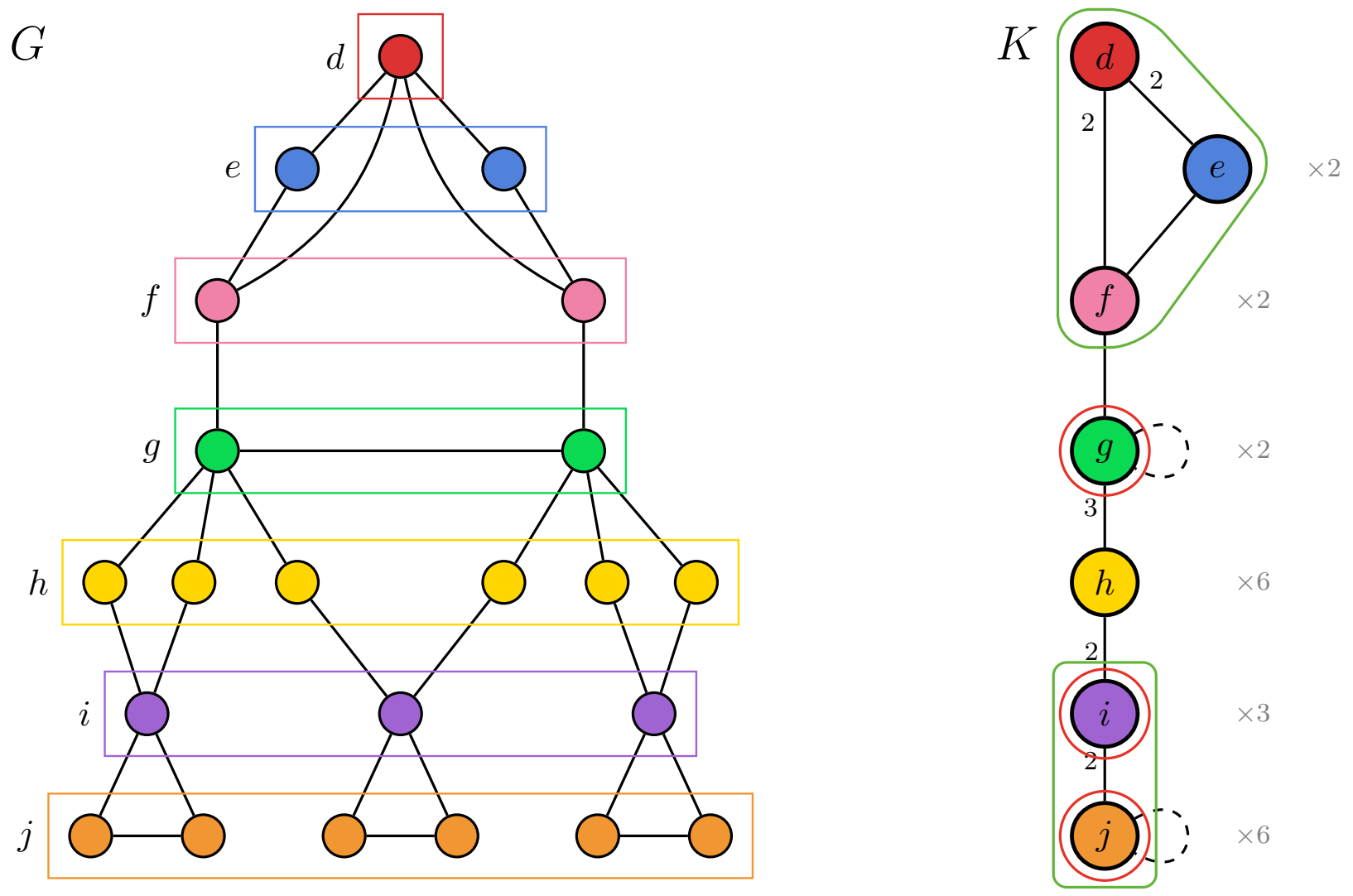

We illustrate these definitions in Figure~\ref{fig:ex free chain}.  Since a chain is determined by its
endpoints (Lemma~\ref{lem:chain_unique}), we name it after them: for
instance, the $di$-chain is the chain with underlying path $(d,f,g,h,i)$.

\begin{itemize}
\item The path $(d,e,f,g)$ is not the underlying path of a chain: $e$ is not the root of the c-block $\{d,e,f\}=\Bup^K(f)$.  The
      $dg$-chain does exist, but its underlying path is $(d,f,g)$.
\item The $dj$-chain ends at a closing color, but is not well-closed, since the
      c-block $\{i,j\}$ is opening.  We will see that $G$ cannot contain a
      block of color~$(d,j)$.
\item The $di$- and $ij$-chains are deepest well-closed chains, and so are the
      $fi$- and $gi$-chains, which are contained in the $di$-chain.  Blocks of
      color $(d,i)$, $(i,j)$, $(f,i)$ and~$(g,i)$ may therefore appear in~$G$.
      We will see, however, that no block of color~$(f,i)$ actually exists, as
      the first c-block $\{f,g\}$ of the $fi$-chain has multiplicity less
      than~$2$.
\item The $dg$-chain is well-closed, but not deepest, since $g$ lies on the underlying
      path of the well-closed $di$-chain.  We will see that despite $g$ being a
      closing color, $G$ cannot contain a block of color~$(d,g)$.
\end{itemize}

\begin{remark}\label{rem:free_ext}
Every well-closed $ij$-chain is contained in a deepest well-closed chain starting
at~$i$, obtained by extending it as long as possible.  Moreover, removing the
first c-blocks of a well-closed (resp.\ deepest well-closed) chain leaves a well-closed (resp.\
deepest well-closed) chain with the same lower endpoint.
\end{remark}

\bigskip

The following observation and lemmas are used throughout this
section.

\begin{observation}[Shared up-block]\label{obs:shared_upblock}
Let $v,w$ be adjacent vertices with $\depth_G(v)<\depth_G(w)$ and
$\ro(\Bup^G(w))\neq v$.  Then $\Bup^G(w)=\Bup^G(v)$.
\end{observation}

\begin{proof}
The edge $\{v,w\}$ lies in some block~$A$.  Since
$\depth_G(v)<\depth_G(w)$, we have $\ro(A)\neq w$
(Lemma~\ref{lem:highest}~(Block root)), so $A=\Bup^G(w)$.
By hypothesis $\ro(\Bup^G(w))\neq v$, hence $\Bup^G(v)=A$.
\end{proof}

\begin{lemma}[Same-c-block shared up-block]
\label{lem:same_cblock}
Let $a,b$ be two distinct colors in a c-block~$B$ such that neither~$a$
nor~$b$ equals $\ro(B)$.  Then for every vertex~$u$ of
color~$a$, there exists a vertex~$v$ of color~$b$ with
$\Bup^G(u)=\Bup^G(v)$.
\end{lemma}

\begin{proof}
Let $u$ be a vertex of color~$a$.
Since $a\in V(B)$ and $|B|\ge 2$, there exists a color $s\in V(B)$
with $s\neq a$ and $K_{as}>0$.
By equitability, $u$ has a neighbor~$w$ of color~$s$, and the
edge $\{u,w\}$ lies in some block~$A$.  Since $a,s\in
V(B)\cap c(A)$ with $a\neq s$,
Lemma~\ref{lem:block_maximality}~(Block maximality) gives
$V(B)\subseteq c(A)$, so there exists a
vertex~$v$ of color~$b$ in~$A$.  Furthermore, there exists a
vertex~$x$ of color~$\ro(B)$ in~$A$.  Since
$a,b\in V(B)\setminus\{\ro(B)\}$,
Lemma~\ref{lem:depth}~(Depth) gives
$\depth_G(u)>\depth_G(x)$ and $\depth_G(v)>\depth_G(x)$.  Hence
neither~$u$ nor~$v$ is the root of~$A$, giving
$\Bup^G(u)=A=\Bup^G(v)$.
\end{proof}

\begin{lemma}[Multiplicity of a c-block]\label{lem:mult_first}
Let $B$ be a c-block with block root~$i$.
\begin{enumerate}
\item If some color of $V(B)\cap N_{\Kt}(i)$ is not solo closing, then
      $K_{is}=\mathrm{mult}(B)$ for every such color~$s$.
\item Otherwise $V(B)=\{i,t\}$ for some color~$t$, the color~$t$ is solo
      closing, and $\mathrm{mult}(B)=K_{it}=1$.
\end{enumerate}
\end{lemma}

\begin{proof}
\emph{Part~(1).}  Let $v$ be a vertex of color~$i$, and let $\mathcal{A}$ be the set of blocks
containing~$v$ and a neighbor of~$v$ whose color lies in
$V(B)\setminus\{i\}$.  By Lemma~\ref{lem:block_maximality}~(Block
maximality), each block of~$\mathcal{A}$ contains every color of~$V(B)$.  We
show that, in each $A\in\mathcal{A}$, the vertex~$v$ has the same
number~$n_A\ge1$ of neighbors of color~$s$, whatever the color
$s\in V(B)\cap N_{\Kt}(i)$ that is not solo closing.
\begin{itemize}
\item If $A$ is discrete, each color of~$V(B)$ occurs exactly once in~$A$, and
      $v$ is its only vertex of color~$i$.  For every
      $s\in V(B)\cap N_{\Kt}(i)$, we have $K_{is}>0$, so
      Lemma~\ref{lem:edge_exists}~(Edges between colors of a block) gives an
      edge between $v$ and the vertex of color~$s$: $n_A=1$.
\item If $A$ is symmetric, a color $s\in V(B)\cap N_{\Kt}(i)$ that is not solo
      closing is neither the color of~$\ro(A)$, which is $i$ or a color of
      smaller depth, nor the color of a unique lowest vertex of~$A$, which is
      solo closing by Lemma~\ref{lem:bottom_symmetric}(2)~(Closing colors and
      symmetric blocks).  So $s$ occurs twice in~$A$, once on each branch, and       
      by Lemma~\ref{lem:sym_structure}(2) and~(4)~(Structure of a symmetric
      block), $v$ is adjacent to both vertices of color~$s$ if $v$ lies on
      both branches of~$A$, and only to the one on its own branch otherwise:
      $n_A=2$ or $n_A=1$ accordingly, and it does not depend on~$s$.   
\end{itemize}
Hence $K_{is}=\sum_{A\in\mathcal{A}}n_A$ is the same for all colors
$s\in V(B)\cap N_{\Kt}(i)$ that are not solo closing.  Let now
$t\in V(B)\cap N_{\Kt}(i)$ be solo closing.  In a block $A\in\mathcal{A}$, a
vertex~$y$ of color~$t$ is not the block root, as $\depthK(i)<\depthK(t)$, so
$A=\Bup^G(y)$ contains no other vertex of color~$t$
(Lemma~\ref{lem:bottom_symmetric}(1)).  So $v$ has at most one neighbor of
color~$t$ in~$A$, and $K_{it}\le\sum_{A\in\mathcal{A}}n_A=K_{is}$. As the
colors of $V(B)\setminus\{i\}$ not adjacent to~$i$ contribute~$0$, the maximum
defining $\mathrm{mult}(B)$ is reached at every color
$s\in V(B)\cap N_{\Kt}(i)$ that is not solo closing.

\medskip
\emph{Part~(2).}  Here every color of $V(B)\cap N_{\Kt}(i)$ is solo closing.  Two
distinct solo closing colors of $V(B)\setminus\{i\}$ would, by
Lemma~\ref{lem:same_cblock}~(Same-c-block shared up-block), have vertices
sharing an up-block~$A$, and each of them would be the color of the unique
vertex of $\Low(A)$ by Lemma~\ref{lem:bottom_symmetric}(1)~(Closing colors and
symmetric blocks), so they would be equal.  Hence $i$ has at most one neighbor
in~$B$.  A c-block is a bridge or is $2$-connected, and in a $2$-connected
graph every vertex has at least two neighbors; so $B$ is a bridge, that is,
$V(B)=\{i,t\}$ with $t$ solo closing.

Suppose $K_{it}\ge2$.  As $\Bup^K(t)=B=\{i,t\}$, the definition of a solo
closing color forces $K_{ti}\ge2$, the color~$j'$ of
Definition~\ref{def:closing} being different from~$t$.
Since $t$ is solo closing, each block contains at most one vertex of color~$t$ (Lemma~\ref{lem:bottom_symmetric}(1) if it is not the block root, Observation~\ref{obs:root-color-unique} otherwise), so a vertex $v$ of color $i$ has at most one neighbor of color $t$ in $\Bup^G(v)$, and thus a neighbor $y$ of color $t$ in some block $A$ rooted at $v$.
By Observation~\ref{obs:i-accessible}~(Root color accessibility) applied to $A$, $v$ is the only
neighbor of $y$ of color $i$, contradicting $K_{it}\ge2$.
Hence $K_{it}=1$, and
$\mathrm{mult}(B)=K_{it}$ as $V(B)\setminus\{i\}=\{t\}$.
\end{proof}

\subsection{Determining Block Type}
\label{subsec:bcl_type}
The goal of this subsection is to prove that for any vertex, the
type of its up-block (discrete or symmetric) and the second
component of its block color can be determined from the color
of the vertex and $K$.
The following lemma captures the most technical part of this.

\begin{lemma}[Well-closed chain descent]\label{lem:free_descent}
Let $C=B_1\cup\dots\cup B_\ell$ be a well-closed $tk$-chain with
underlying path $(t=c_0,c_1,\dots,c_\ell=k)$.
\begin{enumerate}
\item For every color $j\in V(C)\setminus\{t\}$ and every vertex~$u$
      of color~$j$, there exists a vertex~$w$ of color~$k$ with
      $\Bup^G(u)=\Bup^G(w)$.  In particular, $\Bup^G(u)$ is
      symmetric.
\item If $B_1$ is an opening c-block, then for every vertex~$u$
      of color~$t$ and every neighbor~$v$ of~$u$ with
      $c(v)\in V(B_1)\setminus\{t\}$, we have
      $u=\ro(\Bup^G(v))$.
\end{enumerate}
\end{lemma}

\begin{proof}
\textbf{Part~(1).}
By Lemma~\ref{lem:same_cblock}~(Same-c-block shared up-block), it
suffices to prove the result for $j=c_a$ with $a\in[1,\ell]$.  Indeed,
let $j'\in V(C)\setminus\{t\}$ and let $B_a$ be a c-block of the chain
containing~$j'$.  If $j'=\ro(B_a)$, then $j'=c_{a-1}$ by definition, with $a\ge2$ as $j'\neq t$, so $j'$ is itself
one of these colors.  Otherwise $j'$ and~$c_a$ both lie in
$V(B_a)\setminus\{\ro(B_a)\}$, so every vertex of color~$j'$ shares its
up-block with a vertex of color~$c_a$, and the result for~$c_a$ gives
it for~$j'$.  We proceed by backward induction on~$a$.

\emph{Base case} ($a=\ell$, $j=k$): take $w=u$.  Since $k$ is
closing, $\Bup^G(u)$ is symmetric by
Lemma~\ref{lem:bottom_symmetric}~(Closing colors and symmetric
blocks).

\emph{Inductive step}: assume the result holds for all $c_b$
with $b>a$.  Let $u$ be a vertex of color~$c_a$ ($1\le a<\ell$),
and let $A=\Bup^G(u)$.  We show that there exists a neighbor~$v$
of~$u$ of color~$c_{a+1}$ with $\Bup^G(v)=A$, then apply
induction to~$v$.

Suppose for contradiction that every neighbor of~$u$ of
color~$c_{a+1}$ has~$u$ as root of its up-block.  By the
induction hypothesis, each such up-block contains a vertex of
the closing color~$k$ and is therefore symmetric
(Lemma~\ref{lem:bottom_symmetric}~(Closing colors and symmetric
blocks)).

\emph{Case~1: $c_{a+1}$ is not solo closing.}
In each symmetric child block with root~$u$, color~$c_{a+1}$
appears exactly twice (it is neither the root color~$c_a$ nor
the unique bottom of a solo-closing block, so the Dichotomy Lemma
gives two copies).  Both copies are adjacent to~$u$ (the unique
vertex of color~$c_a$ in the block, by
Lemma~\ref{lem:highest}~(Block root) and Lemma~\ref{lem:sym_structure}(2),
Structure of a symmetric block).  Hence each child block consumes exactly two neighbors
of~$u$ of color~$c_{a+1}$, so $K_{c_ac_{a+1}}$ is even.
Since $c_{a+1}\in V(B_{a+1})\cap N_{\Kt}(c_a)$ is not solo closing,
Lemma~\ref{lem:mult_first}(1)~(Multiplicity of a c-block) gives
$\mathrm{mult}(B_{a+1})=K_{c_ac_{a+1}}$, which is even.  So $B_{a+1}$ is
opening, contradicting the well-closed chain condition, since $a+1\ge2$.

\emph{Case~2: $c_{a+1}$ is solo closing.}
By the well-closed chain condition (no intermediate solo closing color),
$a+1=\ell$, so $c_{a+1}=k$, and $c_a=\ro(B_\ell)$
(Lemma~\ref{lem:chain_unique}).  Let $w$ be a neighbor of~$u$ of color~$k$.
By assumption, $\Bup^G(w)$ is rooted at~$u$, so it is a symmetric block of
block color~$(c_a,k)$ (Lemma~\ref{lem:bottom_symmetric}(1), Closing colors and
symmetric blocks), and by Lemma~\ref{lem:color_set}~(Color set) its colors
are those of the $c_ak$-chain, that is, of~$B_\ell$.  Let $t'$ be the color
of the two outer-face neighbors of~$u$ in~$\Bup^G(w)$. These are two
distinct vertices, as $\Bup^G(w)$ is symmetric, hence not a bridge
(Remark~\ref{rem:bridge_discrete}), so $t'$ appears twice in~$\Bup^G(w)$ and
is not solo closing.  Moreover $t'\in V(B_\ell)\cap N_{\Kt}(c_a)$.  By
Lemma~\ref{lem:mult_first}(1)~(Multiplicity of a c-block),
$\mathrm{mult}(B_\ell)=K_{c_at'}$; we next show that it is even, so that
$B_\ell$ is opening, contradicting the well-closed chain condition since $\ell\ge2$.

No neighbor of~$u$ of color~$t'$ lies in~$A$: otherwise $A$ would
contain the colors $c_a$ and~$t'$ of~$B_\ell$, hence a vertex~$w'$ of
color~$k$ by Lemma~\ref{lem:block_maximality}~(Block maximality), and
$A=\Bup^G(w')$ would be symmetric with $\Low(A)=\{w'\}$, as $k$ is solo
closing; since $K_{c_ak}>0$, Lemma~\ref{lem:sym_structure}(2)~(Structure of
a symmetric block) would then make $u$ adjacent to~$w'$, contradicting the
assumption that every neighbor of~$u$ of color~$k$ has~$u$ as root of its
up-block.  So every neighbor of~$u$ of color~$t'$ lies in a block rooted
at~$u$.  Such a block contains the colors $c_a$ and~$t'$ of~$B_\ell$, hence a
vertex~$y$ of color~$k$ (Lemma~\ref{lem:block_maximality}), which is adjacent
to~$u$ (Lemma~\ref{lem:sym_structure}(2)). In each such block, $t'$ is neither the
color~$c_a$ of the root nor the color~$k$ of the unique lowest vertex, so it
appears twice, once on each branch, and $u$ is adjacent to both copies
(Lemma~\ref{lem:sym_structure}(2)).  Hence $\mathrm{mult}(B_\ell)=K_{c_at'}$
is twice the number of these blocks, which is even.

\medskip
In both cases we reach a contradiction, so there exists a
neighbor~$v$ of~$u$ of color~$c_{a+1}$ with
$\ro(\Bup^G(v))\neq u$.  By
Observation~\ref{obs:shared_upblock}~(Shared up-block),
$\Bup^G(v)=\Bup^G(u)=A$.  Applying the induction hypothesis
to~$v$ (color~$c_{a+1}$), there exists a vertex~$w$ of
color~$k$ with $\Bup^G(v)=\Bup^G(w)$, hence
$\Bup^G(u)=\Bup^G(w)$.  Since $k$ is closing, $\Bup^G(w)$ is
symmetric (Lemma~\ref{lem:bottom_symmetric}~(Closing colors and
symmetric blocks)).

\bigskip
\textbf{Part~(2).}
Suppose $B_1$ is opening.  If every color of $V(B_1)\cap N_{\Kt}(t)$ were
solo closing, Lemma~\ref{lem:mult_first}(2)~(Multiplicity of a c-block) would
give $\mathrm{mult}(B_1)=1$, contradicting that $\mathrm{mult}(B_1)$ is even.
So some color $s\in V(B_1)\cap N_{\Kt}(t)$ is not solo closing, and
$K_{ts}=\mathrm{mult}(B_1)$ is even by Lemma~\ref{lem:mult_first}(1).  Fix
such a color~$s$.  \fpass{If $t=1$, every block containing the vertex~$r$ of
color~$t$ is rooted at~$r$ and there is nothing to prove; so assume $t\neq1$.}
Let $u$ be a vertex of color~$t$, and suppose for
contradiction that some neighbor~$v$ of~$u$ with
$c(v)\in V(B_1)\setminus\{t\}$ satisfies $\ro(\Bup^G(v))\neq u$.

Let $A=\Bup^G(u)$.  Since $v$ is a deeper neighbor of~$u$ and
$\ro(\Bup^G(v))\neq u$, Observation~\ref{obs:shared_upblock} (Shared up-block) gives that
$\Bup^G(v)=A$.  Since $c(v)\in V(C)\setminus\{t\}$ (as $V(B_1)\subseteq V(C)$),
Part~(1) gives that $A$ is symmetric.  Since $t,c(v)\in V(B_1)\cap c(A)$,
Lemma~\ref{lem:block_maximality}~(Block maximality) gives
$V(B_1)\subseteq c(A)$; in particular $s\in c(A)$.  We derive a contradiction
by counting the neighbors of~$u$ of color~$s$.

The vertex~$u$ lies on only one branch of~$A$: it is not $\ro(A)$ by
definition of~$A$, and it is not the unique lowest vertex of~$A$, since
otherwise $s$ would appear at a smaller face-depth than~$t$, and
Lemma~\ref{lem:facedepth_upblock}~(Face-depth ordering) would make
$\Bup^K(s)$ an ancestor of or equal to $\Bup^K(t)$ in~$T_{\widetilde{K}}$, whereas
$\Bup^K(s)=B_1$ is a child of~$t$.  By Lemma~\ref{lem:sym_structure}(2)
and~(4)~(Structure of a symmetric block), $u$ has exactly one neighbor of
color~$s$ in~$A$, the one on its own branch.  Every block rooted at~$u$
containing a neighbor of color~$s$ is symmetric (Part~(1), since
$s\in V(B_1)\setminus\{t\}\subseteq V(C)\setminus\{t\}$). In such a block, $s$ is
neither the color~$t$ of the root nor the color of a unique lowest vertex, as
$s$ is not solo closing; so $s$ appears twice, and $u$, which lies on both
branches, is adjacent to both copies (Lemma~\ref{lem:sym_structure}(2)). Hence $K_{ts}=1+2p$, 
for $p\ge 0$ the number of such blocks rooted at~$u$,
which is odd, contradicting $K_{ts}$ even.

Thus no neighbor of~$u$ whose color is in $V(B_1)\setminus\{t\}$ can appear
in $\Bup^G(u)$; the up-blocks of such neighbors are therefore all rooted
at~$u$.

\end{proof}

The following lemma is the main result of this subsection: it characterizes,
from~$K$ alone, the vertices whose up-block is symmetric, based on the
existence of a well-closed chain, and determines the second component of the
block color.  This only depends on the color of the vertex.

\begin{lemma}[Symmetry and well-closed chains]\label{lem:sym_free_chain}
Let $j\neq1$ be a color and $v$ a vertex of color~$j$.
\begin{enumerate}
\item If $\Bup^G(v)$ is symmetric, of block color~$(i,k)$, then there
      exists a deepest well-closed $ik$-chain whose color set
      contains~$j$.
\item If $j\neq t$ lies in the color set of a deepest well-closed
      $tk$-chain, then $\Bup^G(v)$ is symmetric
      and its block color is $(c(\ro(\Bup^G(v))), k)$.
\end{enumerate}
\end{lemma}

Note that in the second part of the lemma, the existence of the deepest well-closed
$tk$-chain gives $k$ as the second component of the block color, but does not
give anything about the first component of the block color (in particular, it
is not necessarily $t$).

\begin{proof}
We first prove a claim used in both parts.

\begin{claim}\label{cl:closing_in_block}
Let $A$ be a symmetric block, and let $k$ be a closing color appearing at a
vertex of~$A$ other than~$\ro(A)$.  If $c(\Low(A))\neq\{k\}$, then $k$ is
paired closing and separates $c(\ro(A))$ from the color of~$\Low(A)$
in~$\widetilde{K}$.
\end{claim}

\begin{proof}[Proof of the claim]
Let $u\neq\ro(A)$ be a vertex of color~$k$ in~$A$, so that $A=\Bup^G(u)$.  If
$k$ were solo closing, Lemma~\ref{lem:bottom_symmetric}(1)~(Closing colors and
symmetric blocks) would give $\Low(A)=\{u\}$; so $k$ is paired closing.  By the
Dichotomy Lemma, the two vertices of color~$k$ in~$A$ lie at the same
face-depth, one on each branch, strictly between the face-depths of~$\ro(A)$
and of~$\Low(A)$ (as $k$ is the color of neither), and since $K_{kk}>0$ they
are joined by a chord (Lemma~\ref{lem:sym_structure}(4), Structure of a
symmetric block).  By outerplanarity, this chord separates $\ro(A)$
from~$\Low(A)$ in~$A$: every path in~$A$ from~$\ro(A)$ to a vertex
of~$\Low(A)$ goes through a vertex of color~$k$.  \fpass{In particular, $k$ lies on
the color sequence of a shortest path from~$\ro(A)$ to a vertex of~$\Low(A)$,
which is the underlying path of the chain of~$A$
(Remark~\ref{rem:chain_shortest_path}), and it is neither its first nor its
last color.  By Lemma~\ref{lem:chain_unique}(2)~(Chains
in~$T_{\widetilde{K}}$), $k$ separates the next color of this path from~$1$
in~$\widetilde{K}$, hence also the color of~$\Low(A)$, which that color
reaches along the end of the path, whose colors are deeper than~$k$.  As
$\depthK(c(\ro(A)))<\depthK(k)$, a shortest path from~$1$ to~$c(\ro(A))$
avoids~$k$; so $k$ separates $c(\ro(A))$ from the color of~$\Low(A)$.}
\end{proof}

\textbf{Part~(1).}
Let $A=\Bup^G(v)$ (since $j\neq 1$, $v\neq r$, so $A$ is well-defined),
of block color $(i,k)$, so that $i=c(\ro(A))$ and
$\{k\}=c(\Low(A))$.  Since $v\neq\ro(A)$, we have $i\neq j$ (by
definition of a block root and the Depth Lemma \ref{lem:depth}).
Lemma~\ref{lem:color_set} (Color Set) and Remark \ref{rem:chain_shortest_path}
show that the color sequence of a shortest path
from $\ro(A)$ to a vertex in $\Low(A)$ is the underlying path
$(i=c_0,c_1,\dots,c_\ell=k)$ of an $ik$-chain
$C=B_1\cup\dots\cup B_\ell$ with $V(C)=c(A)$; in particular
$j\in V(C)$.  By Lemma~\ref{lem:bottom_symmetric}(2), its lower endpoint~$k$ is
closing.  It remains to show that no~$c_a$ with $a\in[1,\ell-1]$ is solo
closing (Step~1) and no~$B_a$ with $a\in[2,\ell]$ is opening (Step~2), so that
the chain is well-closed, and that it is deepest (Step~3).

\medskip
\emph{Step~1: no intermediate solo closing color.}
By Lemma~\ref{lem:bottom_symmetric}(1) (Closing colors and symmetric blocks), a
vertex of solo closing color in~$A$ must either be $\ro(A)$ or the unique
vertex in~$\Low(A)$. As the underlying path of our $ik$-chain is the
color sequence of a shortest path from $\ro(A)$ to a vertex in $\Low(A)$, only
$c_0$ and $c_\ell$ can be solo closing, and in particular no~$c_a$ with
$a\in[1,\ell-1]$ is solo closing.

\medskip
\emph{Step~2: no opening c-block.}
Suppose for contradiction that $B_{a}$ is opening for some
$a\in[2,\ell]$; take $a_0$ the maximal such $a$.  The
$c_{a_0-1}k$-chain $B_{a_0}\cup\dots\cup B_\ell$, with underlying path
$(c_{a_0-1},c_{a_0},\dots,c_\ell=k)$ (Lemma~\ref{lem:chain_unique}), is well-closed:
the c-blocks $B_a$ for all $a>a_0$ are non-opening by maximality of $a_0$, and
there are no intermediate solo closing colors by Step~1.  Its first
c-block~$B_{a_0}$ is opening.

Since $a_0\ge 2$, the color $c_{a_0-1}\neq i$, so there exists a
vertex~$v'$ of color~$c_{a_0-1}$ in~$A$ with $\Bup^G(v')=A$.  Since
$K_{c_{a_0-1}c_{a_0}}>0$ and both
colors appear in~$A$, Lemma~\ref{lem:sym_structure}(2)~(Structure of a
symmetric block) gives that $v'$
has a neighbor~$w'$ of color~$c_{a_0}$ in~$A$.  Since
$c_{a_0}\in B_{a_0}\setminus\{c_{a_0-1}\}$, and $\Bup^G(w')=A$
is rooted at $\ro(A)\neq v'$, this contradicts
Lemma~\ref{lem:free_descent}(2) (Well-closed chain descent) applied to the
$c_{a_0-1}k$-chain.

\medskip
\emph{Step~3: the chain is deepest.}
Suppose for contradiction that a well-closed $ik'$-chain with $k'\neq k$ has
$k$ on its underlying path $(i=c'_0,c'_1,\dots,c'_{\ell'}=k')$, say $k=c'_a$
with $a\in[1,\ell'-1]$; in particular $\depthK(k)<\depthK(k')$, as the
underlying path is descending.  Let $u\in\Low(A)$, of color~$k$.  As $k$ lies
in the color set of the $ik'$-chain and differs from~$i$,
Lemma~\ref{lem:free_descent}(1)~(Well-closed chain descent) applied to that chain
gives a vertex~$u'$ of color~$k'$ with $\Bup^G(u')=\Bup^G(u)=A$, and
$k'$ is closing by definition of a well-closed chain. As $c(\Low(A))=\{k\}\neq\{k'\}$, Claim~\ref{cl:closing_in_block} shows that
$k'$ separates~$i$ from~$k$ in~$\widetilde{K}$.  As $k'$ appears in~$A$ below
its root, $\depthK(i)<\depthK(k')$, so a shortest path from~$i$ to~$1$
avoids~$k'$, and $k'$ separates~$k$ from~$1$.  Hence a shortest path from~$1$
to~$k$ goes through~$k'$, and $\depthK(k')<\depthK(k)$, contradicting
$\depthK(k)<\depthK(k')$.

\bigskip
\textbf{Part~(2).}
Let $C=B_1\cup\dots\cup B_\ell$ be a deepest well-closed $tk$-chain with
underlying path $(t=c_0,c_1,\dots,c_\ell=k)$, with
$j\in V(C)\setminus\{t\}$, and $v$ a vertex of color $j$.  By
Lemma~\ref{lem:free_descent}(1),
there exists a vertex~$w$ of color~$k$ with
$\Bup^G(v)=\Bup^G(w)=:A$.  Since $k$ is closing,
Lemma~\ref{lem:bottom_symmetric} gives that $A$ is symmetric.

It remains to show $c(\Low(A))=\{k\}$.  Suppose for
contradiction that $c(\Low(A))=\{k'\}$ with $k'\neq k$, and let
$t'=c(\ro(A))$.  As $w\neq\ro(A)$, Claim~\ref{cl:closing_in_block} shows that
$k$ is paired closing and separates~$t'$ from~$k'$ in~$\widetilde{K}$.
We show that the $tk$-chain is not deepest.
Apply Part~(1) to~$w$ ($A$ is symmetric of block color $(t',k')$):
there exists a deepest well-closed $t'k'$-chain
$B'_1\cup\dots\cup B'_{\ell'}$ whose color set contains~$k$.
As $k$ separates~$t'$ from~$k'$ in~$\widetilde{K}$, it lies on any path from $t'$ to $k'$ 
and in particular the underlying path $(t'=c'_0,\dots,c'_{\ell'}=k')$ of this chain, say $k=c'_a$
with $a\in[1,\ell'-1]$.

Concatenating the underlying path of the $tk$-chain with the end
$(c'_a,\dots,c'_{\ell'})$ of that of the $t'k'$-chain gives a path
$(t=c_0,\dots,c_\ell=k=c'_a,c'_{a+1},\dots,c'_{\ell'}=k')$, descending as both
parts are.  Each of its edges lies in a c-block of one of these two chains,
whose block root is its upper end, so this path is the underlying path of the
$tk'$-chain $B_1\cup\dots\cup B_\ell\cup B'_{a+1}\cup\dots\cup B'_{\ell'}$.
This $tk'$-chain is well-closed: $k'$ is closing; its c-blocks of index at
least~$2$ are c-blocks of index at least~$2$ of the $tk$-chain or of the
$t'k'$-chain, hence non-opening; and its intermediate colors are intermediate
colors of one of these two chains, or~$k$, which is not solo closing as it is
paired closing (Remark~\ref{rem:solo_paired}).

So $k$ lies on the underlying path of a well-closed $tk'$-chain with
$k'\neq k$, contradicting that the $tk$-chain is deepest.
\end{proof}

\bigskip
\begin{corollary}[Color determines up-block type]
\label{cor:color_type}
For any color~$j\neq 1$, either every vertex of color~$j$ has a
symmetric up-block, or every vertex of color~$j$ has a discrete
up-block.
\end{corollary}

\begin{proof}
By Lemma~\ref{lem:sym_free_chain}~(Symmetry and well-closed chains),
$\Bup^G(v)$ is symmetric if and only if there exists a deepest
well-closed $tk$-chain with $t\neq j$ whose color set contains~$j$.
This condition depends only on~$j$, not on~$v$.
\end{proof}

\begin{corollary}[Block type determines appearance in well-closed chain]
\label{cor:block_type_conditions}
If a block of color~$(i,j)$ is symmetric, then there exists a deepest well-closed
$ij$-chain. If it is discrete, then there is no well-closed $tk$-chain, for any colors $t,k$
with $t\neq j$, whose color set contains~$j$.
\end{corollary}

\begin{proof}
Let $A$ be a block of block color~$(i,j)$.  If $A$ is symmetric,
Lemma~\ref{lem:sym_free_chain}(1)~(Symmetry and well-closed chains), applied to a
vertex of $\Low(A)$, gives a deepest well-closed $ij$-chain.  If $A$ is discrete,
let~$u$ be its vertex of color~$j$, so that $A=\Bup^G(u)$.  A well-closed $tk$-chain
with $t\neq j$ whose color set contains~$j$ would be contained in a deepest
well-closed chain with the same upper endpoint (Remark~\ref{rem:free_ext}), and
Lemma~\ref{lem:sym_free_chain}(2) (Symmetry and well-closed chains) would make $A$ symmetric, contradicting the
Dichotomy Lemma~\ref{lem:dichotomy}.
\end{proof}

\begin{lemma}[Common lower endpoint]\label{lem:one_family}
Two deepest well-closed chains with a c-block in common have the same lower
endpoint.
\end{lemma}

\begin{proof}
Let $B$ be a c-block common to two deepest well-closed chains, with lower
endpoints $k$ and~$k'$.  Pick $p\in V(B)\setminus\{\ro(B)\}$ and a
vertex~$v$ of color~$p$, and let $A=\Bup^G(v)$, which is well defined (since $p\neq \ro(B)$ implies $p\neq 1$) and unique.  For each of the two
chains, $p$ lies in its color set and differs from its upper endpoint, which
is either $\ro(B)$ or a color of strictly smaller depth than every color
of~$B$.  Lemma~\ref{lem:sym_free_chain}(2)~(Symmetry and well-closed chains)
applied to each chain therefore gives that the second component of
$\tilde{c}(A)$ is~$k$, and also~$k'$.  Hence $k=k'$.
\end{proof}

\subsection{Counting}\label{subsec:bcl_count}

\begin{lemma}[Counting discrete blocks]\label{lem:counting_disc}
Suppose no well-closed $tk$-chain with $t\neq j$ has~$j$ in its color
set, $i=\ro(\Bup^K(j))$, and $j$ is the minimum neighbor of~$i$
in~$\Bup^K(j)$.  Then for any vertex~$v$ of color~$i$, the
number of blocks of color~$(i,j)$ with root~$v$ equals~$K_{ij}$, which
is also $\mathrm{mult}(\Bup^K(j))$.
\end{lemma}

\begin{proof}
By the contrapositive of
Lemma~\ref{lem:sym_free_chain}~(Symmetry and well-closed chains),
Part~(1), since no well-closed $tk$-chain with $t\neq j$ has~$j$ in
its color set, every vertex of color~$j$ has a discrete up-block.
We show that the blocks of color~$(i,j)$ with root~$v$ are
exactly the up-blocks of the neighbors of~$v$ of color~$j$, each
neighbor giving a distinct block.

On one hand, let~$A$ be a block of color~$(i,j)$ with root~$v$.
Then $A=\Bup^G(w)$ for some vertex~$w$ of color~$j$, and since
$A$ is discrete, $w$ is the unique vertex of color~$j$ in~$A$.
By definition of~$\tilde{c}$, $j$ is a neighbor of~$i$
in~$\Bup^K(j)$, so Lemma~\ref{lem:asym_iso}~(Discrete block isomorphism) gives that $w$ is a
neighbor of~$v$ in~$A$ ($A$ is color isomorphic to $\Bup^K(j)$).

On the other hand, let~$w$ be a neighbor of~$v$ of color~$j$.
Its up-block $\Bup^G(w)$ is discrete and isomorphic
to~$\Bup^K(j)$ (Lemma~\ref{lem:asym_iso}, Discrete block
isomorphism). Since $j$
is the minimum neighbor of~$i$ in~$\Bup^K(j)$, we get
\fpass{$\tilde{c}(\Bup^G(w))=(i,j)$. Now if $v\neq \ro(\Bup^G(w))$, then
$v\notin\Bup^G(w)$, whose block root is its only vertex of color~$i$, so}
the edge $\{v,w\}$ lies in some block~$A'\neq \Bup^G(w)$ and thus with
$\ro(A')=w$, giving by definition of a block root $\depthK(j)<\depthK(i)$. Since
$i=\ro(\Bup^K(j))$ gives
$\depthK(i)<\depthK(j)$, we get a contradiction. Thus $\Bup^G(w)$ is rooted in
$v$ and its color is $(i,j)$.

Since each block of color $(i,j)$ rooted at $v$ contributes exactly one
neighbor of $v$ of color $j$ and each
neighbor determines exactly one such block, the count
equals~$K_{ij}$.

Finally, $j$ is not closing, as otherwise the up-blocks of the vertices
of color~$j$ would be symmetric (Lemma~\ref{lem:bottom_symmetric}(1), Closing
colors and symmetric blocks), whereas they are discrete.  So
Lemma~\ref{lem:mult_first}(1)~(Multiplicity of a c-block) gives
$K_{ij}=\mathrm{mult}(\Bup^K(j))$.
\end{proof}

\begin{lemma}[Counting symmetric blocks]\label{lem:counting_sym}
Suppose there exists a deepest
well-closed $ij$-chain with first c-block~$B_1$.  Then for any vertex~$v$ of
color~$i$, the number of blocks of color~$(i,j)$ with root~$v$
equals $\lfloor\mathrm{mult}(B_1)/2\rfloor$.
\end{lemma}

\begin{proof}

Let $(i=c_0,c_1,\dots,c_\ell=j)$ be the underlying path of the chain.
If every color of $V(B_1)\cap N_{\Kt}(i)$ is solo closing, then
$V(B_1)=\{i,c_1\}$ with $c_1$ solo closing and $\mathrm{mult}(B_1)=1$, by
Lemma~\ref{lem:mult_first}(2)~(Multiplicity of a c-block); as a solo closing
color cannot be an intermediate color of a well-closed chain, $\ell=1$ and $c_1=j$. 
No block of block color~$(i,j)$ then
exists.  Indeed, such a block~$A$ would be symmetric, $j$ being closing
(Corollary~\ref{cor:block_type}, Block type determination), with color set
$V(B_1)=\{i,j\}$ by Lemma~\ref{lem:color_set}~(Color set); the two outer-face
neighbors of~$\ro(A)$, which are distinct as $A$ is not a bridge
(bridges are considered discrete), would then both have color~$j$, whereas
$j$, being solo closing, occurs only once in~$A$
(Lemma~\ref{lem:bottom_symmetric}(1), Closing colors and symmetric blocks).
Otherwise, let $s$ be a color of $V(B_1)\cap N_{\Kt}(i)$ that is not solo closing;
then $K_{is}=\mathrm{mult}(B_1)$ by Lemma~\ref{lem:mult_first}(1), and we show
that the number of blocks is $\lfloor K_{is}/2\rfloor$.

Let $W=\{w\in N(v):c(w)=s\}$; by equitability $|W|=K_{is}$.
Let $p$ denote the number of blocks of color~$(i,j)$ with
root~$v$, and let
$q=|\{w\in W:\ro(\Bup^G(w))\neq v\}|$.

\medskip\noindent
\textit{(1) Each block of color~$(i,j)$ with root~$v$ contains
exactly two elements of~$W$, and distinct blocks contain distinct
elements.}

Let $A$ be such a block. The existence of a well-closed $ij$-chain gives that $j$ is
closing, and thus $A$ is symmetric
(Lemma~\ref{lem:bottom_symmetric}).  By
Lemma~\ref{lem:color_set} (color set of a symmetric block),
$B_1\subseteq c(A)$, so
$s\in c(A)$.  Since $s\neq i$ and $s$ is
not solo closing, Lemma~\ref{lem:sym_structure}(2) and~(3)~(Structure of a symmetric
block) gives that $s$ appears exactly twice in~$A$ (once on
each branch).
The vertex~$v$ is the unique vertex of color~$i$ in~$A$
(Lemma~\ref{lem:highest}, Block root), and $s$ is a neighbor of $i$, so Lemma~\ref{lem:sym_structure}(2) gives that both vertices of color~$s$ are
adjacent to~$v$.  Hence they both belong to~$W$.
Distinct blocks contain distinct elements: if $w\in W$ lies in
two blocks $A_1,A_2$ with root~$v$, then
$A_1=\Bup^G(w)=A_2$ by uniqueness of the up-block.

\medskip\noindent
\textit{(2) For each $w\in W$ with $\ro(\Bup^G(w))=v$, the
block $\Bup^G(w)$ has color~$(i,j)$.}

Since $s$ is in the color set of a deepest well-closed $ij$-chain by definition,
Lemma~\ref{lem:sym_free_chain}(2) gives that the second color of any
$\Bup^G(w)$ with $w$ of color $s$ is $j$. If, in addition, $\ro(\Bup^G(w))=v$,
then $\tilde{c}(\Bup^G(w))=(i,j)$.

\medskip\noindent
\textit{(3) At most one neighbor of~$v$ of color~$s$ does not lie in a child
block of~$v$, that is, $q\le 1$.}

Suppose for contradiction that distinct $w_1,w_2\in W$ both
satisfy $\ro(\Bup^G(w_x))\neq v$ for $x \in\{1,2\}$.  \fpass{As $\depth_G(v)<\depth_G(w_x)$,
the edge $\{v,w_x\}$ lies in $\Bup^G(w_x)$, which is not rooted at~$v$; so
$v\neq r$ and $\Bup^G(w_x)=\Bup^G(v)=:A'$ for $x\in\{1,2\}$.} Thus
$w_1,w_2\in A'$.

Since $A'$ contains two vertices of color~$s$, the Dichotomy
Lemma gives that $A'$ is symmetric and $s$ appears exactly twice.
The vertex~$v$ is adjacent to both~$w_1$ and~$w_2$ inside~$A'$;
by Lemma~\ref{lem:sym_structure}(2) and~(4), only $\ro(A')$ or the unique vertex
in~$\Low(A')$ can be adjacent to both copies of another color.
Since $v\neq\ro(A')$, we must have $v\in\Low(A')$ with
$|\Low(A')|=1$.

Then $\facedepth_{A'}(i)>\facedepth_{A'}(s)$, so by
Lemma~\ref{lem:facedepth_upblock}~(Face-depth ordering), $\Bup^K(s)$ is an ancestor
of or equal to~$\Bup^K(i)$ in~$T_{\widetilde{K}}$.  But
$s\in B_1\setminus\{i\}$ gives $\Bup^K(s)=B_1$, which is a
child of~$i$ in~$T_{\widetilde{K}}$ (since $i=\ro(B_1)$)---a contradiction.

An example for which $q=1$ is obtained by taking for~$v$ one of the vertices of
color~$g$, and for~$W$ the set of its yellow neighbors, of color~$h$, in
Figure~\ref{fig:ex free chain}.  Since $K_{gh}=3$, two of these neighbors lie in
the child block of~$v$ of block color~$(g,i)$, and the third one must lie in the
up-block of~$v$, of block color~$(d,i)$.

\medskip\noindent
\textbf{Conclusion.}
By~(1), each of the $p$ blocks consumes exactly two elements
of~$W$, and by~(2), every element of~$W$ with root~$v$ belongs
to one of these blocks.  Hence $K_{is}=2p+q$.  By~(3),
$q\le 1$, so $0\le K_{is}-2p\le 1$, giving
$p=\lfloor K_{is}/2\rfloor=\lfloor\mathrm{mult}(B_1)/2\rfloor$.
\end{proof}

\subsection{Summary: the Block Count Lemma}\label{subsec:bcl_proof}

We gather the results of this section in the following lemma, which gives, for
every pair of colors, the number of blocks of that block color hanging below a
vertex.

\begin{lemma}[Block Count]\label{lem:block-count}
For a color~$i$, let $F_K(i)$ be the multiset of colors in which each
color~$j$ occurs:
\begin{itemize}
\item \emph{(Case~A):} $\lfloor\mathrm{mult}(B_1)/2\rfloor$ times, if
      there exists a deepest well-closed $ij$-chain, with first
      c-block~$B_1$;
\item \emph{(Case~B):} $\mathrm{mult}(\Bup^K(j))$ times, if no
      well-closed $tk$-chain with $t\neq j$ has~$j$ in its color set,
      $i=\ro(\Bup^K(j))$, and $j$ is the minimum color among the neighbors
      of~$i$ in~$\Bup^K(j)$;
\item $0$ times otherwise.
\end{itemize}
Then for every vertex~$v$ of color~$i$ and every color~$j$, the number
of blocks of block color~$(i,j)$ with block root~$v$ is the multiplicity
of~$j$ in~$F_K(i)$.
\end{lemma}

\begin{remark}\label{rem:f_cases}
Cases~A and~B are mutually exclusive: Case~A requires a well-closed
$ij$-chain (whose color set contains~$j$), while Case~B requires
that no well-closed chain contains~$j$.
\end{remark}

\begin{proof}[Proof of the Block Count Lemma
(Lemma~\ref{lem:block-count})]
Fix colors $i,j$ and a vertex~$v$ of color~$i$.

\medskip\noindent
\textbf{Case~A.}
Suppose the conditions of Case~A hold.  By
Lemma~\ref{lem:counting_sym}~(Counting symmetric blocks), the
count equals $\lfloor\mathrm{mult}(B_1)/2\rfloor$.

\medskip\noindent
\textbf{Case~B.}
Suppose the conditions of Case~B hold.  By
Lemma~\ref{lem:counting_disc}~(Counting discrete blocks), the
count equals $\mathrm{mult}(\Bup^K(j))$.

\medskip\noindent
\textbf{Otherwise.}
Suppose neither Case~A nor Case~B holds.  If a
block~$A$ of color~$(i,j)$ with root~$v$ existed, it would be
symmetric or discrete (Dichotomy Lemma).
Let us prove that if $A$ is symmetric then Case~A holds and if
$A$ is discrete then Case~B holds, giving a contradiction.

\emph{If $A$ is symmetric:}
Corollary~\ref{cor:block_type_conditions}~(Block type determines
appearance in well-closed chain) gives a deepest well-closed $ij$-chain, which is the
condition of Case~A.

\emph{If $A$ is discrete:}
Corollary~\ref{cor:block_type_conditions} gives that no well-closed
$tk$-chain with $t\neq j$ has~$j$ in its color set.  By Lemma~\ref{lem:asym_iso}~(Discrete block isomorphism), applied to the
vertex of color~$j$ in~$A$, whose up-block is~$A$, the block $A$ is isomorphic
to~$\Bup^K(j)$, so $i=\ro(\Bup^K(j))$.  By
definition of~$\tilde{c}$ for discrete blocks, $j$ is the
minimum neighbor of~$i$ in~$\Bup^K(j)$.  These are the
conditions of Case~B.

Hence no block of color~$(i,j)$ with root~$v$ exists if neither Case~A nor
Case~B holds.
\end{proof}

\begin{corollary}[Existence of a symmetric block]\label{cor:exists_sym_block}
Let $i,j$ be two colors.  A symmetric block of block color~$(i,j)$ exists
in~$G$ if and only if there is a deepest well-closed $ij$-chain
whose multiplicity is at least~$2$.
\end{corollary}

\begin{proof}
If there is a deepest well-closed $ij$-chain with first c-block~$B_1$ and
$\mathrm{mult}(B_1)\ge2$, then by Case~A of Lemma~\ref{lem:block-count}~(Block
Count), every vertex of color~$i$ is the block root of
$\lfloor\mathrm{mult}(B_1)/2\rfloor\ge1$ blocks of block color~$(i,j)$, which
are symmetric as $j$ is closing (Corollary~\ref{cor:block_type}, Block type
determination).  Conversely, if $A$ is a symmetric block of block
color~$(i,j)$, Corollary~\ref{cor:block_type_conditions}~(Block type
determines appearance in well-closed chain) gives a deepest well-closed $ij$-chain, with
first c-block~$B_1$ say, and by Case~A, the block root of~$A$ is the block root
of exactly $\lfloor\mathrm{mult}(B_1)/2\rfloor$ blocks of block
color~$(i,j)$, one of which is~$A$; so $\mathrm{mult}(B_1)\ge2$.
\end{proof}

\subsection{The dichotomy for c-blocks}\label{subsec:families}

We can now prove the dichotomy for c-blocks stated in
Section~\ref{sec:details}.

\cblockdichotomy*

\begin{proof}
Let $B$ be a c-block with block root~$i$.
Suppose first that $B$ belongs to some well-closed chain.  By
Remark~\ref{rem:free_ext}, this chain is contained in a deepest well-closed chain, and
removing the c-blocks preceding~$B$ in the latter leaves a deepest well-closed chain
whose first c-block is~$B$, hence whose upper endpoint is $i=\ro(B)$.  Let $j$
be its lower endpoint; it is unique by Lemma~\ref{lem:one_family}~(Common lower
endpoint).  By Case~A of the Block Count Lemma~\ref{lem:block-count}, every
vertex of color~$i$ is the block root of exactly
$\lfloor\mathrm{mult}(B)/2\rfloor$ blocks of block color~$(i,j)$, and by the
Block Structure Lemma~\ref{lem:block-structure} each of them is
color-preserving isomorphic to $\B_K(i,j)=\unfold(i,j,K)$, as $j$ is
closing.
 
Suppose now that $B$ belongs to no well-closed chain, and let
$j=\min(V(B)\cap N_{\Kt}(i))$.  Then $B=\Bup^K(j)$ and $i=\ro(\Bup^K(j))$.  No
well-closed $tk$-chain with $t\neq j$ has $j$ in its color set: $j$ would lie in some
c-block~$B_a$ of such a chain, and either $B_a=\Bup^K(j)=B$, or
$j=\ro(B_a)=c_{a-1}$ with $a\ge2$ and $B_{a-1}=\Bup^K(j)=B$
(Lemma~\ref{lem:chain_unique}, Chains in~$T_{\widetilde{K}}$); either way $B$ would belong
to a well-closed chain.  The conditions of Case~B of the Block Count Lemma are
therefore satisfied: every vertex of color~$i$ is the block root of exactly
$\mathrm{mult}(B)$ blocks of block color~$(i,j)$.  The color~$j$ is not
closing, since otherwise $B$, which is an $ij$-chain with underlying path
$(i,j)$, would be a well-closed chain; so these blocks are color-preserving isomorphic
to $\B_K(i,j)=B$ by the Block Structure Lemma.
 
The two cases are mutually exclusive and exhaustive.
\end{proof}

\section{Putting it all together: the inversion algorithm}\label{app:inversion}

\subsection{Phase 1: the bottom-up scan}\label{app:phase1}


Phase~1 computes the multisets $F_K(i)$ of Lemma~\ref{lem:block-count}.
Evaluating the conditions of the Block Count Lemma separately for each
pair of colors would be wasteful: deciding whether a deepest well-closed
$ij$-chain exists requires the whole descent below~$j$, and the same c-block
may belong to several chains.  We therefore compute all the chains at once, by
a single traversal of~$T_{\widetilde{K}}$ starting from its leaves.

By the dichotomy for c-blocks (Lemma~\ref{lem:c-block-dichotomy}), it
suffices to decide, for each c-block~$B$ with block root~$i$, whether $B$ is
the first c-block of a deepest well-closed chain, and if so, to find the lower
endpoint~$j$ of this chain, which is unique (Lemma~\ref{lem:one_family}, Common
lower endpoint).  If $B$ is such a first c-block, Case~A of the Block Count
Lemma gives $\lfloor\mathrm{mult}(B)/2\rfloor$ blocks of block color~$(i,j)$
below each vertex of color~$i$.  Otherwise, $B$ belongs to no well-closed
chain (Lemma~\ref{lem:c-block-dichotomy}), and Case~B gives $\mathrm{mult}(B)$ blocks of block color~$(i,j_0)$,
where $j_0=\min(V(B)\cap N_{\Kt}(i))$.  To find~$j$, note that a deepest
well-closed $ij$-chain with first c-block~$B$ is either $B$ alone, $j$ being
then a closing color of $V(B)\cap N_{\Kt}(i)$ as the lower endpoint of this
well-closed chain, or $B$ followed by a deepest well-closed chain starting at
a color $t\in V(B)\cap N_{\Kt}(i)$, whose first c-block~$B'$ is a child c-block
of~$t$ in~$T_{\widetilde{K}}$ and is not opening.

The algorithm processes the c-blocks of~$\widetilde{K}$ starting from the deepest ones, moving towards the root. 
When it processes a
c-block~$B$ with block root $i=\ro(B)$, the algorithm does two things.  It
adds a color to~$F_K(i)$, and it fills in $\texttt{end}(B)$, which is what
the c-blocks above~$B$ will read. The array \texttt{end} stores a single piece of information
per c-block~$B$: the lower endpoint of the deepest well-closed chain
with first c-block~$B$, if there is one and $B$ is not opening (so that this
chain can be continued above~$B$) and~$\bot$ otherwise.  

The algorithm first reads the entries $\texttt{end}(B')$ of the child
c-blocks~$B'$ of the colors of $V(B)\cap N_{\Kt}(i)$ (Line~\ref{line:below}); if
all of them are~$\bot$, it looks for a closing color in $V(B)\cap N_{\Kt}(i)$.  If it finds~$j$ in one of these two ways, it
adds $j$ to $F_K(i)$ with multiplicity described by Case~A, and sets $\texttt{end}(B)$ to~$j$ exactly
when $\mathrm{mult}(B)$ is odd, that is, when $B$ is not opening
(Definition~\ref{def:opening}).  Otherwise, it adds $j$ to $F_K(i)$ with multiplicity described by
Case~B, and $\texttt{end}(B)$ remains~$\bot$.  


\begin{algorithm}[t]
\LinesNumbered
\caption{Phase~1: computing the multisets $F_K$}
\label{alg:phase1}
\KwIn{a color graph~$K$ with root color $1$}
\KwOut{the multisets $F_K(i)$ for every color~$i$}
initialize $F_K(i)\leftarrow\emptyset$ for every color~$i$, and
  $\texttt{end}(B)\leftarrow\bot$ for every c-block~$B$\;
\ForEach{c-block $B$ of $\widetilde{K}$, by decreasing depth in
  $T_{\widetilde{K}}$}{
  $i\leftarrow\ro(B)$; $j\leftarrow\bot$\;
  \ForEach{color $t\in V(B)\cap N_{\Kt}(i)$}{
    \ForEach{c-block $B'$ child of $t$ in $T_{\widetilde{K}}$}{
      \lIf*{$\texttt{end}(B')\neq\bot$}{$j\leftarrow\texttt{end}(B')$%
        \label{line:below}}
      \Comment*[r]{at most once, by Remark~\ref{rem:phase1_unique}}
    }
  }
  \lIf{$j=\bot$ \textnormal{and some} $t\in V(B)\cap N_{\Kt}(i)$
       \textnormal{is closing}}{$j\leftarrow t$\label{line:closing}}
  \eIf{$j\neq\bot$}{
    \tcp{symmetric block case}
    add $j$ to $F_K(i)$ with multiplicity
      $\lfloor\mathrm{mult}(B)/2\rfloor$\;
    \lIf*{$\mathrm{mult}(B)$ is odd}{$\texttt{end}(B)\leftarrow j$}
    \Comment*[r]{$B$ is not opening}
  }{
    \tcp{discrete block case}
    $j_0\leftarrow\min\bigl(V(B)\cap N_{\Kt}(i)\bigr)$\;
    add $j_0$ to $F_K(i)$ with multiplicity $\mathrm{mult}(B)$\;
  }
}
\Return{$(F_K(i))_i$}
\end{algorithm}


\textbf{Running time.}
Each c-block is processed once, and the work done at a c-block~$B$ is linear in
$|V(B)|$, every test ranging over the
colors of~$B$, and so does the computation of $\mathrm{mult}(B)$, once
$T_{\widetilde{K}}$, the c-blocks $\Bup^K(s)$ and the depths of the colors have
been computed, together with the closing colors, which take linear time
by Definition~\ref{def:closing}.  Every c-block is read by
exactly one parent, so the second term contributes once per c-block over the
whole run; and every edge of~$\widetilde{K}$ lies in exactly one c-block, so
$\sum_B|V(B)|$ is linear in the size of~$K$.  The total running time is
therefore linear in the size of~$K$.

\subsection*{Correctness}\label{subsec:phase1_correctness}

\begin{lemma}[Invariant of the scan]\label{lem:phase1_invariant}
Let $B$ be a c-block of~$\widetilde{K}$ and $i=\ro(B)$.  When
Algorithm~\ref{alg:phase1} processes~$B$, the variable~$j$ takes a value
different from~$\bot$ if and only if $B$ is the first c-block of a
deepest well-closed chain, and $j$ is then the lower endpoint of this chain.
Moreover, $\texttt{end}(B)\neq\bot$ at the end of that iteration if and only if
such a chain exists and $B$ is not opening.
\end{lemma}

\begin{proof}
We argue by induction on the processing order, that is, on the c-blocks
of~$T_{\widetilde{K}}$ from the deepest to the shallowest; the induction
hypothesis is the whole statement of the lemma, for every c-block processed
before~$B$.  Throughout, $B'$ denotes a c-block child of a color
$t\in V(B)\cap N_{\Kt}(i)$ in~$T_{\widetilde{K}}$; these are exactly the c-blocks
whose entries of \texttt{end} are read when processing~$B$.  Steps~1
and~2 show that every value given to~$j$ is the lower endpoint of a deepest
well-closed chain with first c-block~$B$, and Step~3 that $j$ is given a value
whenever such a chain exists.  As the lower endpoint of such a chain is unique
(Lemma~\ref{lem:one_family}, Common lower endpoint), this proves the first
assertion.  Step~4 deduces the second assertion from the first.

\medskip
\textbf{Step 1: if $j$ takes the value of $\texttt{end}(B')$ (Line~\ref{line:below}), then it is the
lower endpoint of a deepest well-closed chain with first c-block~$B$.}
Suppose $j$ is set to $\texttt{end}(B')$, where $B'$ is a child c-block of the
color~$t$.  As $\texttt{end}(B')$ was set to the value of the
variable~$j$ when $B'$ was processed, the induction hypothesis applied to~$B'$
shows that $B'$ is the first c-block of a deepest well-closed chain~$C'$ with
lower endpoint~$j$ and underlying path $(t=c_1,c_2,\dots,c_\ell=j)$, and $B'$
is not opening.  We show that $B\cup C'$ is a deepest well-closed
$ij$-chain with first c-block~$B$, with underlying path
$(i,t,c_2,\dots,c_\ell)$.  It is an $ij$-chain, as $B$ has block
root~$i$ and contains the edge~$it$, and $C'$ satisfies the other conditions
of Definition~\ref{def:ij-chain}.  It is well-closed: $j$ is closing; its c-blocks other than the first are $B'$ and the c-blocks of~$C'$ other than its first, all of which are non-opening; and its intermediate colors
are~$t$, which is not solo closing by Claim~\ref{cl:t_not_solo} below, and
those of~$C'$, which are not solo closing as $C'$ is well-closed.
It is deepest: suppose that a well-closed $ik$-chain with $k\neq j$
has~$j$ on its underlying path.  By Lemma~\ref{lem:chain_unique}(3)~(Chains
in~$T_{\widetilde{K}}$), its c-blocks up to~$j$ form an $ij$-chain, which is
$B\cup C'$ by uniqueness; so its first c-block is~$B$, and removing it leaves
a well-closed $tk$-chain (Remark~\ref{rem:free_ext}) with~$j$ on its
underlying path, contradicting that~$C'$ is deepest.

\begin{claim}\label{cl:t_not_solo}
The color~$t$ is not solo closing.
\end{claim}

\begin{proof}[Proof of the claim]
Suppose it is, and let $v$ be a vertex of color~$t$, with $A=\Bup^G(v)$.  By
Lemma~\ref{lem:bottom_symmetric}(1)~(Closing colors and symmetric blocks), $A$
is symmetric and $\Low(A)=\{v\}$, so $c(\Low(A))=\{t\}$.

\fpass{Let $c_2$ be the color following~$t$ on the underlying path of~$C'$.  The
argument of} the inductive step in the proof of
Lemma~\ref{lem:free_descent}(1)~(Well-closed chain \fpass{descent), applied to~$v$
and~$c_2$, gives a neighbor~$w$ of~$v$ with $c(w)=c_2\in V(B')\setminus\{t\}$
and $\Bup^G(w)=\Bup^G(v)=A$: this argument only uses Part~(1) for the colors
of~$C'$, which is well-closed, the fact that $B'=\Bup^K(c_2)$ is not opening,
and the fact that $c_2$ is not solo closing unless $c_2=j$.}  As $c(w)$ belongs to the color set of~$C'$ and
differs from~$t$, Lemma~\ref{lem:sym_free_chain}(2)~(Symmetry and well-closed
chains) applied to~$C'$ gives $c(\Low(A))=\{j\}$.  Hence $t=j$, contradicting
the condition that the two endpoints of the $tj$-chain are distinct.
\end{proof}

\medskip
\textbf{Step 2: if $j$ is set by the closing test (Line~\ref{line:closing}), it is the lower endpoint of
a deepest well-closed chain with first c-block~$B$.}
Here $\texttt{end}(B')$ is $\bot$ for all the child c-blocks $B'$ of $t$ and $j\in V(B)\cap N_{\Kt}(i)$ is
closing.  Then $B$ itself is an
$ij$-chain, with underlying path $(i,j)$, and it is well-closed.  It is deepest: otherwise some well-closed chain starting at~$i$ would have~$j$ as an intermediate color; extending it as long as possible (Remark~\ref{rem:free_ext}), we obtain a deepest well-closed chain in which $j$ is still an intermediate color.
By Lemma~\ref{lem:chain_unique}, its first c-block is~$B$ and its second color
is~$j$, so removing~$B$ leaves a deepest well-closed chain
(Remark~\ref{rem:free_ext}) whose first c-block~$B'$ is a child c-block
of~$j$.  As $B'$ is the second c-block of the longer chain, it is not opening.  The
induction hypothesis applied to~$B'$ then gives $\texttt{end}(B')\neq\bot$,
contradicting our assumption.

\medskip
\textbf{Step 3: if a deepest well-closed chain has $B$ as its first c-block, then
$j\neq\bot$.}
Let $(i=c_0,c_1,\dots,c_\ell=k)$ be the underlying path of such a chain; note
that $c_1$ lies in $V(B)\cap N_{\Kt}(i)$, so the c-blocks below~$c_1$ are  read when processing~$B$.
If $\ell=1$, then $k=c_1$ is a closing color of $V(B)\cap N_{\Kt}(i)$, so
$j$ is set, from below or by the closing test.  If $\ell\ge2$, removing~$B$
leaves a deepest well-closed $c_1k$-chain (Remark~\ref{rem:free_ext}) whose
first c-block~$B_2$ is not opening, by
Definition~\ref{def:free-chain} applied to the original chain.  The induction
hypothesis applied to~$B_2$ gives $\texttt{end}(B_2)\neq\bot$, so $j$ is set
when processing~$B$.

\medskip
\textbf{Step 4: the second assertion.}
The algorithm sets $\texttt{end}(B)$ exactly when $j\neq\bot$, that is, when a
deepest well-closed chain with first c-block~$B$ exists (by the first
assertion of the invariant, proved in Steps~1 to~3), and when $\mathrm{mult}(B)$ is
odd, that is, when $B$ is not opening (Definition~\ref{def:opening}).  This is
the second assertion.
\end{proof}

\begin{remark}\label{rem:phase1_unique}
When the algorithm processes a c-block~$B$ with $i=\ro(B)$, at most one of the
entries $\texttt{end}(B')$ read at Line~\ref{line:below} differs from~$\bot$.
Indeed, by Step~1 of the proof of Lemma~\ref{lem:phase1_invariant}, two such
entries would give two deepest well-closed chains with first c-block~$B$. By
Lemma~\ref{lem:one_family}~(Common lower endpoint), they have the same lower
endpoint, hence are equal (Lemma~\ref{lem:chain_unique}, Chains
in~$T_{\widetilde{K}}$), and the two c-blocks~$B'$ are both its second
c-block.  Likewise, the closing color chosen at Line~\ref{line:closing} is
unique: by Step~2 of the same proof, two such colors would both be lower
endpoints of the deepest well-closed chain consisting of~$B$ alone, hence
equal.
\end{remark}

\begin{proposition}\label{prop:phase1_correct}
Algorithm~\ref{alg:phase1} returns, for every color~$i$, the multiset
$F_K(i)$ of Lemma~\ref{lem:block-count} (Block Count).
\end{proposition}

\begin{proof}
Each c-block~$B$, with $i=\ro(B)$, adds a single color to~$F_K(i)$, with some multiplicity. We first check that the color added by each c-block comes with the
multiplicity prescribed by the Block Count Lemma, then that each color is added
by the right c-block.

If $j\neq\bot$, then by Lemma~\ref{lem:phase1_invariant}~(Invariant of the
scan) there is a deepest well-closed $ij$-chain with first c-block~$B$.  The
conditions of Case~A of the Block Count Lemma~\ref{lem:block-count} are therefore satisfied,
and the recorded multiplicity $\lfloor\mathrm{mult}(B)/2\rfloor$ is exactly the
number of blocks of block color~$(i,j)$ below a vertex of color~$i$.

If $j=\bot$, then by Lemma~\ref{lem:phase1_invariant} no deepest
well-closed chain has first c-block~$B$. So $B$ is in the second case
of the dichotomy for c-blocks (Lemma~\ref{lem:c-block-dichotomy}): it belongs
to no well-closed chain.  As shown in its proof, Case~B of the Block Count
Lemma then applies to~$(i,j_0)$, where $j_0=\min(V(B)\cap N_{\Kt}(i))$, so the
recorded multiplicity $\mathrm{mult}(B)$ is the number of blocks of block
color~$(i,j_0)$ below a vertex of color~$i$.

It remains to see that each color~$j$ is added to~$F_K(i)$ by exactly one c-block when the Block Count Lemma gives it a positive multiplicity, and by none otherwise.  A color is only added to~$F_K(i)$ by c-blocks with block root~$i$.  It is added by at most one of them: a color added in the symmetric case comes from a deepest well-closed $ij$-chain, whose first c-block is unique by Lemma~\ref{lem:chain_unique}~(Chains in~$T_{\widetilde{K}}$); a color added in the discrete case comes from $\Bup^K(j)$; and the two cases cannot both occur, Cases~A and~B being exclusive (Remark~\ref{rem:f_cases}). Furthermore, a c-block only adds a color to which Case~A or Case~B applies, so
it adds no color of multiplicity~$0$ in~$F_K(i)$ (in the case $j\neq\bot$,
when $\lfloor\mathrm{mult}(B)/2\rfloor=0$, nothing is added).  
Conversely, if the multiplicity of~$j$ is positive by Case~A, the first
c-block of the deepest well-closed $ij$-chain adds~$j$, by the case
$j\neq\bot$ just above.  If it is positive by Case~B, the c-block $\Bup^K(j)$ has
block root~$i$, and it is the first c-block of no deepest well-closed chain,
since the color set of such a chain would contain~$j$, contrary to Case~B.  As $j$ is moreover the smallest neighbor of~$i$ in $\Bup^K(j)$, this c-block
adds~$j$, by the case $j=\bot$ just above.  Lemma~\ref{lem:block-count}~(Block
Count) then gives the conclusion.
\end{proof}

\subsection{Phase 2: the top-down construction of $G^*$}\label{app:phase2}


Given a color graph~$K$, Algorithm~\ref{alg:inversion} computes the
graph $(G^*,c^*)=\mathrm{Inv}(K,1)$ of Theorem~\ref{thm:inversion} by starting
from a single root vertex and recursively attaching copies of~$\B_K(i,j)$ for
appropriate $i,j$. The multisets $F_K(i)$, which give the pairs $(i,j)$ and the number of
copies of each $\B_K(i,j)$ to attach, are computed in Phase~1
(Appendix~\ref{app:phase1}).  The blocks $\B_K(i,j)$, defined in the Block
Structure Lemma~\ref{lem:block-structure}, are computed as described in
Section~\ref{subsec:compute_B}.

\begin{algorithm}[H]
\caption{Reconstruction of a rooted outerplanar graph from its color graph}
\label{alg:inversion}
\SetKwFunction{Recon}{Reconstruct}
\KwIn{a color graph~$K$ with root color $1$.}
\KwOut{a rooted colored graph~$(G^*, c^*)$.}

Initialize~$G^*$ as a single vertex~$r^*$ of color~$1$\;
\Recon{$r^*,\,1$}\;

\BlankLine
\SetKwProg{Fn}{Function}{:}{}
\Fn{\Recon{$v,\,i$}}{
  \ForEach{$j \in F_K(i)$ \textnormal{(with multiplicity, in increasing
  order)}}{
    Create a fresh copy~$A$ of the colored block~$\B_K(i,j)$%
      \;\label{ln:copy}
    Identify~$v$ with the root of~$A$; add all remaining vertices
      and edges of~$A$ to~$G^*$ and color those vertices in $c^*$
      with the color they have in~$A$\;\label{ln:attach}
    \ForEach{$u \in V(A)\setminus\{v\}$, in canonical order
      \textnormal{(see below)}}{
      \Recon{$u,\,c^*(u)$}\;\label{ln:recurse}
    }
  }
}
\end{algorithm}

\noindent\textbf{Canonical order.}
Within each block copy~$A$, the non-block-root vertices are traversed along the
Hamiltonian path obtained by removing~$\ro(A)$ from the outer cycle
of~$A$.  For a single-edge block (bridge), there is exactly one non-block-root
vertex and no ordering choice arises.  When~$A$ has three or more vertices, the
path admits two orientations; we start from the neighbor of~$\ro(A)$ with the
smaller color.  If both neighbors of~$\ro(A)$
share the same color, the block is symmetric and both orientations produce
isomorphic results, so the
choice does not affect~$G^*$ up to color-preserving isomorphism. This allows
us to get a canonical representation of the class of graphs that are
color-isomorphic to~$G$.  Processing the entries of~$F_K(i)$ in
increasing order and numbering the vertices of~$G^*$ by their color and the
order in which they are created, the labeled output depends only on~$K$: this
is the canonical labeling of Section~\ref{sec:intuition}.

\medskip
When~$F_K(i) = \emptyset$, the call~$\textsc{Reconstruct}(v,i)$ performs no
action and so~$v$ is not a cut vertex in~$G^*$.  Termination is guaranteed
because~$\depthK$ increases strictly from the block root to every other
vertex within each block copy: for $j\in F_K(i)$, the colors
of~$\B_K(i,j)$ are those of~$\Bup^K(j)$ or of the $ij$-chain, and all of them
except~$i$ lie strictly below~$i$ in~$T_{\widetilde{K}}$
(Remark~\ref{rem:K_rooted}); and~$\depthK$ is bounded by the number of colors.

\subsection*{Correctness}\label{subsec:correctness}

We show that Algorithm~\ref{alg:inversion} run with the multisets~$F_K$ computed in
Phase~1 (Proposition~\ref{prop:phase1_correct}), and with the
blocks~$\B_K(i,j)$ computed as in Section~\ref{subsec:compute_B} satisfies the conditions of Theorem~\ref{thm:inversion}.

\begin{proof}[Proof of Theorem~\ref{thm:inversion}]
Let $(G^*,c^*)$ be the output of Algorithm~\ref{alg:inversion} on the color graph $K$ with root $1$.
We construct a color-preserving isomorphism
$\pi\colon V(G)\to V(G^*)$ by induction on~$\depthK$, processing vertices
from the root downward.

\emph{Root.} Set $\pi(r)=r^*$. Both have color~$1$.

\emph{Inductive step.} Suppose $\pi$ has been defined on a vertex~$v$ of
color~$i$, with $\pi(v)=v^*$. We show that $\pi$ extends consistently to all
vertices ``below''~$v$ --- that is, to all blocks of~$G$ having~$v$ as their
block root, and recursively to all their descendants.

By the Block Count Lemma~\ref{lem:block-count}, the sorted multiset of
second colors of child blocks of~$v$ in~$G$ equals~$F_K(i)$. By construction,
Algorithm~\ref{alg:inversion} creates, at~$v^*$, child blocks with second
colors forming the same multiset~$F_K(i)$. We pair child blocks of~$v$
and~$v^*$ by corresponding entries of~$F_K(i)$. When several child blocks
share the same block color~$(i,j)$, each child block $A$ is
color-isomorphic to~$\B_K(i,j)$ by the Block Structure
Lemma~\ref{lem:block-structure}, so a color-preserving isomorphism
from $A$ to the corresponding copy of $\B_K(i,j)$ allows to produce a
valid extension of~$\pi$ as follows.
Consider a matched pair: a block~$A$ of~$G$ with
$\tilde{c}(A)=(i,j)$ and the corresponding copy~$A^*$
of~$\B_K(i,j)$ in~$G^*$. By the Block Structure Lemma, $A$ is
color-preserving isomorphic to~$\B_K(i,j)\iso A^*$; let
$\phi\colon V(A)\to V(A^*)$ be such an isomorphism. Since $v$
is the unique vertex of color~$i$ in~$A$ (Lemmas~\ref{lem:highest},
Block root, and~\ref{lem:depth}, Depth), we have $\phi(v)=v^*=\pi(v)$, so
$\phi$ is compatible with~$\pi$ on the already-mapped vertex. Set
$\pi(u)=\phi(u)$ for every $u\in V(A)\setminus\{v\}$. $\pi$ is then a
color-preserving isomorphism from $A$ to $A^*$.
For each vertex $u\in V(A)\setminus\{v\}$, we then recurse on~$u$
and~$\pi(u)$; this is valid since $\depthK(c(u))>\depthK(i)$
(Lemma~\ref{lem:highest}, Block root). This gives a well-defined function $\pi$: every non-root vertex of~$G$ belongs to exactly one
up-block (Lemma~\ref{lem:up_block}, Unique up-block), so $\pi(u)$ is assigned
exactly once.
We conclude by showing that $\pi$ has the following properties.

\emph{Injectivity.} Suppose $u\neq u'$ are non-root vertices of~$G$. If
they belong to the same up-block~$A$, then $\pi(u)\neq\pi(u')$
because~$\phi$ is injective on~$V(A)$. If they belong to distinct
up-blocks, their images lie in~$V(A^*_1)\setminus\{\ro(A^*_1)\}$
and~$V(A^*_2)\setminus\{\ro(A^*_2)\}$ for distinct
blocks~$A^*_1,A^*_2$ of~$G^*$ by definition of $\pi$. These
sets are disjoint: if there exists a vertex
$x \in V(A^*_1) \cap V(A^*_2)$, then $x$ must be the block root
of at least one of those two blocks by construction, hence
$\big(V(A^*_1)\setminus\{\ro(A^*_1)\}\big)\cap\big(V(A^*_2)\setminus\{\ro(A^*_2)\}\big)=\emptyset$.
Then $\pi(u) \neq \pi(u')$.

\emph{Surjectivity.} Every vertex of~$G^*$ is either the root (in the image
by construction) or belongs to $V(A^*)\setminus\{\ro(A^*)\}$
for some block~$A^*$ created by the algorithm. By induction, each such
block is matched with a block of~$G$, and $\phi$ is a bijection, so every
vertex of~$G^*$ is in the image of~$\pi$.

\emph{Edges.} Every edge of~$G$ lies in a unique block~$A$. The
restriction of~$\pi$ to~$V(A)$ coincides with~$\phi$ (for
$V(A)\setminus\{v\}$ by definition, and for~$v$ since
$\phi(v)=v^*=\pi(v)$), and
$\phi$ preserves edges. Conversely, every edge of~$G^*$ lies in a
block~$A^*$ matched with some block~$A$ of~$G$,
and~$\phi^{-1}$ maps it to an edge of~$A$. Hence~$\pi$ preserves and
reflects edges.

\emph{Color preservation.} The root is mapped to a vertex of the same color,
and each~$\phi$ is color-preserving by Lemma~\ref{lem:block-structure},
so~$\pi$ preserves colors.

\medskip
Thus~$\pi$ is a color-preserving graph isomorphism from~$(G,c)$
to~$(G^*,c^*)$.
\end{proof}

\subsection*{Time complexity}

Phase~1 runs in time linear in the size of~$K$
(Appendix~\ref{app:phase1}). Sorting the multisets $F_K(i)$, as Algorithm~\ref{alg:inversion} requires,
takes time linear in the size of~$K$, by a single bucket sort of all the
pairs $(i,j)$ at once. In Phase~2, each call
$\textsc{Reconstruct}(v,i)$ creates the copies of $\B_K(i,j)$ for
$j\in F_K(i)$, each in time linear in its size (Section~\ref{subsec:compute_B}).
The blocks created are pairwise edge-disjoint and share only their block roots,
so their total size is linear in the size of the output~$G^*$.  Phase~2
therefore runs in time linear in the size of~$G^*$, once the c-block
structure of~$K$ has been computed, which takes time linear in the size of~$K$.
Finally, the size of~$K$ is at most that of~$G^*$: as $(G^*,c^*)$ is
isomorphic to~$(G,c)$, $K$ is also its color graph, so every color class
of~$c^*$ is non-empty, and every non-zero entry~$K_{st}$ is witnessed by an
edge of~$G^*$ between colors $s$ and~$t$.  The whole inversion therefore runs
in time linear in the size of its output, as stated in
Theorem~\ref{thm:inversion}.

\begin{remark}
Given $(G,c)$ and the output $(G^*,c^*)$ of Algorithm~\ref{alg:inversion},
the isomorphism~$\pi$ of the proof can be computed in linear time, by the
same top-down traversal of~$T_G$.  The block-cut tree~$T_G$ and the outer
faces of the blocks can be computed in linear time~\cite{HopcroftTarjan73, HopcroftTarjan74},
and the isomorphism~$\phi$ between a block~$A$ and its copy~$A^*$ is obtained
by following their outer faces in parallel from their roots.  If $A$ is
discrete, only one orientation of the outer face of~$A^*$ matches the colors
of~$A$; if $A$ is symmetric, both orientations give an isomorphism.
Consequently, $\pi$ gives the canonical labeling of~$G$ of
Section~\ref{sec:intuition}, and for two connected rooted outerplanar graphs
that CR does not distinguish, composing the isomorphism of one to~$G^*$ with
the inverse of the other's gives an isomorphism between them, in linear time
once CR has been run.
\end{remark}
\section{Statement of the Dichotomy Lemma proven in Rocq}\label{app:rocq-dicho-lem}

\coqlibrary{DichoLemma.Check}{Library }{DichoLemma.Check}

\begin{coqdoccode}
\coqdocemptyline
\coqdocnoindent
\coqdockw{From} \coqdocvar{mathcomp} \coqdockw{Require} \coqdockw{Import} \coqdoclibrary{all\_boot}.\coqdoceol
\coqdocnoindent
\coqdockw{From} \coqdocvar{GraphTheory} \coqdockw{Require} \coqdockw{Import} \coqdoclibrary{sgraph} \coqdoclibrary{minor} \coqdoclibrary{treewidth} \coqdoclibrary{connectivity} \coqdoclibrary{excluded}.\coqdoceol
\coqdocemptyline
\coqdocnoindent
\coqdockw{Set Implicit Arguments}.\coqdoceol
\coqdocnoindent
\coqdockw{Unset} \coqdockw{Strict} \coqdockw{Implicit}.\coqdoceol
\coqdocnoindent
\coqdockw{Unset} \coqdockw{Printing} \coqdockw{Implicit} \coqdocvar{Defensive}.\coqdoceol
\coqdocemptyline
\coqdocnoindent
\coqdockw{Module} \coqdef{DichoLemma.Check.Spec}{Spec}{\coqdocmodule{Spec}}.\coqdoceol
\coqdocemptyline
\coqdocnoindent
\coqdockw{Definition} \coqdef{DichoLemma.Check.Spec.hamiltonian cycle}{hamiltonian\_cycle}{\coqdocdefinition{hamiltonian\_cycle}} (\coqdef{DichoLemma.Check.H:1}{H}{\coqdocbinder{H}} : \coqdocrecord{sgraph}) (\coqdef{DichoLemma.Check.p:2}{p}{\coqdocbinder{p}} : \coqdocabbreviation{seq} (\coqdocprojection{svertex} \coqref{DichoLemma.Check.H:1}{\coqdocvariable{H}})) := \coqdoceol
\coqdocindent{2.00em}
\coqdocdefinition{ucycle} \coqdocprojection{sedge} \coqref{DichoLemma.Check.p:2}{\coqdocvariable{p}} \coqexternalref{::type scope:x '/x5C' x}{http://rocq-prover.org/doc/V9.1.1/stdlib//Corelib.Init.Logic}{\coqdocnotation{\ensuremath{\land}}} \coqdocnotation{\#|}\coqref{DichoLemma.Check.p:2}{\coqdocvariable{p}}\coqdocnotation{\ensuremath{|}} \coqexternalref{::type scope:x '=' x}{http://rocq-prover.org/doc/V9.1.1/stdlib//Corelib.Init.Logic}{\coqdocnotation{=}} \coqdocnotation{\#|}\coqdocprojection{svertex} \coqref{DichoLemma.Check.H:1}{\coqdocvariable{H}}\coqdocnotation{\ensuremath{|}}.\coqdoceol
\coqdocemptyline
\coqdocnoindent
\coqdockw{Section} \coqdef{DichoLemma.Check.Spec.Outerplanar}{Outerplanar}{\coqdocsection{Outerplanar}}.\coqdoceol
\coqdocemptyline
\coqdocnoindent
\coqdockw{Variable} \coqdef{DichoLemma.Check.Spec.Outerplanar.G}{G}{\coqdocvariable{G}} : \coqdocrecord{sgraph}.\coqdoceol
\coqdocnoindent
\coqdockw{Implicit} \coqdockw{Types} \coqdocvar{u} \coqdocvar{v} \coqdocvar{x} \coqdocvar{y} : (\coqdocprojection{svertex} \coqref{DichoLemma.Check.Spec.Outerplanar.G}{\coqdocvariable{G}}).\coqdoceol
\coqdocemptyline
\coqdocnoindent
\coqdockw{Definition} \coqdef{DichoLemma.Check.Spec.K4 free}{K4\_free}{\coqdocdefinition{K4\_free}} := \coqexternalref{::type scope:'x7E' x}{http://rocq-prover.org/doc/V9.1.1/stdlib//Corelib.Init.Logic}{\coqdocnotation{\ensuremath{\lnot}}} \coqdocdefinition{minor} \coqref{DichoLemma.Check.Spec.Outerplanar.G}{\coqdocvariable{G}} \coqdocnotation{'}\coqdocnotation{K\_4}.\coqdoceol
\coqdocemptyline
\coqdocnoindent
\coqdockw{Definition} \coqdef{DichoLemma.Check.Spec.K23 free}{K23\_free}{\coqdocdefinition{K23\_free}} := \coqexternalref{::type scope:'x7E' x}{http://rocq-prover.org/doc/V9.1.1/stdlib//Corelib.Init.Logic}{\coqdocnotation{\ensuremath{\lnot}}} \coqdocdefinition{minor} \coqref{DichoLemma.Check.Spec.Outerplanar.G}{\coqdocvariable{G}} \coqdocnotation{'}\coqdocnotation{K\_2}\coqdocnotation{,}3.\coqdoceol
\coqdocemptyline
\coqdocnoindent
\coqdockw{Definition} \coqdef{DichoLemma.Check.Spec.outerplanar}{outerplanar}{\coqdocdefinition{outerplanar}} := \coqref{DichoLemma.Check.Spec.K4 free}{\coqdocdefinition{K4\_free}} \coqexternalref{::type scope:x '/x5C' x}{http://rocq-prover.org/doc/V9.1.1/stdlib//Corelib.Init.Logic}{\coqdocnotation{\ensuremath{\land}}} \coqref{DichoLemma.Check.Spec.K23 free}{\coqdocdefinition{K23\_free}}.\coqdoceol
\coqdocemptyline
\coqdocnoindent
\coqdockw{Definition} \coqdef{DichoLemma.Check.Spec.outerplanar hamiltonian stmt}{outerplanar\_hamiltonian\_stmt}{\coqdocdefinition{outerplanar\_hamiltonian\_stmt}} := \coqdoceol
\coqdocindent{2.00em}
\coqref{DichoLemma.Check.Spec.outerplanar}{\coqdocdefinition{outerplanar}}  \coqexternalref{::type scope:x '->' x}{http://rocq-prover.org/doc/V9.1.1/stdlib//Corelib.Init.Logic}{\coqdocnotation{\ensuremath{\rightarrow}}} 2\coqdocnotation{.-}\coqdocnotation{connected} \coqdocnotation{G}\coqdoceol
\coqdocindent{2.00em}
\coqexternalref{::type scope:x '->' x}{http://rocq-prover.org/doc/V9.1.1/stdlib//Corelib.Init.Logic}{\coqdocnotation{\ensuremath{\rightarrow}}} \coqexternalref{::type scope:'exists' x '..' x ',' x}{http://rocq-prover.org/doc/V9.1.1/stdlib//Corelib.Init.Logic}{\coqdocnotation{\ensuremath{\exists}}} \coqdef{DichoLemma.Check.p:5}{p}{\coqdocbinder{p}} : \coqdocabbreviation{seq} (\coqdocprojection{svertex} \coqref{DichoLemma.Check.Spec.Outerplanar.G}{\coqdocvariable{G}})\coqexternalref{::type scope:'exists' x '..' x ',' x}{http://rocq-prover.org/doc/V9.1.1/stdlib//Corelib.Init.Logic}{\coqdocnotation{,}} @\coqref{DichoLemma.Check.Spec.hamiltonian cycle}{\coqdocdefinition{hamiltonian\_cycle}} \coqref{DichoLemma.Check.Spec.Outerplanar.G}{\coqdocvariable{G}} \coqref{DichoLemma.Check.p:5}{\coqdocvariable{p}}.\coqdoceol
\coqdocemptyline
\coqdocnoindent
\coqdockw{Definition} \coqdef{DichoLemma.Check.Spec.equitable sets}{equitable\_sets}{\coqdocdefinition{equitable\_sets}} (\coqdef{DichoLemma.Check.A:6}{A}{\coqdocbinder{A}} \coqdef{DichoLemma.Check.B:7}{B}{\coqdocbinder{B}} : \coqdocnotation{\{}\coqdocnotation{set} \coqdocnotation{(}\coqdocprojection{svertex} \coqref{DichoLemma.Check.Spec.Outerplanar.G}{\coqdocvariable{G}}\coqdocnotation{)\}}) := \coqdoceol
\coqdocindent{2.00em}
\coqdocnotation{[}\coqdocnotation{\ensuremath{\forall}} \coqdef{DichoLemma.Check.u:8}{u}{\coqdocbinder{u}} \coqdocnotation{in} \coqref{DichoLemma.Check.A:6}{\coqdocvariable{A}}\coqdocnotation{,} \coqdocnotation{[}\coqdocnotation{\ensuremath{\forall}} \coqdef{DichoLemma.Check.v:11}{v}{\coqdocbinder{v}} \coqdocnotation{in} \coqref{DichoLemma.Check.A:6}{\coqdocvariable{A}}\coqdocnotation{,}\coqdoceol
\coqdocindent{2.00em}
\coqdocnotation{\#|}\coqexternalref{::function scope:'[' 'pred' x '|' x ']'}{http://rocq-prover.org/doc/V9.1.1/stdlib//Corelib.ssr.ssrbool}{\coqdocnotation{[}}\coqexternalref{::function scope:'[' 'pred' x '|' x ']'}{http://rocq-prover.org/doc/V9.1.1/stdlib//Corelib.ssr.ssrbool}{\coqdocnotation{pred}} \coqdef{DichoLemma.Check.x:14}{x}{\coqdocbinder{x}} \coqexternalref{::function scope:'[' 'pred' x '|' x ']'}{http://rocq-prover.org/doc/V9.1.1/stdlib//Corelib.ssr.ssrbool}{\coqdocnotation{\ensuremath{|}}} \coqexternalref{::bool scope:x 'x26x26' x}{http://rocq-prover.org/doc/V9.1.1/stdlib//Corelib.Init.Datatypes}{\coqdocnotation{(}}\coqref{DichoLemma.Check.x:14}{\coqdocvariable{x}} \coqexternalref{::bool scope:x 'x5Cin' x}{http://rocq-prover.org/doc/V9.1.1/stdlib//Corelib.ssr.ssrbool}{\coqdocnotation{\symbol{92}}}\coqexternalref{::bool scope:x 'x5Cin' x}{http://rocq-prover.org/doc/V9.1.1/stdlib//Corelib.ssr.ssrbool}{\coqdocnotation{in}} \coqref{DichoLemma.Check.B:7}{\coqdocvariable{B}}\coqexternalref{::bool scope:x 'x26x26' x}{http://rocq-prover.org/doc/V9.1.1/stdlib//Corelib.Init.Datatypes}{\coqdocnotation{)}} \coqexternalref{::bool scope:x 'x26x26' x}{http://rocq-prover.org/doc/V9.1.1/stdlib//Corelib.Init.Datatypes}{\coqdocnotation{\&\&}} \coqdocprojection{sedge} \coqref{DichoLemma.Check.u:8}{\coqdocvariable{u}} \coqref{DichoLemma.Check.x:14}{\coqdocvariable{x}}\coqexternalref{::function scope:'[' 'pred' x '|' x ']'}{http://rocq-prover.org/doc/V9.1.1/stdlib//Corelib.ssr.ssrbool}{\coqdocnotation{]}}\coqdocnotation{\ensuremath{|}} \coqdocnotation{==} \coqdocnotation{\#|}\coqexternalref{::function scope:'[' 'pred' x '|' x ']'}{http://rocq-prover.org/doc/V9.1.1/stdlib//Corelib.ssr.ssrbool}{\coqdocnotation{[}}\coqexternalref{::function scope:'[' 'pred' x '|' x ']'}{http://rocq-prover.org/doc/V9.1.1/stdlib//Corelib.ssr.ssrbool}{\coqdocnotation{pred}} \coqdef{DichoLemma.Check.y:15}{y}{\coqdocbinder{y}} \coqexternalref{::function scope:'[' 'pred' x '|' x ']'}{http://rocq-prover.org/doc/V9.1.1/stdlib//Corelib.ssr.ssrbool}{\coqdocnotation{\ensuremath{|}}} \coqexternalref{::bool scope:x 'x26x26' x}{http://rocq-prover.org/doc/V9.1.1/stdlib//Corelib.Init.Datatypes}{\coqdocnotation{(}}\coqref{DichoLemma.Check.y:15}{\coqdocvariable{y}} \coqexternalref{::bool scope:x 'x5Cin' x}{http://rocq-prover.org/doc/V9.1.1/stdlib//Corelib.ssr.ssrbool}{\coqdocnotation{\symbol{92}}}\coqexternalref{::bool scope:x 'x5Cin' x}{http://rocq-prover.org/doc/V9.1.1/stdlib//Corelib.ssr.ssrbool}{\coqdocnotation{in}} \coqref{DichoLemma.Check.B:7}{\coqdocvariable{B}}\coqexternalref{::bool scope:x 'x26x26' x}{http://rocq-prover.org/doc/V9.1.1/stdlib//Corelib.Init.Datatypes}{\coqdocnotation{)}} \coqexternalref{::bool scope:x 'x26x26' x}{http://rocq-prover.org/doc/V9.1.1/stdlib//Corelib.Init.Datatypes}{\coqdocnotation{\&\&}} \coqdocprojection{sedge} \coqref{DichoLemma.Check.v:11}{\coqdocvariable{v}} \coqref{DichoLemma.Check.y:15}{\coqdocvariable{y}}\coqexternalref{::function scope:'[' 'pred' x '|' x ']'}{http://rocq-prover.org/doc/V9.1.1/stdlib//Corelib.ssr.ssrbool}{\coqdocnotation{]}}\coqdocnotation{\ensuremath{|}}\coqdoceol
\coqdocindent{2.00em}
\coqdocnotation{]]}.\coqdoceol
\coqdocemptyline
\coqdocnoindent
\coqdockw{Definition} \coqdef{DichoLemma.Check.Spec.rooted equitable partition}{rooted\_equitable\_partition}{\coqdocdefinition{rooted\_equitable\_partition}} (\coqdef{DichoLemma.Check.P:20}{P}{\coqdocbinder{P}} : \coqdocnotation{\{}\coqdocnotation{set} \coqdocnotation{\{}\coqdocnotation{set} \coqdocnotation{(}\coqdocprojection{svertex} \coqref{DichoLemma.Check.Spec.Outerplanar.G}{\coqdocvariable{G}}\coqdocnotation{)\}\}}) := \coqdoceol
\coqdocindent{2.00em}
\coqdocdefinition{partition} \coqref{DichoLemma.Check.P:20}{\coqdocvariable{P}} \coqdocnotation{[}\coqdocnotation{set}\coqdocnotation{:} \coqdocprojection{svertex} \coqref{DichoLemma.Check.Spec.Outerplanar.G}{\coqdocvariable{G}}\coqdocnotation{]}\coqdoceol
\coqdocindent{2.00em}
\coqexternalref{::type scope:x '/x5C' x}{http://rocq-prover.org/doc/V9.1.1/stdlib//Corelib.Init.Logic}{\coqdocnotation{\ensuremath{\land}}} \coqdocnotation{[}\coqdocnotation{\ensuremath{\forall}} \coqdef{DichoLemma.Check.A:21}{A}{\coqdocbinder{A}} \coqdocnotation{in} \coqref{DichoLemma.Check.P:20}{\coqdocvariable{P}}\coqdocnotation{,} \coqdocnotation{[}\coqdocnotation{\ensuremath{\forall}} \coqdef{DichoLemma.Check.B:24}{B}{\coqdocbinder{B}} \coqdocnotation{in} \coqref{DichoLemma.Check.P:20}{\coqdocvariable{P}}\coqdocnotation{,} \coqref{DichoLemma.Check.Spec.equitable sets}{\coqdocdefinition{equitable\_sets}} \coqref{DichoLemma.Check.A:21}{\coqdocvariable{A}} \coqref{DichoLemma.Check.B:24}{\coqdocvariable{B}}\coqdocnotation{]]} \coqdoceol
\coqdocindent{2.00em}
\coqexternalref{::type scope:x '/x5C' x}{http://rocq-prover.org/doc/V9.1.1/stdlib//Corelib.Init.Logic}{\coqdocnotation{\ensuremath{\land}}} \coqdocnotation{[}\coqdocnotation{\ensuremath{\exists}} \coqdef{DichoLemma.Check.A:31}{A}{\coqdocbinder{A}} \coqdocnotation{in} \coqref{DichoLemma.Check.P:20}{\coqdocvariable{P}}\coqdocnotation{,} \coqdocnotation{\#|}\coqref{DichoLemma.Check.A:31}{\coqdocvariable{A}}\coqdocnotation{\ensuremath{|}} \coqdocnotation{==} 1\coqdocnotation{]}. \coqdocemptyline
\coqdocnoindent
\coqdockw{Definition} \coqdef{DichoLemma.Check.Spec.is block}{is\_block}{\coqdocdefinition{is\_block}} (\coqdef{DichoLemma.Check.S:36}{S}{\coqdocbinder{S}} : \coqdocnotation{\{}\coqdocnotation{set} \coqdocnotation{(}\coqdocprojection{svertex} \coqref{DichoLemma.Check.Spec.Outerplanar.G}{\coqdocvariable{G}}\coqdocnotation{)\}}) : \coqexternalref{bool}{http://rocq-prover.org/doc/V9.1.1/stdlib//Corelib.Init.Datatypes}{\coqdocinductive{bool}} := \coqdoceol
\coqdocindent{1.00em}
\coqdocdefinition{maxset} (\coqdockw{fun} \coqdef{DichoLemma.Check.S:37}{S}{\coqdocbinder{S}} \ensuremath{\Rightarrow} \coqdocdefinition{kconnectedb} 2 (\coqdocdefinition{induced} \coqref{DichoLemma.Check.S:37}{\coqdocvariable{S}})) \coqref{DichoLemma.Check.S:36}{\coqdocvariable{S}}.\coqdoceol
\coqdocemptyline
\coqdocnoindent
\coqdockw{Definition} \coqdef{DichoLemma.Check.Spec.discrete block}{discrete\_block}{\coqdocdefinition{discrete\_block}} (\coqdef{DichoLemma.Check.P:38}{P}{\coqdocbinder{P}} : \coqdocnotation{\{}\coqdocnotation{set} \coqdocnotation{\{}\coqdocnotation{set} \coqdocnotation{(}\coqdocprojection{svertex} \coqref{DichoLemma.Check.Spec.Outerplanar.G}{\coqdocvariable{G}}\coqdocnotation{)\}\}}) (\coqdef{DichoLemma.Check.S:39}{S}{\coqdocbinder{S}} : \coqdocnotation{\{}\coqdocnotation{set} \coqdocnotation{(}\coqdocprojection{svertex} \coqref{DichoLemma.Check.Spec.Outerplanar.G}{\coqdocvariable{G}}\coqdocnotation{)\}}) : \coqexternalref{bool}{http://rocq-prover.org/doc/V9.1.1/stdlib//Corelib.Init.Datatypes}{\coqdocinductive{bool}} :=\coqdoceol
\coqdocindent{1.00em}
\coqdocnotation{[}\coqdocnotation{\ensuremath{\forall}} \coqdef{DichoLemma.Check.A:40}{A}{\coqdocbinder{A}} \coqdocnotation{in} \coqref{DichoLemma.Check.P:38}{\coqdocvariable{P}}\coqdocnotation{,} \coqdocnotation{\#|}\coqref{DichoLemma.Check.A:40}{\coqdocvariable{A}} \coqdocnotation{:\&:} \coqref{DichoLemma.Check.S:39}{\coqdocvariable{S}}\coqdocnotation{\ensuremath{|}} \coqdocnotation{\ensuremath{\le}} 1\coqdocnotation{]}.\coqdoceol
\coqdocemptyline
\coqdocnoindent
\coqdockw{Definition} \coqdef{DichoLemma.Check.Spec.symmetric block}{symmetric\_block}{\coqdocdefinition{symmetric\_block}} (\coqdef{DichoLemma.Check.P:45}{P}{\coqdocbinder{P}} : \coqdocnotation{\{}\coqdocnotation{set} \coqdocnotation{\{}\coqdocnotation{set} \coqdocnotation{(}\coqdocprojection{svertex} \coqref{DichoLemma.Check.Spec.Outerplanar.G}{\coqdocvariable{G}}\coqdocnotation{)\}\}}) (\coqdef{DichoLemma.Check.S:46}{S}{\coqdocbinder{S}} : \coqdocnotation{\{}\coqdocnotation{set} \coqdocnotation{(}\coqdocprojection{svertex} \coqref{DichoLemma.Check.Spec.Outerplanar.G}{\coqdocvariable{G}}\coqdocnotation{)\}}) :=\coqdoceol
\coqdocindent{1.00em}
\coqexternalref{::type scope:x '/x5C' x}{http://rocq-prover.org/doc/V9.1.1/stdlib//Corelib.Init.Logic}{\coqdocnotation{(}}\coqdockw{\ensuremath{\forall}} \coqdef{DichoLemma.Check.A:47}{A}{\coqdocbinder{A}} : \coqdocnotation{\{}\coqdocnotation{set} \coqdocnotation{(}\coqdocprojection{svertex} \coqref{DichoLemma.Check.Spec.Outerplanar.G}{\coqdocvariable{G}}\coqdocnotation{)\}}, \coqref{DichoLemma.Check.A:47}{\coqdocvariable{A}} \coqexternalref{::bool scope:x 'x5Cin' x}{http://rocq-prover.org/doc/V9.1.1/stdlib//Corelib.ssr.ssrbool}{\coqdocnotation{\symbol{92}}}\coqexternalref{::bool scope:x 'x5Cin' x}{http://rocq-prover.org/doc/V9.1.1/stdlib//Corelib.ssr.ssrbool}{\coqdocnotation{in}} \coqref{DichoLemma.Check.P:45}{\coqdocvariable{P}} \coqexternalref{::type scope:x '->' x}{http://rocq-prover.org/doc/V9.1.1/stdlib//Corelib.Init.Logic}{\coqdocnotation{\ensuremath{\rightarrow}}} \coqdocnotation{\#|}\coqref{DichoLemma.Check.A:47}{\coqdocvariable{A}} \coqdocnotation{:\&:} \coqref{DichoLemma.Check.S:46}{\coqdocvariable{S}}\coqdocnotation{\ensuremath{|}} \coqdocnotation{\ensuremath{\le}} 2\coqexternalref{::type scope:x '/x5C' x}{http://rocq-prover.org/doc/V9.1.1/stdlib//Corelib.Init.Logic}{\coqdocnotation{)}}\coqdoceol
\coqdocindent{1.00em}
\coqexternalref{::type scope:x '/x5C' x}{http://rocq-prover.org/doc/V9.1.1/stdlib//Corelib.Init.Logic}{\coqdocnotation{\ensuremath{\land}}} \coqdockw{let} \coqdef{DichoLemma.Check.H:48}{H}{\coqdocbinder{H}} := \coqdocdefinition{induced} \coqref{DichoLemma.Check.S:46}{\coqdocvariable{S}} \coqdoctac{in} \coqdoceol
\coqdocindent{2.50em}
\coqexternalref{::type scope:'exists' x '..' x ',' x}{http://rocq-prover.org/doc/V9.1.1/stdlib//Corelib.Init.Logic}{\coqdocnotation{\ensuremath{\exists}}} \coqdef{DichoLemma.Check.r:49}{r}{\coqdocbinder{r}} : \coqdocprojection{svertex} \coqref{DichoLemma.Check.H:48}{\coqdocvariable{H}}\coqexternalref{::type scope:'exists' x '..' x ',' x}{http://rocq-prover.org/doc/V9.1.1/stdlib//Corelib.Init.Logic}{\coqdocnotation{,}} \coqexternalref{::type scope:'exists' x '..' x ',' x}{http://rocq-prover.org/doc/V9.1.1/stdlib//Corelib.Init.Logic}{\coqdocnotation{\ensuremath{\exists}}} \coqdef{DichoLemma.Check.R:50}{R}{\coqdocbinder{R}} : \coqdocnotation{\{}\coqdocnotation{set} \coqdocnotation{(}\coqdocprojection{svertex} \coqref{DichoLemma.Check.Spec.Outerplanar.G}{\coqdocvariable{G}}\coqdocnotation{)\}}\coqexternalref{::type scope:'exists' x '..' x ',' x}{http://rocq-prover.org/doc/V9.1.1/stdlib//Corelib.Init.Logic}{\coqdocnotation{,}} \coqdoceol
\coqdocindent{2.50em}
\coqref{DichoLemma.Check.R:50}{\coqdocvariable{R}} \coqexternalref{::bool scope:x 'x5Cin' x}{http://rocq-prover.org/doc/V9.1.1/stdlib//Corelib.ssr.ssrbool}{\coqdocnotation{\symbol{92}}}\coqexternalref{::bool scope:x 'x5Cin' x}{http://rocq-prover.org/doc/V9.1.1/stdlib//Corelib.ssr.ssrbool}{\coqdocnotation{in}} \coqref{DichoLemma.Check.P:45}{\coqdocvariable{P}} \coqexternalref{::type scope:x '/x5C' x}{http://rocq-prover.org/doc/V9.1.1/stdlib//Corelib.Init.Logic}{\coqdocnotation{\ensuremath{\land}}} \coqref{DichoLemma.Check.R:50}{\coqdocvariable{R}} \coqdocnotation{:\&:} \coqref{DichoLemma.Check.S:46}{\coqdocvariable{S}} \coqexternalref{::type scope:x '=' x}{http://rocq-prover.org/doc/V9.1.1/stdlib//Corelib.Init.Logic}{\coqdocnotation{=}} \coqdocnotation{[}\coqdocnotation{set} \coqdocabbreviation{val} \coqref{DichoLemma.Check.r:49}{\coqdocvariable{r}}\coqdocnotation{]}\coqdoceol
\coqdocindent{2.50em}
\coqexternalref{::type scope:x '/x5C' x}{http://rocq-prover.org/doc/V9.1.1/stdlib//Corelib.Init.Logic}{\coqdocnotation{\ensuremath{\land}}} \coqexternalref{::type scope:'exists' x '..' x ',' x}{http://rocq-prover.org/doc/V9.1.1/stdlib//Corelib.Init.Logic}{\coqdocnotation{\ensuremath{\exists}}} \coqdef{DichoLemma.Check.p:51}{p}{\coqdocbinder{p}} : \coqdocabbreviation{seq} (\coqdocprojection{svertex} \coqref{DichoLemma.Check.H:48}{\coqdocvariable{H}})\coqexternalref{::type scope:'exists' x '..' x ',' x}{http://rocq-prover.org/doc/V9.1.1/stdlib//Corelib.Init.Logic}{\coqdocnotation{,}} \coqdoceol
\coqdocindent{4.00em}
\coqdockw{let} \coqdef{DichoLemma.Check.f:53}{f}{\coqdocbinder{f}} : \coqdocnotation{\{}\coqdocnotation{ffun} \coqdocprojection{svertex} \coqref{DichoLemma.Check.H:48}{\coqdocvariable{H}} \coqexternalref{::type scope:x '->' x}{http://rocq-prover.org/doc/V9.1.1/stdlib//Corelib.Init.Logic}{\coqdocnotation{\ensuremath{\rightarrow}}} \coqdocprojection{svertex} \coqref{DichoLemma.Check.H:48}{\coqdocvariable{H}}\coqdocnotation{\}} := \coqdoceol
\coqdocindent{6.00em}
\coqdocnotation{[}\coqdocnotation{ffun} \coqdef{DichoLemma.Check.v:52}{v}{\coqdocbinder{v}} \coqdocnotation{\ensuremath{\Rightarrow}} \coqexternalref{::boolean if scope:'if' x 'then' x 'else' x}{http://rocq-prover.org/doc/V9.1.1/stdlib//Corelib.ssr.ssreflect}{\coqdocnotation{if}} \coqref{DichoLemma.Check.v:52}{\coqdocvariable{v}} \coqdocnotation{==} \coqref{DichoLemma.Check.r:49}{\coqdocvariable{r}} \coqexternalref{::boolean if scope:'if' x 'then' x 'else' x}{http://rocq-prover.org/doc/V9.1.1/stdlib//Corelib.ssr.ssreflect}{\coqdocnotation{then}} \coqref{DichoLemma.Check.v:52}{\coqdocvariable{v}} \coqexternalref{::boolean if scope:'if' x 'then' x 'else' x}{http://rocq-prover.org/doc/V9.1.1/stdlib//Corelib.ssr.ssreflect}{\coqdocnotation{else}} \coqdocdefinition{nth} \coqref{DichoLemma.Check.v:52}{\coqdocvariable{v}} \coqref{DichoLemma.Check.p:51}{\coqdocvariable{p}} (\coqdocdefinition{size} \coqref{DichoLemma.Check.p:51}{\coqdocvariable{p}} \coqdocnotation{-} \coqdocdefinition{index} \coqref{DichoLemma.Check.v:52}{\coqdocvariable{v}} \coqref{DichoLemma.Check.p:51}{\coqdocvariable{p}})\coqdocnotation{]} \coqdoctac{in}\coqdoceol
\coqdocindent{4.00em}
\coqdocdefinition{head} \coqref{DichoLemma.Check.r:49}{\coqdocvariable{r}} \coqref{DichoLemma.Check.p:51}{\coqdocvariable{p}} \coqexternalref{::type scope:x '=' x}{http://rocq-prover.org/doc/V9.1.1/stdlib//Corelib.Init.Logic}{\coqdocnotation{=}} \coqref{DichoLemma.Check.r:49}{\coqdocvariable{r}} \coqexternalref{::type scope:x '/x5C' x}{http://rocq-prover.org/doc/V9.1.1/stdlib//Corelib.Init.Logic}{\coqdocnotation{\ensuremath{\land}}} @\coqref{DichoLemma.Check.Spec.hamiltonian cycle}{\coqdocdefinition{hamiltonian\_cycle}} \coqref{DichoLemma.Check.H:48}{\coqdocvariable{H}} \coqref{DichoLemma.Check.p:51}{\coqdocvariable{p}}\coqdoceol
\coqdocindent{4.00em}
\coqexternalref{::type scope:x '/x5C' x}{http://rocq-prover.org/doc/V9.1.1/stdlib//Corelib.Init.Logic}{\coqdocnotation{\ensuremath{\land}}} \coqexternalref{::type scope:'x7B' 'mono' x ':' x x '/' x 'x7D'}{http://rocq-prover.org/doc/V9.1.1/stdlib//Corelib.ssr.ssrfun}{\coqdocnotation{\{}}\coqexternalref{::type scope:'x7B' 'mono' x ':' x x '/' x 'x7D'}{http://rocq-prover.org/doc/V9.1.1/stdlib//Corelib.ssr.ssrfun}{\coqdocnotation{mono}} \coqref{DichoLemma.Check.f:53}{\coqdocvariable{f}} \coqexternalref{::type scope:'x7B' 'mono' x ':' x x '/' x 'x7D'}{http://rocq-prover.org/doc/V9.1.1/stdlib//Corelib.ssr.ssrfun}{\coqdocnotation{:}} \coqdef{DichoLemma.Check.x:54}{x}{\coqdocbinder{x}} \coqdef{DichoLemma.Check.y:55}{y}{\coqdocbinder{y}} \coqexternalref{::type scope:'x7B' 'mono' x ':' x x '/' x 'x7D'}{http://rocq-prover.org/doc/V9.1.1/stdlib//Corelib.ssr.ssrfun}{\coqdocnotation{/}} \coqdocprojection{sedge} \coqref{DichoLemma.Check.x:54}{\coqdocvariable{x}} \coqref{DichoLemma.Check.y:55}{\coqdocvariable{y}}\coqexternalref{::type scope:'x7B' 'mono' x ':' x x '/' x 'x7D'}{http://rocq-prover.org/doc/V9.1.1/stdlib//Corelib.ssr.ssrfun}{\coqdocnotation{\}}} \coqdoceol
\coqdocindent{4.00em}
\coqexternalref{::type scope:x '/x5C' x}{http://rocq-prover.org/doc/V9.1.1/stdlib//Corelib.Init.Logic}{\coqdocnotation{\ensuremath{\land}}} \coqdockw{\ensuremath{\forall}} \coqdef{DichoLemma.Check.x:58}{x}{\coqdocbinder{x}} : \coqdocprojection{svertex} \coqref{DichoLemma.Check.H:48}{\coqdocvariable{H}}, \coqdocdefinition{pblock} \coqref{DichoLemma.Check.P:45}{\coqdocvariable{P}} (\coqdocabbreviation{val} (\coqref{DichoLemma.Check.f:53}{\coqdocvariable{f}} \coqref{DichoLemma.Check.x:58}{\coqdocvariable{x}})) \coqexternalref{::type scope:x '=' x}{http://rocq-prover.org/doc/V9.1.1/stdlib//Corelib.Init.Logic}{\coqdocnotation{=}} \coqdocdefinition{pblock} \coqref{DichoLemma.Check.P:45}{\coqdocvariable{P}} (\coqdocabbreviation{val} \coqref{DichoLemma.Check.x:58}{\coqdocvariable{x}}). \coqdocemptyline
\coqdocnoindent
\coqdockw{Definition} \coqdef{DichoLemma.Check.Spec.dichotomy lemma stmt}{dichotomy\_lemma\_stmt}{\coqdocdefinition{dichotomy\_lemma\_stmt}} :=\coqdoceol
\coqdocindent{1.00em}
\coqref{DichoLemma.Check.Spec.outerplanar}{\coqdocdefinition{outerplanar}} \coqexternalref{::type scope:x '->' x}{http://rocq-prover.org/doc/V9.1.1/stdlib//Corelib.Init.Logic}{\coqdocnotation{\ensuremath{\rightarrow}}} \coqdocdefinition{connected} \coqdocnotation{[}\coqdocnotation{set}\coqdocnotation{:} \coqdocprojection{svertex} \coqref{DichoLemma.Check.Spec.Outerplanar.G}{\coqdocvariable{G}}\coqdocnotation{]} \coqexternalref{::type scope:x '->' x}{http://rocq-prover.org/doc/V9.1.1/stdlib//Corelib.Init.Logic}{\coqdocnotation{\ensuremath{\rightarrow}}}\coqdoceol
\coqdocindent{1.00em}
\coqdockw{\ensuremath{\forall}} \coqdef{DichoLemma.Check.P:59}{P}{\coqdocbinder{P}} : \coqdocnotation{\{}\coqdocnotation{set} \coqdocnotation{\{}\coqdocnotation{set} \coqdocnotation{(}\coqdocprojection{svertex} \coqref{DichoLemma.Check.Spec.Outerplanar.G}{\coqdocvariable{G}}\coqdocnotation{)\}\}},\coqdoceol
\coqdocindent{1.00em}
\coqref{DichoLemma.Check.Spec.rooted equitable partition}{\coqdocdefinition{rooted\_equitable\_partition}} \coqref{DichoLemma.Check.P:59}{\coqdocvariable{P}} \coqexternalref{::type scope:x '->' x}{http://rocq-prover.org/doc/V9.1.1/stdlib//Corelib.Init.Logic}{\coqdocnotation{\ensuremath{\rightarrow}}}\coqdoceol
\coqdocindent{1.00em}
\coqdockw{\ensuremath{\forall}} \coqdef{DichoLemma.Check.S:60}{S}{\coqdocbinder{S}} : \coqdocnotation{\{}\coqdocnotation{set} \coqdocnotation{(}\coqdocprojection{svertex} \coqref{DichoLemma.Check.Spec.Outerplanar.G}{\coqdocvariable{G}}\coqdocnotation{)\}}, \coqdoceol
\coqdocindent{1.00em}
\coqref{DichoLemma.Check.Spec.is block}{\coqdocdefinition{is\_block}} \coqref{DichoLemma.Check.S:60}{\coqdocvariable{S}} \coqexternalref{::type scope:x '=' x}{http://rocq-prover.org/doc/V9.1.1/stdlib//Corelib.Init.Logic}{\coqdocnotation{=}} \coqexternalref{true}{http://rocq-prover.org/doc/V9.1.1/stdlib//Corelib.Init.Datatypes}{\coqdocconstructor{true}} \coqexternalref{::type scope:x '->' x}{http://rocq-prover.org/doc/V9.1.1/stdlib//Corelib.Init.Logic}{\coqdocnotation{\ensuremath{\rightarrow}}} \coqref{DichoLemma.Check.Spec.discrete block}{\coqdocdefinition{discrete\_block}} \coqref{DichoLemma.Check.P:59}{\coqdocvariable{P}} \coqref{DichoLemma.Check.S:60}{\coqdocvariable{S}} \coqexternalref{::type scope:x '=' x}{http://rocq-prover.org/doc/V9.1.1/stdlib//Corelib.Init.Logic}{\coqdocnotation{=}} \coqexternalref{true}{http://rocq-prover.org/doc/V9.1.1/stdlib//Corelib.Init.Datatypes}{\coqdocconstructor{true}} \coqexternalref{::type scope:x 'x5C/' x}{http://rocq-prover.org/doc/V9.1.1/stdlib//Corelib.Init.Logic}{\coqdocnotation{\ensuremath{\lor}}} \coqref{DichoLemma.Check.Spec.symmetric block}{\coqdocdefinition{symmetric\_block}} \coqref{DichoLemma.Check.P:59}{\coqdocvariable{P}} \coqref{DichoLemma.Check.S:60}{\coqdocvariable{S}}.\coqdoceol
\coqdocemptyline
\coqdocnoindent
\coqdockw{End} \coqref{DichoLemma.Check.Spec.Outerplanar}{\coqdocsection{Outerplanar}}.\coqdoceol
\coqdocemptyline
\end{coqdoccode}
\begin{coqdoccode}
\coqdocnoindent
\coqdockw{Definition} \coqdef{DichoLemma.Check.Spec.Target}{Target}{\coqdocdefinition{Target}} := \coqdockw{\ensuremath{\forall}} \coqdef{DichoLemma.Check.G:61}{G}{\coqdocbinder{G}} : \coqdocrecord{sgraph}, \coqref{DichoLemma.Check.Spec.dichotomy lemma stmt}{\coqdocdefinition{dichotomy\_lemma\_stmt}} \coqref{DichoLemma.Check.G:61}{\coqdocvariable{G}}.\coqdoceol
\coqdocemptyline
\coqdocnoindent
\coqdockw{End} \coqref{DichoLemma.Check.Spec}{\coqdocmodule{Spec}}.\coqdoceol
\coqdocemptyline
\end{coqdoccode}

\begin{coqdoccode}
\coqdocnoindent
\coqdockw{Require} \coqdocvar{DichoLemma.dicho.dichotomy}.\coqdoceol
\coqdocemptyline
\end{coqdoccode}
\begin{coqdoccode}
\coqdocnoindent
\coqdockw{Theorem} \coqdocvar{dichotomy\_lemma\_holds} : \coqdocvar{DichoLemma.Check.Spec.Target}.\coqdoceol
\coqdocnoindent
\coqdockw{Proof}. \coqdoctac{exact}: \coqdocvar{DichoLemma.dicho.dichotomy.dichotomy\_lemma}. \coqdockw{Qed}.\coqdoceol
\coqdocemptyline
\coqdocnoindent
\coqdockw{Print} \coqdockw{Assumptions} \coqdocvar{dichotomy\_lemma\_holds}.\coqdoceol
\end{coqdoccode}

\end{document}